\documentclass[12pt]{article}        

\usepackage{amsmath}
\usepackage{amssymb}

\usepackage{epsfig}

\usepackage{cite}

\usepackage{epic}

\begin{document}

\begin{titlepage}


\begin{flushright}
{\bf  CERN-TH/2002-059}
\end{flushright}

\vspace{6mm}
\begin{center}
{\LARGE\bf
  Foam: A General-Purpose Cellular\\
  Monte Carlo Event Generator$^{\star}$
}
\end{center}

\vspace{1mm}
\begin{center}
  \large\bf S. Jadach
\end{center}

\vspace{1mm}
\begin{center}
  {\em Institute of Nuclear Physics,
    ul. Kawiory 26a, 30-055 Cracow, Poland}\\
  { and}\\
  {\em CERN Theory Division, CH-1211 Geneva 23, Switzerland}
\end{center}

\vspace{5mm}
\begin{abstract}
A general-purpose, self-adapting Monte Carlo (MC) event generator (simulator)
{\tt Foam} is described.
The high efficiency of the MC, that is small maximum weight or variance
of the MC weight is achieved by means
of dividing the integration domain into small cells.
The cells can be $n$-dimensional simplices,
hyperrectangles or a Cartesian product of them.
The grid of cells, called ``foam'', is produced in the process of the binary
split of the cells.
The choice of the next cell to be divided and the position/direction
of the division hyperplane
is driven by the algorithm which optimizes
the ratio of the maximum weight to the average weight or (optionally) the total variance.
The algorithm is able to deal, in principle, with an arbitrary pattern
of the singularities in the distribution.
As any MC generator, {\tt Foam}  can also be used for the MC integration.
With the typical personal computer CPU, the program
is able to perform adaptive integration/simulation
of a relatively small number of dimensions ($\leq 16$).
With the continuing progress in CPU power, this limit will inevitably get shifted
to ever higher dimensions.
{\tt Foam} program is aimed (and already tested) as a component of the MC event generators
for the high energy physics experiments.
A few simple examples of the related applications are presented.
{\tt Foam} code is written in fully object-oriented style, in the C++ language.
Two other versions with a slightly limited functionality,
are available in the Fortran77 language.
The source codes are available from {http://jadach.home.cern.ch/jadach/}.
\end{abstract}

\vspace{4mm}
\begin{center}
  {\em Submitted to Comput. Phys. Commun.}
\end{center}

\vspace{4mm}
\footnoterule
\noindent
{\footnotesize
\begin{itemize}
\item[${\star}$]
Work supported in part by the European Community's Human Potential
Programme under contract HPRN-CT-2000-00149 ``Physics at Colliders'',
by Polish Government grant
KBN 5P03B09320, 
and by NATO grant PST.CLG.977751.
\end{itemize}
}

\vspace{1mm}
\begin{flushleft}
{\bf 
  CERN-TH/2002-059\\
  March 2002}
\end{flushleft}

\end{titlepage}

\tableofcontents
\newpage

\noindent{\bf PROGRAM SUMMARY}
\vspace{10pt}

\noindent{\sl Title of the program:}\\
 {\tt Foam}, version 2.05.

\noindent{\sl Computer:}\\
any computer with C++ or Fortran 77 compilers and UNIX operating system

\noindent{\sl Operating system:}\\
UNIX; program was tested under Linux 6.x.

\noindent{\sl Programming languages used:}\\
ANSI C++ and FORTRAN 77 with popular extensions such as long names, long code lines, etc.

\noindent{\sl High-speed storage required:}\\
  $<$ 50 MB

\noindent{\sl No. of lines in combined program and test deck:}\\
4235 lines of C++ code and 9826 lines of F77 code.


\noindent{\sl Nature of the physical problem:}\\
Monte Carlo simulation or generation of unweighted (weight equal 1) events
is a standard problem in many areas of research.
It is highly desirable to have in the program library a general-purpose numerical tool (program)
with a MC generation algorithm featuring built-in capability of adjusting
automatically the generation procedure to an arbitrary
pattern of singularities in the probability distribution.
Our primary goal is simulation of the differential distribution in the
multiparticle Lorentz invariant phase space for the purpose of comparison between
Quantum Field Theory prediction and experiments in the high energy experiments.

\noindent{\sl Method of solution:}\\
In the algorithm, a grid (foam) of cells is built in the process of the binary split of the cells.
The resulting foam is adapted automatically to the shape 
of the integrand in such a way that the resulting ratio
of average weight to maximum weight or variance to average weight is arbitrarily good.
The above algorithm, is a substantial improvement on the previous version in Ref.~[1].
The division of the cell is improved and, in addition to cells of a simplical shape of Ref.~[1],
a hyperrectangular cells are also used.

\noindent{\sl Restrictions on the complexity of the problem:}\\
The program is memory-hungry and therefore
limited, at present, to relatively small dimensions $\leq 16$.
(In {\tt Foam} 1.x of Ref.~[1] the dimension was limited to $n\leq 6$.)

\noindent{\sl Typical running time:}\\
The CPU time necessary to build up a foam of cells depends strongly on the number
of dimensions and the requested size of the grid.
On the PC with a 550~MHz Intel chip,
it takes about 30 seconds
to build a hyperrectangular grid of 10000 cells
for a simple 3-dimensional distribution.

\vspace{1mm}
\noindent
[1] S.~Jadach, Comput. Phys. Commun. {\bf 130},  244  (2000).

\newpage
\section{Introduction}
This work describe a new version of an algorithm
for producing random points according to an arbitrary, user-defined distribution
in the $n$-dimensional space --
much improved with respect to the original version of Ref.~\cite{foam1:2000}.
The new implementation is realized in the C++ programming language
in a fully object-oriented manner%
\footnote{ The early C++ version of {\tt Foam}
  was coded by M. Ciesla and M. Slusarczyk~\cite{CieslaSlusarczyk}.
  It was a translation of a version 1.x from Fortran77 to C++.
  Analogous translation to the JAVA language was also done.}.
Since the changes, with respect to Ref.~\cite{foam1:2000},
both in the algorithm and in the implementation
are quite essential,
a complete description of the method and the new code is provided,
instead of only an update of Ref.~\cite{foam1:2000}.

For the problem of function minimization, integration (quadrature),
there are plenty of {\em general-purpose} programs, which can be applied
to an arbitrary user-defined function.
``General-purpose'' means that all these tools work, in principle,
for a very wide range of user-defined functions.
For the multidimensional Monte Carlo simulation problem,
that is for the problem of randomly generating points according to a given
$n$-dimensional distribution, there is precious few
examples of the {\em General-Purpose Monte Carlo Simulators} (GPMCS),
that is programs that work (in principle) for arbitrary 
distribution~\cite{Lepage:1978sw,Kawabata:1995th,Ohl:1998jn,Manankova:1995xe}.
As in these works,
here we are concerned mainly with the MC applications to high energy physics,
that is to simulation of the differential distributions in the 
multi-particle Lorentz invariant phase space
provided the Quantum Field Theory,
see also classic Ref.~\cite{James:1980} on this subject.
An example of the work on GPMCS applied to other fields see the interesting works%
\footnote{In these works there is far more emphasis on the parallel computing aspects
  of the integration (not simulation) than on the cell geometry, as compared with our work.}
of Refs.~\cite{Doncker:1998,Doncker:1999,Doncker:parint1}.

Let us also note that the two-dimensional cellular MC sampler VESKO2
with the primitive binary split was already included in the program LESKO-F
of Ref.~\cite{Jadach:1992ty} long time ago.
Very similar programs were also described in Refs.~\cite{dice:1992,dice:1996}%
\footnote{We thank S. Kawabata for drawing our attention to these works.}.
Still another very interesing class of GPMCSs for the high energy physics,
based of the Metropolis algorith~\cite{Metropolis:1953am},
is described in Ref.~\cite{moretti:2000}. 

Two essential reasons for a realtive scarcity of GPMCSs,
as compared to other programming tools,
are the lack of novel ideas for an efficient algorithm
and the need of much CPU power and memory -- only recently available or affordable.

GPMCS is essentially a random-number generator of
points in multidimensional space with a non-uniform 
user-defined probability distribution.
Inevitably the GPMCS works in two stages: {\em exploration} and {\em generation}%
\footnote{Exploration and generation could be done simultaneously,
  at the expense of complications in the algorithm and the code.}.
During an {\em exploration} phase, the GPMCS is ``digesting''
the entire shape of the $n$-dimensional distribution $\rho(\vec x)$ to be generated,
memorizing it as efficiently as possible,
using all available CPU processing power and memory%
\footnote{The procedure of memorizing multidimensional distribution $\rho(\vec x)\geq 0$ is a kind
  of interpolation, in which the grid of cells is denser in places where the distribution peaks
  and/or varies strongly.}.
In {\tt Foam}, the exploration phase is the phase
of the {\em build-up of the system of cells}
covering entirely the integration space, which will be
called ``foam'', produced in the process of the binary split of the cells.
In the {\em generation} phase, GPMCS provides
a method of the MC generation of the random points $\vec x$
{\em exactly} according to $\rho(\vec x)$.
The vector $x=\vec x=(x_1,x_2,\dots,x_n)$ will also be called in the following
a {\em Monte Carlo event}.
In {\tt Foam}, the MC generation is very simple:
a cell is chosen randomly, and next, a point is generated within the cell with
uniform distribution; see below for more details.
The value of the integrand is already estimated in the exploration;
it can be calculated with an arbitrary precision in the generation phase.

During the exploration {\tt Foam} constructs a distribution $\rho'(x)$,
which is uniform within each cell, and is used for the MC generation.
Events are weighted with the weight $w=\rho/\rho'$.
The quality of the distribution of this weight,
measured in terms of the weight {\em distribution parameters}, such as
variance and ratio of maximum to average, 
is determined by the quality of the exploration.
The basic principle of the {\tt Foam} algorithm is that the parameters
of the anticipated ``target weight distribution'' in the MC generation phase are
used as a driving force guiding the cell build-up (exploration).
In the case of a successful exploration,
weighted MC events can be turned efficiently into unweighted ones
with the usual rejection method, that is with a small rejection rate.

Since the exploration phase may be CPU-time consuming,
it is natural to expect that GPMCS has a built-in mechanism of {\em persistency},
i.e., there is a mechanism of writing into a mass storage (computer disk)
the whole information on the memorized shape of the distribution
obtained from the exploration phase,
such that the generation of the MC events can be (re)started at any later time,
without any need of repeating the time-consuming exploration.
One small step further is to require that the generation of events with GPMCS
can be stopped at any time, the entire status of the GPMCS can be written on the disk,
and the generation of the next event can be resumed at any later time
upon reading the stored information;
the next generated event will be such as if there had been no break in the generation process.
In fact, this is what we shall really mean in the following as a {\em persistency}
mechanism for GPMCS, and what is actually implemented in the {\tt Foam}.
There, the persistency is realized using the ROOT%
\footnote{
  The use of ROOT is optional in {\tt Foam}. 
  However, version of {\tt Foam} without ROOT does not feature any kind of persistency.}
package~\cite{root}.

The GPMCS programs will always be limited to ``small dimensions''.
With currently available computers, ``small'' means in practice $n\leq 10$,
up to $n\leq 16$ for ``mildly'' singular distributions.
This is already quite satisfactory, especially if we remember that this limit will pushed
higher, as the available hardware gets more powerful,
without any need of modifying the existing code%
\footnote{The present implementation of {\tt Foam} is 
  fully based on the dynamic allocation of the memory
  and the space dimension is a user-defined parameter.}.
Twelve years from now, with portable computer featuring a 100~GHz processor and 1~TByte disk
the same version of {\tt Foam} will work efficiently for even higher dimensions.

{\tt Foam} has been developed having in mind that it will
be used as part of a bigger MC program, typically, 
to generate a subset of variables in which a model distribution is the most singular
(has strong peaks).
This is why we are not so much concerned by the fact that the cellular algorithm
of {\tt Foam} is inefficient for, say, 150 variables.
The user is supposed to select $n\leq 16$ 
``wild variables''~\cite{Kawabata:1995th} and apply {\tt Foam} to them.
For the remaining ``mild variables'', {\tt Foam} may merely serve as a provider of the uniform
random numbers, if the user of {\tt Foam} wishes to exploit that option.
On the other hand, for smaller MC problems, {\tt Foam} may
play the role of a ``stand-alone MC generator'' or ``stand-alone MC integrator''.
Also, from the following description of the various modes of the use of {\tt Foam}
it will be clear that the subprogram providing the model distribution 
to {\tt Foam} can have quite a complicated structure.
Nevertheless, this user-provided part of the program will be smaller
than a solution without {\tt Foam}, because
{\tt Foam} provides for essential functionalities concerning the optimization
of the MC weight distribution.
This remark is especially true for the case of implementation of multibranching
with the help of {\tt Foam}.

It is worth mentioning that {\tt Foam} is not based
on the ``principle of factorizability'' of the integrand distribution,
$\rho(\vec x)=\prod_1^n \rho_i(x_i)$,
on which {\tt VEGAS}-family programs are 
built~\cite{Lepage:1978sw,Kawabata:1995th,Ohl:1998jn}.

The outline of the paper is the following: 
in Section~2 we describe the cellular {\tt Foam} algorithm, delegating
the description of the cell division procedure to Section~3.
Section~4 is devoted to the description of the {\tt Foam} code in C++.
The use of {\tt Foam} is described in Section~5 and examples of the numerical
results (MC efficiency) are given in Section~6.
Conclusions and Appendices on the variance minimization
finalize the paper.

\begin{figure}[!ht]
\begin{center}
{\epsfig{file=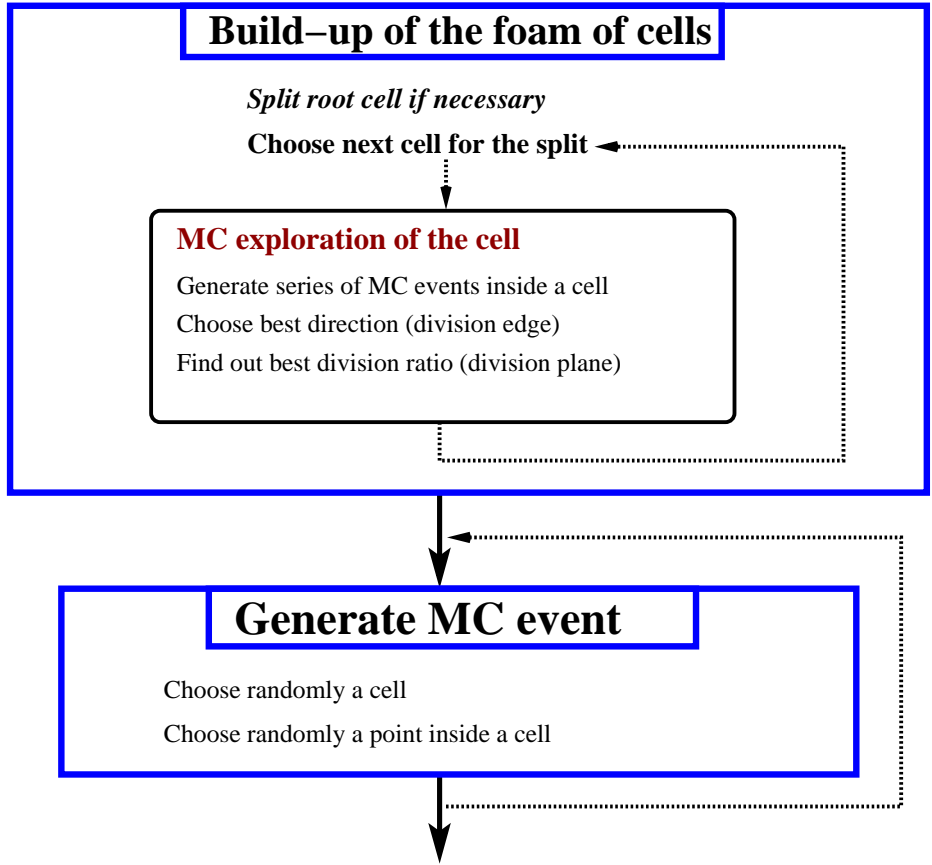,height=150mm,angle=270}}
\end{center}
\caption{\small\sf
  Two stages in the cellular algorithm of {\tt Foam}.
}
\label{fig:stages}
\end{figure}

\section{The {\tt Foam} algorithm}
As already mentioned, the execution of the {\tt Foam} algorithm is clearly separated into
the first stage of the ``distribution exploration'' consisting of the
``build-up of the foam of cells'', which in a sense memorizes
the $n$-dimensional shape of the distribution,
and the second stage of the actual ``MC generation'',
see Fig.~\ref{fig:stages}.
The most essential part of the present {\tt Foam} algorithm is the procedure of the binary
split of the cell, in which it is decided which cell is picked up for the next split
and the necessary parameters of the geometry of the cell split are determined.
This part of the {\tt Foam} algorithm description is delegated to the next section.
In the present section we describe other, more general, aspects of the {\tt Foam} algorithm.

\subsection{Cellular exploration of the distribution}

The most obvious method to minimize
the variance or maximum weight of the target weight distribution in the MC generation,
already considered some 40 years ago,
is to split the integration domain into many cells, 
so that the distribution $\rho(\vec x)$ is approximated by
$\rho'(\vec x)$, which is constant within each cell%
\footnote{
  For the MC simulation, our main aim, a more sophisticated interpolation 
  of $\rho(\vec x)$ within a cell does not seem
  to be worth the effort -- 
  it would be interesting if our main aim was the integration of $\rho(\vec x)$.}.
This is a {\em cellular class} of the general-purpose MC algorithms%
\footnote{The term ``stratified sampling'', 
  used in the literature, has in our opinion a narrower meaning than ``cellular class''.}.

The immediate questions are:
What kind or shapes of the cells to use and
how to cover the integration domain with cells?
The reader may find in Ref.~\cite{Manankova:1995xe}
an example of a rather general discussion of these questions.
In the {\tt Foam} program the user may opt for one of the three geometries of the cells:
(1) simplices, (2) hyperrectangles
and (3) Cartesian products of simplices and hyperrectangles.
For these particular types of cells there exists
an efficient method of parametrizing them in the computer memory
and handling their geometry.

The system of many cells can be created and reorganized all at once,
as in {\tt VEGAS}-type programs~\cite{Lepage:1978sw,Kawabata:1995th,Ohl:1998jn},
or in a more evolutionary way, as the cell split process of this work.
In the {\tt Foam} algorithm we rely on the {\em binary split} of cells.
Starting from the entire integration domain (unit hyperrectangle or simplex)
cells are split into two daughter cells, step by step,
until the user-defined memory limit is reached.
The choice of next cell to be split and the geometry of the split
in the exploration phase
are driven by the ``target weight distribution'' of the generation process,
see Section~\ref{sec:cell_split_geom}.
The important advantage of any cell split algorithm
is that it assures automatically the {\em full coverage} of the integration domain --
simply because the primary {\em root cell} is identical to the entire integration domain
and the two daughter cells always cover the parent cell entirely.
The problem of {\em blind spots} discussed in Ref.~\cite{Manankova:1995xe}
is avoided by construction.

In the early version of {\tt Foam}~\cite{foam1:2000},
there was a possibility in the algorithm that the ``unsuccessful'' branch
in the tree of all cells can be erased and rebuild.
This was called ``collapse'' and ``rebuild''.
In the present version this option was removed%
\footnote{A ``flush method'', which erases the entire foam of cells from the computer memory
  and allows for its reinitialization is, however, available.},
because the experience
with many testing functions has shown that the algorithm of the cell
build-up is rather ``deterministic'' and the ``rebuild'' procedure was usually
leading to a new branch of foam with about the same features as the old one.

Let us finally remark that the version of the cellular algorithm
presented in this paper is, in fact, result of many experiments
with different variants of the cellular algorithm.
The presented version is the best one out of several development versions.
In the code one may still see some ``hooks'' and unused features
(class members or methods) related to these alternative variants.
We have left them just in the case that some new idea of improving the algorithm
emerges, or for certain kinds of debugging/testing.

\subsection{Variance reduction versus maximum weight reduction}

In the construction of the {\tt Foam} algorithm, most effort was invested into
a minimization of the ratio of the maximum weight to the average weight
$w_{\max}/\langle w \rangle$.
This parameter is essential, if we want to transform
variable-weight events into $w=1$ events, at the latter stage of the MC generation%
\footnote{We provide optionally in {\tt Foam} for the rejection leading to $w=1$ events.}.

Minimizing the maximum weight $w_{\max}$
is not the same as minimizing the variance
$\sigma=\sqrt{ \langle w^2 \rangle -\langle w \rangle^2 }$.
Minimizing $w_{\max}$ may be more difficult --
but it is worth an effort because {\tt Foam} is really meant
to be a part of a bigger MC program, where it is usually essential
that the ``inner part'' of the program provides events with an excellent
weight distribution, or even $w=1$ events.
Nevertheless, minimizing the variance is also implemented in {\tt Foam}
and available optionally.
It can be useful if one is satisfied with the variable-weight events,
and/or if the main aim is the evaluation of the integral and not the MC simulation.

\begin{figure}[!ht]
\centering
\setlength{\unitlength}{0.1mm}
\begin{picture}(1600,1100)
\put( 450,950){\makebox(0,0)[b]{\large (a)}}
\put(1000,950){\makebox(0,0)[b]{\large (b)}}
\put(1550,950){\makebox(0,0)[b]{\large (c)}}
\put(    0, 550){\makebox(0,0)[lb]{\epsfig{file=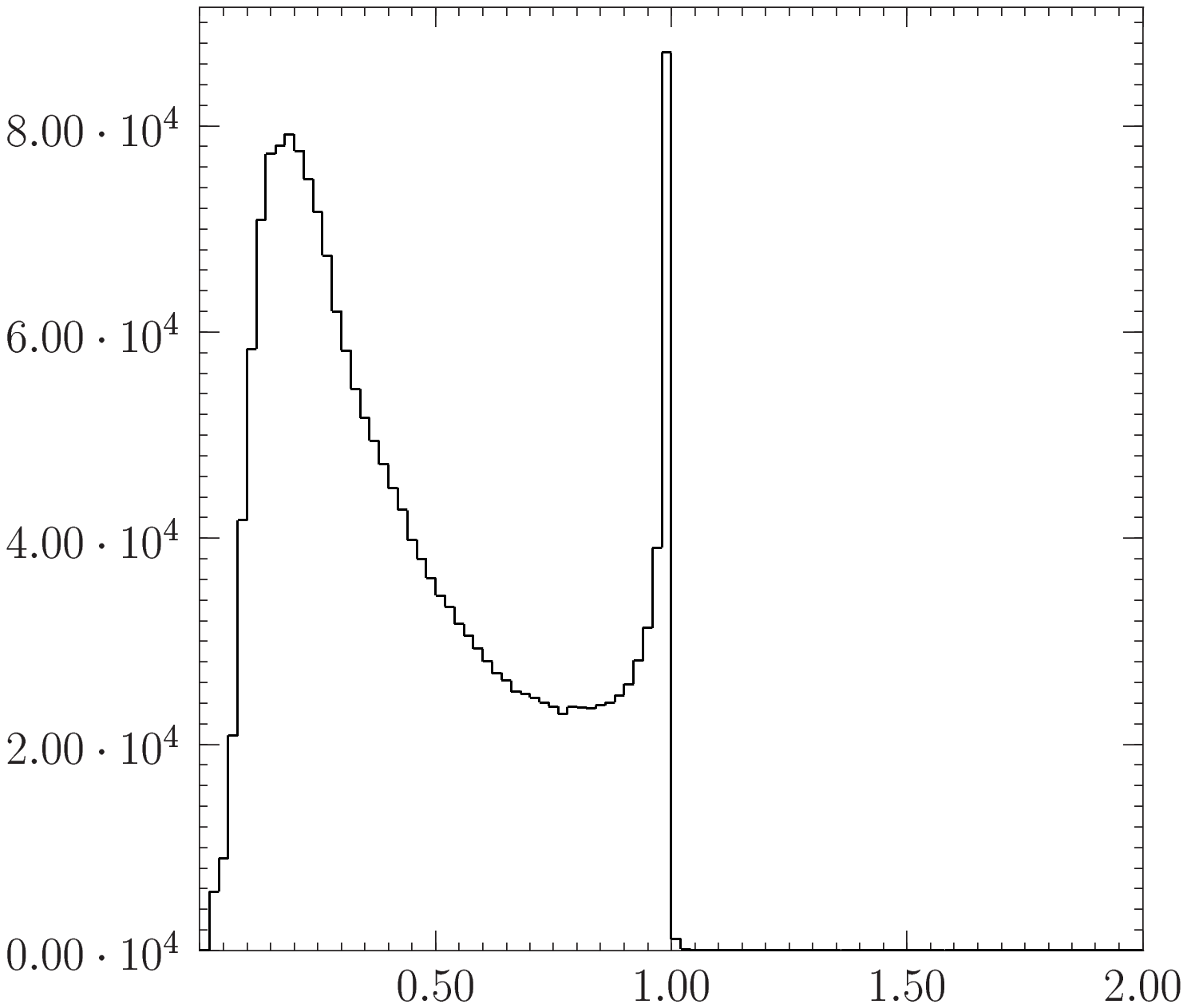,width=55mm}}}
\put(  550, 550){\makebox(0,0)[lb]{\epsfig{file=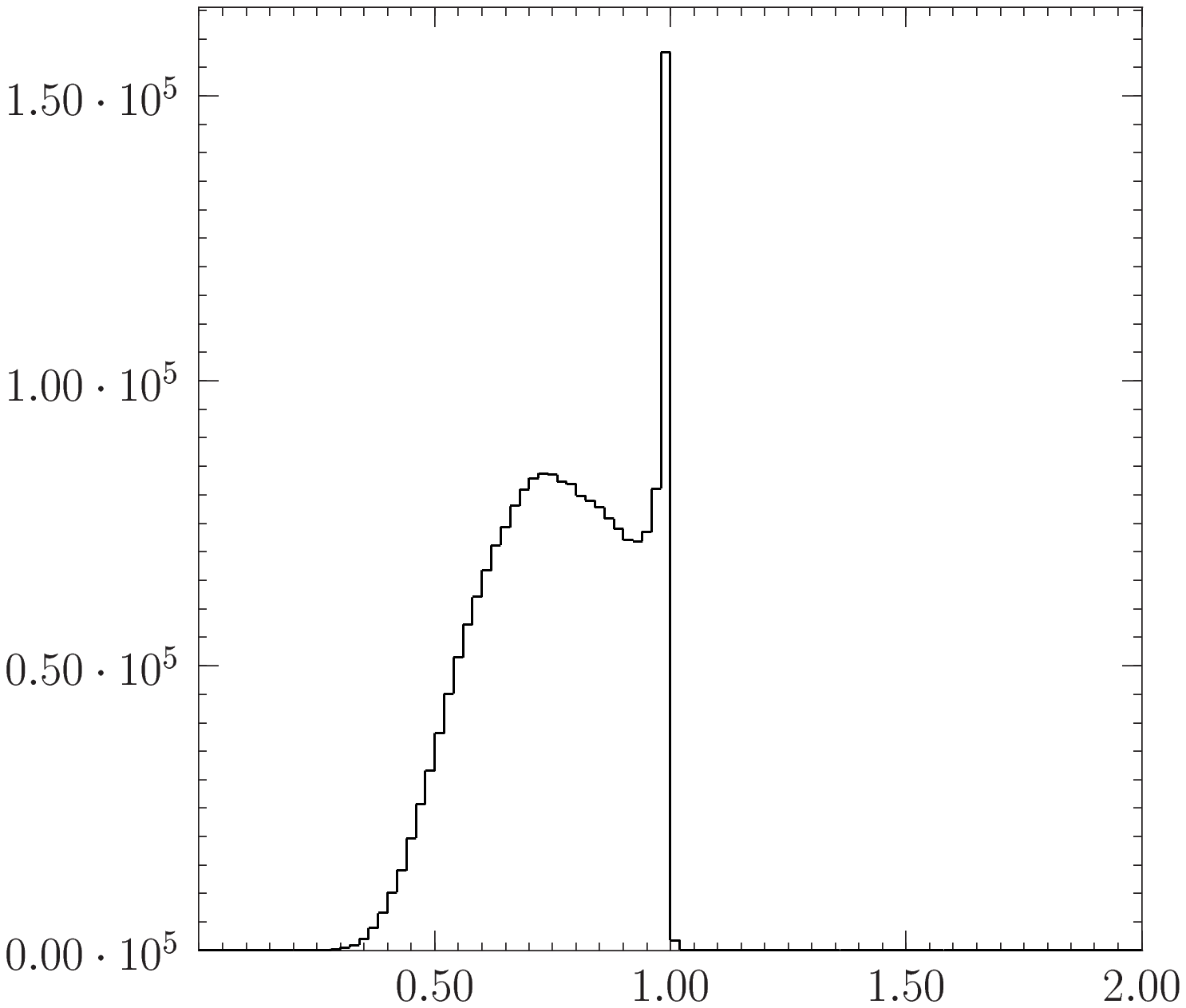,width=55mm}}}
\put( 1100, 550){\makebox(0,0)[lb]{\epsfig{file=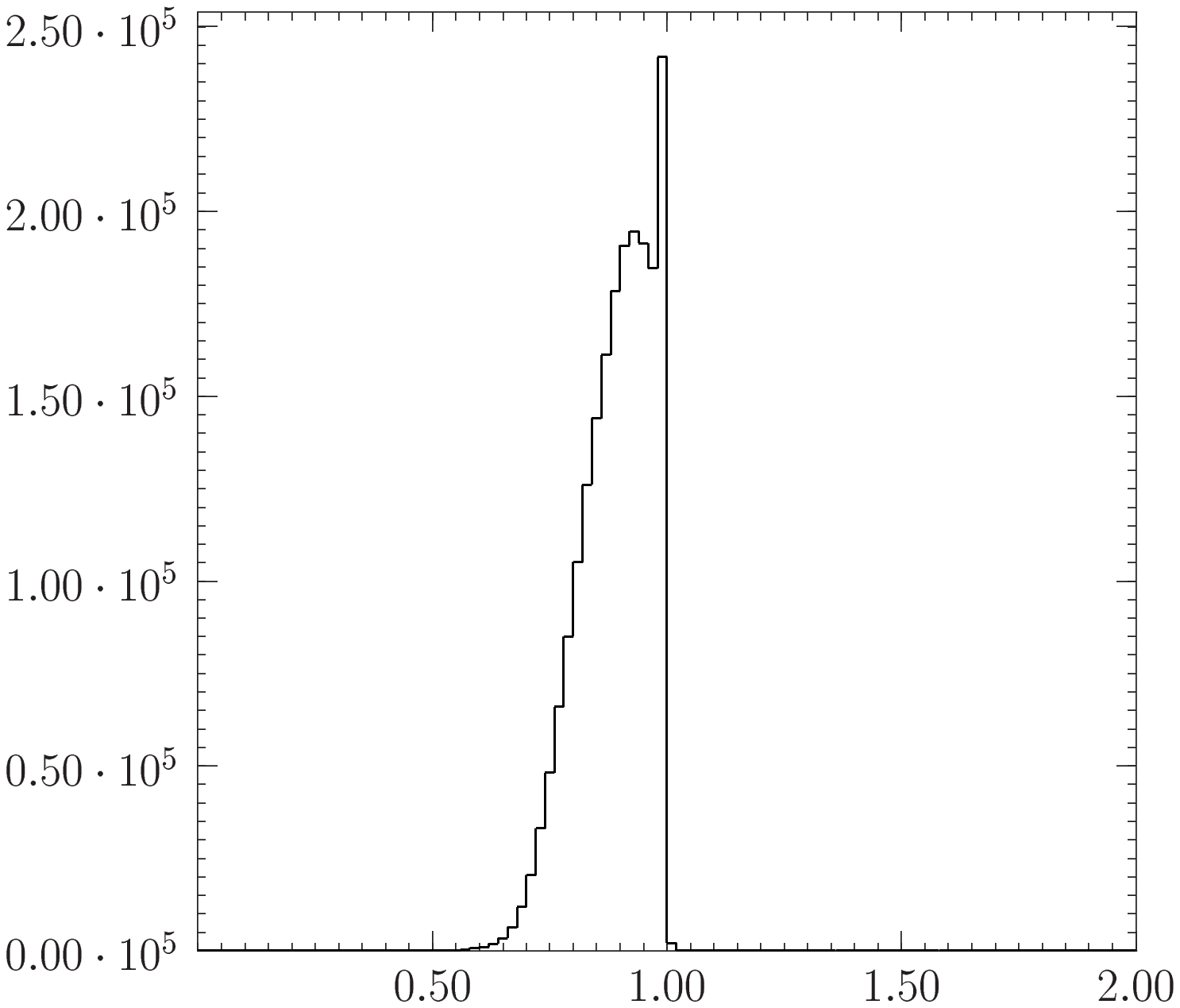,width=55mm}}}
\put( 450,400){\makebox(0,0)[b]{\large (d)}}
\put(1000,400){\makebox(0,0)[b]{\large (e)}}
\put(1550,400){\makebox(0,0)[b]{\large (f)}}
\put(    0,   0){\makebox(0,0)[lb]{\epsfig{file=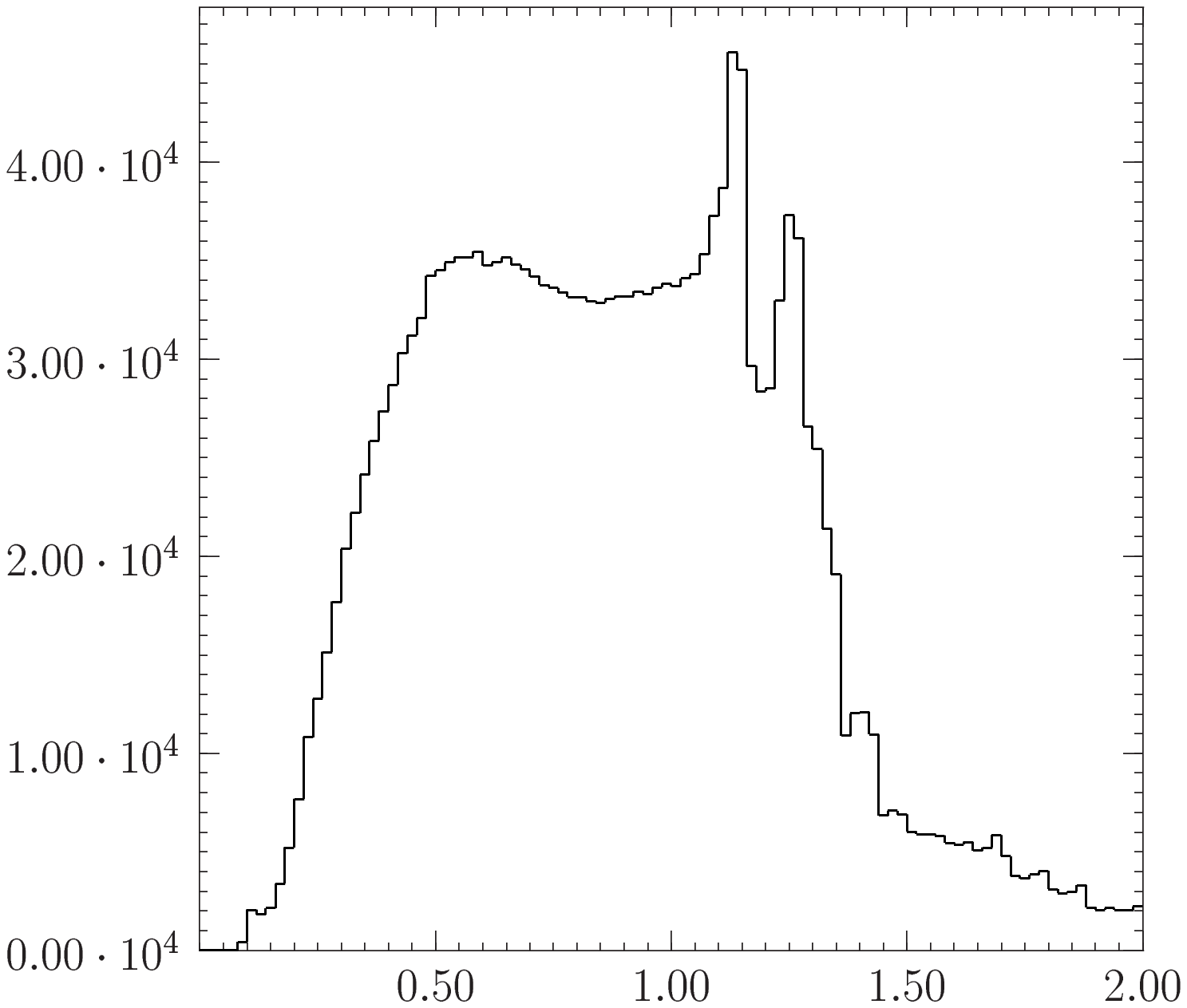,width=55mm}}}
\put(  550,   0){\makebox(0,0)[lb]{\epsfig{file=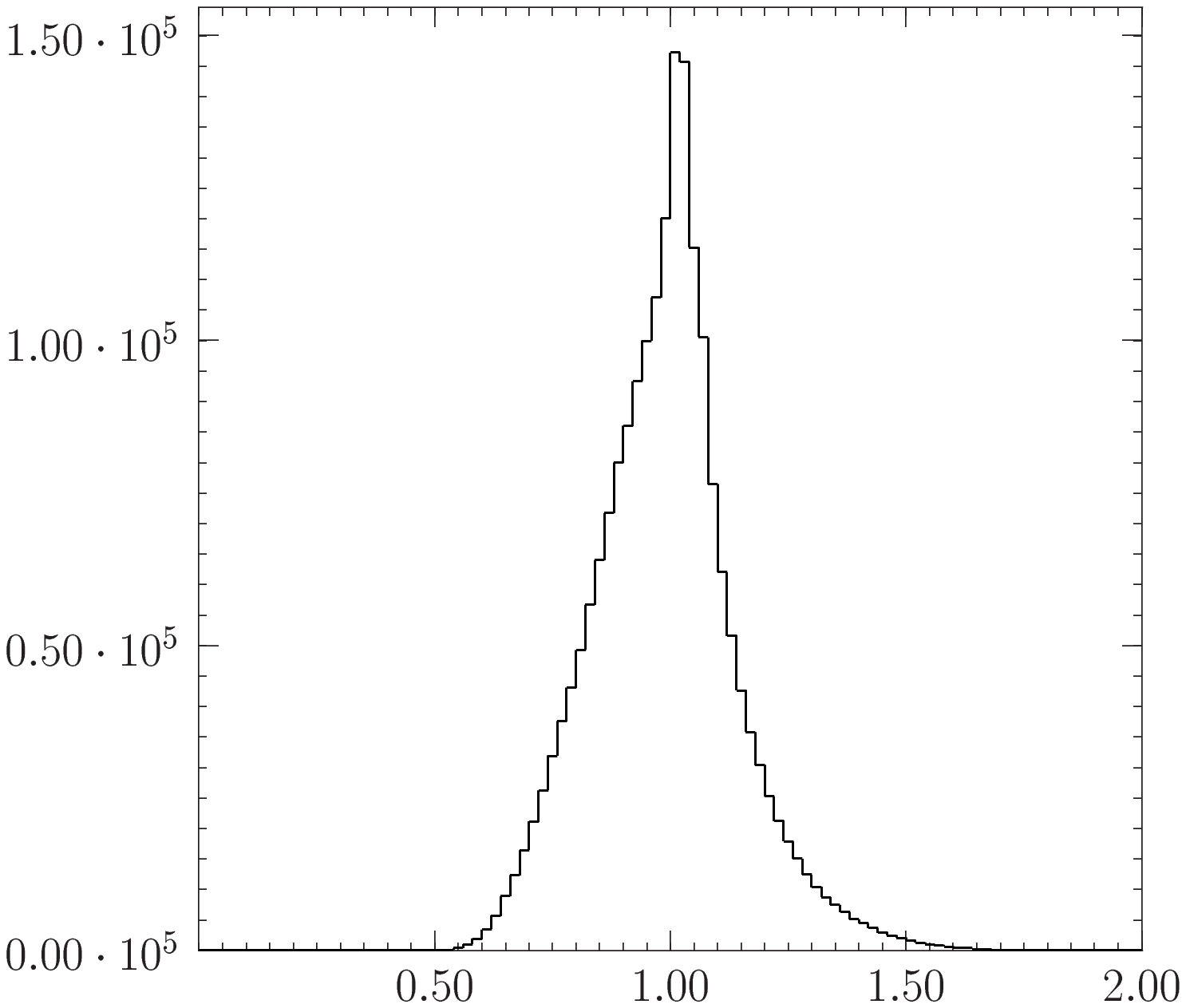,width=55mm}}}
\put( 1100,   0){\makebox(0,0)[lb]{\epsfig{file=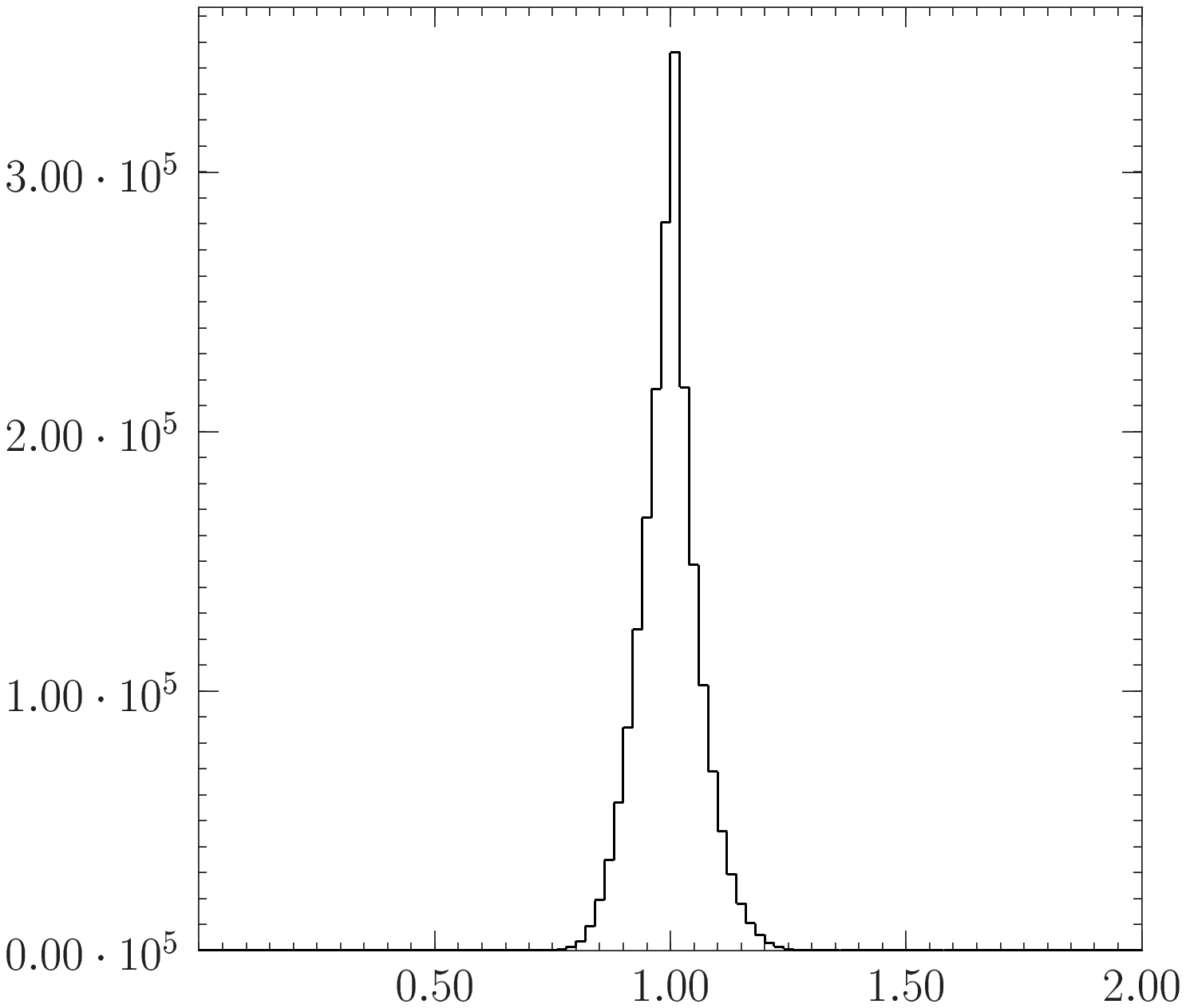,width=55mm}}}
\end{picture}
\caption{\small\sf
  Weight distribution of the {\tt Foam} for the default option with the maximum weight optimization
  (a-c) compared to analogous distributions obtained for an option
  with the variance optimization (d-f).
  Number of cells is 200, 2000 and 20000 for
  (a-c) and (d-f), correspondingly.
}
\label{fig:wtdist1}
\end{figure}

The difference between the above two options is well illustrated in Fig.~\ref{fig:wtdist1},
which shows two examples of the evolution of the MC weight distribution
due to a gradual increase of the number of cells.
For the default configuration, {\tt Foam} optimises the ratio $w_{\max}/\langle w \rangle$.
This case is shown in plots (a--c) in Fig.~\ref{fig:wtdist1}.
Here, the weight distribution features a sharper and sharper drop of the weight distribution
at $w=1$, while increasing the number of cells.
Also, the average weight increases gradually and the weight distribution gets narrower.
The optional case of the optimization of $\sigma/\langle w \rangle$
is shown in plots (d--f) of Fig.~\ref{fig:wtdist1}.
In this case the variance is decreasing with the growing numbers of cells.
On the other hand, the maximum weight is noticably higher than before.
All weight distributions were obtained for the same 2-dimensional
testing function $\rho_b(x)$, used also in Section~\ref{sec:NumeResults}.

\subsection{Hyperrectangles or simplices?}
\label{sec:hr_vs_simp}
In Ref.~\cite{foam1:2000} simplical cells have been chosen
instead of simpler hyperrectangles, mainly because
of the author's ``prejudice'' that simplices
may adapt more efficiently to complicated singularities
in the distribution $\rho(x)$ spanned along subspaces,
not necessarily parallel to the axes of the global reference frame.
Hyperrectangles tend to remember the orientation of the parent hyperrectangle,
while simplices feature, in principle,  a kind of ``angular mobility'',
i.e. they can forget the orientation of great-grand-parents, and adapt to the orientation
of the singularity in $\rho(x)$.
However, an experience with tens of testing functions 
has shown that in many cases hyperrectangles provide the same or even
better final MC efficiency than simplices, for the same number of cells.
Moreover, simplices have certain additional disadvantages.
At present, {\tt Foam} with simplices is practically limited to rather low dimensions $n\leq 5$,
because in most cases the entire integration domain is a unit hyperrectangle,
which has to be divided into $n!$ simplices,
where $n!$ quickly becomes a large number%
\footnote{
  Mapping of the hyperrectangle into a simplex is possible,
  however, one should use transformation with $|\partial x / \partial y| \ne 0$,
  in order to avoid producing nasty singularities located
  at the vertices, edges and walls of the simplex.}.
This limitation is, of course, not valid, if the entire integration domain is actually
a simplex of high dimensionality instead of a hyperrectangle.
({\tt Foam} can be configured to start cell evolution from a simplex or Cartesian product of
a simplex and a hyperrectangle.)
Furthermore,
geometry manipulations in the simplical case require calculation of many determinants --
this slows down the program execution at higher dimensions.
In addition, in the present implementation, the memory consumption in
a simplical foam build-up is $\sim 16\times n$~Bytes/cell,
while for hyperrectangles we have found a method 
that limits memory consumption  to below $\sim 80$~Bytes/cell
{\em independently} of $n$, see Section~\ref{sec:save_memory}.
We can therefore reach easily the level of $10^6$ hyperrectangular cells
at any dimension (in practice $n\leq 16$)
and about $50,000$ simplical cells, for $n\leq 5$.
As we see, hyperrectangular foam seems to win on many fronts.
Nevertheless, we keep simplical foam as an option, because
in certain applications one will definitely encounter
distributions for which it turns out to be more efficient to use simplices,
in spite of all their limitations, at least for a subset of the integration
variables.

\subsection{Build up of the foam and data organization}
\label{sec:data_org}
The foam of cells is built-up by starting from the {\em root cell},
which is the entire integration domain, through a process of binary split of a parent
cell into two daughter cells.
The root cell is either a unit hypercube $0\leq x_i \leq 1$ (default)
or a simplex $0\leq x_1\leq x_2\leq x_3\leq \dots \leq x_n\leq 1$.
Also a Cartesian product of these two shapes available on option.
Any cell being a product of the cell split can be
also a hyperrectangle, a simplex or Cartesian product
of the $k$-dimensional hyperrectangle and $n$-dimensional simplex,
with the total dimensionality $k+n$.
If the starting root cell is a hypercube and cells are simplical (or of mixed type)
then the root cell is immediately divided into $n!$ simplical (or mixed type) cells.

Each cell is {\em explored} immediately after its creation.
In the exploration of the cell, about $100$-$1000$ MC events
(the user may define this number) are generated
inside the cell with flat (uniform)
distribution and using MC weight equal to $\rho(x)$;
certain averages and certain integrals over the cell are estimated.
Also, the best geometry of the binary split of the cell is established
and recorded for future use.
In this way, every created cell is ready for an immediate split.
The determination of the best split is described
in fine detail in Section~\ref{sec:cell_split_geom}.
In the exploration the estimate of the integral $R_I=\int_{\omega_I} \rho(x) dx^n$
is calculated for each cell $\omega_I$.
Far more important is another functional
$R_{loss}|_I=\int_{\omega_I} \rho_{loss}(x) dx^n$,
see Section~\ref{sec:cell_split_geom} for its definition,
which determines the evolution of the foam and the split of the cell.
Next cell to be divided into two is a cell chosen
randomly, according to a probability proportional to $R_{loss}|_I$
or, optionally, a cell with the largest $R_{loss}|_I$.

The process of cell division continues until the user-defined
maximum number $N_c$ of cells is reached.
$N_c$ includes,
in addition to the normal cells called {\em active} cell,
also all cells that were split, i.e. all parents,
grand-parents and great-grand-parents inluding root cell,
which we shall call {\em inactive} cells.
Usually, when we refer to cells, we mean both active and inactive ones.
Keeping inactive cells in the record may look like a waste of the memory, 
but owing to the binary character of the cell split,
the loss is only a mere factor of 2.
There are many reasons,
which make it profitable to keep all inactive cells (including the root cell) in the memory;
in particular, as we shall see in Section~\ref{sec:save_memory},
keeping all cells in the record will help us encode all cells in the memory in an economic way,
that at higher dimensions we finally gain in terms of total consumption of the CPU memory.
Furthermore, for certain quantities that are the integrals over the cell,
such as $R_I$, we do the following: just after the split, when a new, more precise
value of $R_I$ is known for the daughter cells
-- the value of the $R_I$ of the parent cell is updated
with the sum of the contributions from two daughter cells.
This correcting procedure is repeated for all ancestor cells up to the root cell.
In this way, the root cell (and any other inactive cell)
always keeps track of the actual value of the total $R_I$
during the whole foam build-up process.
This can be done for any other integral quantity as well,
and can be exploited for various purposes.

Since the maximum number of cells $N_c$ is defined at the beginning of the foam build-up,
all the cell objects and/or other related objects (vertices) are
allocated in the computer memory at once, at the very beginning of the cell build-up.
On the other hand, the cell objects are organized as a multiply-linked list,
with pointers towards parents and daughters.
In addition, an array of pointers to all active cells is created
at the end of the foam build-up, in order to speed-up the MC generation.

Let us now briefly explain how the geometry of an individual cell is parametrized 
and stored in the memory.
It is relatively easy to parametrize an $n$-dimensional hyperrectangle or simplex
in a way that does not require much computer memory.
An $n$-dimensional simplex is fully determined by its $n+1$ vertices.
Since most of the vertices are shared by the adjacent simplices, the most efficient
method is to build an array of {\em all} vertices%
\footnote{In the foam build-up every new vertex is appended at the end of the array.}
$\vec V_K, K=1,2,...,N_V$, each of them being an $n$-component vector, and to define
every simplex as a list of $n+1$ vertex indices (integers or pointers) $K_1,K_2,...,K_{n+1}$.
For $N_c$ simplical cells resulting from the binary split of a single ``root'' simplex cell,
the number of vertices is $n+1+N_c$, because each binary split adds one new vertex.
(We include in $N_c$ also cells that have got split).
The interior points of the simplex are parametrized as follows
\begin{equation}
  \label{eq:munbs}
  \vec x= \sum_{i\neq p}^n  \lambda_i (\vec V_{K_i} -\vec V_{K_p}),
  \quad \lambda_i>0,\quad \sum_{i\neq p}\lambda_i<1,\quad i=1,2,\dots,n,
\end{equation}
using basis vectors relative to the $p$-th vertex.
The above method would be inefficient for
$n$-dimensional hyperrectangles,
because memorizing all $2^n$ vertices would require too much memory at higher dimensions.
Instead, we use another way of parametrization:
each hyperrectangle is defined by the $n$-dimensional
vector $\vec q$ defining the origin of the cell
and another vector $\vec h=(h_1,h_2,...,h_n)$,
where each component $h_i$ is the length of the hyperrectangle
along the $i$-th direction.
This is even clearer from the explicit
parametrization of the interior of the hyperrectangle:
\begin{equation}
  \label{eq:munbh}
  x_i= q_i+ \lambda_i h_i,\quad  0<\lambda_i<1,\quad i=1,2,\dots,k.
\end{equation}
For cells with mixed topology, we apply Eq.~(\ref{eq:munbh})
for $i=1,2,\dots,k$ and Eq.~(\ref{eq:munbs}) for $i=k+1,k+2,\dots,k+n$.
In Section~\ref{sec:save_memory} we describe an optional
method of storing hyperrectangular cells, in which just two integer numbers are recorded
instead of two vectors $\vec q$ and $\vec h$
(two 2-Byte integers instead of $2n$ of 8-Byte floating-point numbers).
This method is implemented for the hyperrectangular part of the space only.

\subsection{Monte Carlo generation}
Once the build-up of the cells is finished, the Monte Carlo generation takes place.
There is no need for any reorganization of the cells.
MC generation can be started immediately.
The only thing, that is done at the very end of the foam build-up is the preparation
of the list of pointers to all active cells and the array of corresponding $R'_I$.

The MC point is generated in two steps.
First, a cell is chosen with a probability proportional to
$R'_I = \int_{x\in Cell_I} \rho'(x)$
and next a MC point $x$ is chosen with uniform probability inside the cell.
The MC weight $w=\rho'(x)/\rho(x)$ is associated with the event.
For a successful foam of cells, the MC weight is close to 1 and the user
may turn weighted events into $w=1$ event by means of the rejection method
(with the acceptance rate $\sim \langle w \rangle / w_{\max}$).
{\tt Foam} can do this also for the user.
However, the user can sometimes organize
the calculation of the $\langle w \rangle$ and bookkeeping of other parameters of the weight better,
in a way that best fits his own aims.
This is why the mode of variable weights MC events is also available.
The total integral, usually necessary for the proper normalization of the MC sample,
is calculated using $R=R'\langle w \rangle$.
{\tt Foam} program provides both the exact value of the $R'$ and 
the MC estimate of the integral $R$.

\subsection{Economic use of the computer memory}
\label{sec:save_memory}
The actual implementation of the single cell object occupies about 80~Bytes
(it could be shrinked to about 40~Bytes, if necessary)
of various integer and double precision attributes,  plus the dimension-dependent part.
In the case of a simplical cell, each new cell adds one $n$-component double-precision
vector (vertex) and the total memory consumption is therefore 
of order $(80+8\times n)$~Bytes/cell.
For $n=5$ and 100K cells it is therefore $\sim$~15~MBytes of the memory,
still an affordable amount.
For the hyperrectangle cells we have to count two $n$-component double-precision vectors
per cell, that is about $(80+16\times n)$~Bytes/cell.
For the $10^6$ cells and $n=15$ that would mean $\sim$~340~MBytes for the entire foam of cells,
which could be annoying.
Fortunately, we have found a method of substantially reducing the memory consumption
for a hyperrectangular foam.
As discussed in Section~\ref{sec:cell_split_geom} the geometry of the cell
division is fully determined in terms of two integers: one of them is
the index of an edge to which the division plane is perpendicular and the other
one defines the position of a division plane.
The position parameter is a rational number, and 
only the integer numerator has to be remembered, while the denominator is common to all cells.
The above two integers define uniquely the position of the two daughter cells
relative to a parent cell.
With this method the memory consumption is down to about 80~Bytes/cell,
{\em independently} of $n$ in the present implementation%
\footnote{In fact, it can be reduced below 40~Bytes/cell, if really necessary.}.
There is, however, a price to be payed in terms of CPU time.
For the generation of the point inside a cell, or even the evaluation of the weight,
we need the ``absolute'' components of $x$, that is in the reference frame of the root cell,
not relative to vertices of the cell.
It is therefore necessary to use a procedure (a method in the class of cells)
to construct the absolute position of a given cell ``in flight''.
This is done by means of tracing all ancestors of a given cell up to the root
cell and translating position and size with respect to its parent into absolute ones,
step by step, finally relative to the root cell.
It is implemented exploiting the organization of cell objects into a linked binary tree.
The average number of ancestor cells to be traced back from a given active cell up to the root
cell for $N_c=10^6$ cells is on the average about $\ln_2 N_c \sim 20$.
This may cause about $20\%$ increase in the CPU time of the MC generation
-- an affordable price,
if we remember that the MC efficiency improves mainly with the increase of $N_c$.
In principle, this kind of memory-saving arrangement is also possible for
simplical cells; however, in this case the CPU time overhead would be larger,
because of the necessity of the full linear transformation at each step,
on the way from a given cell up to the root cell.
In the case of hyperrectangular cells the transformation is much simpler (and faster);
it is the translation and/or dilatation along a {\em single} spatial
direction at each step.

\begin{figure}[!ht]
\begin{center}
{\epsfig{file=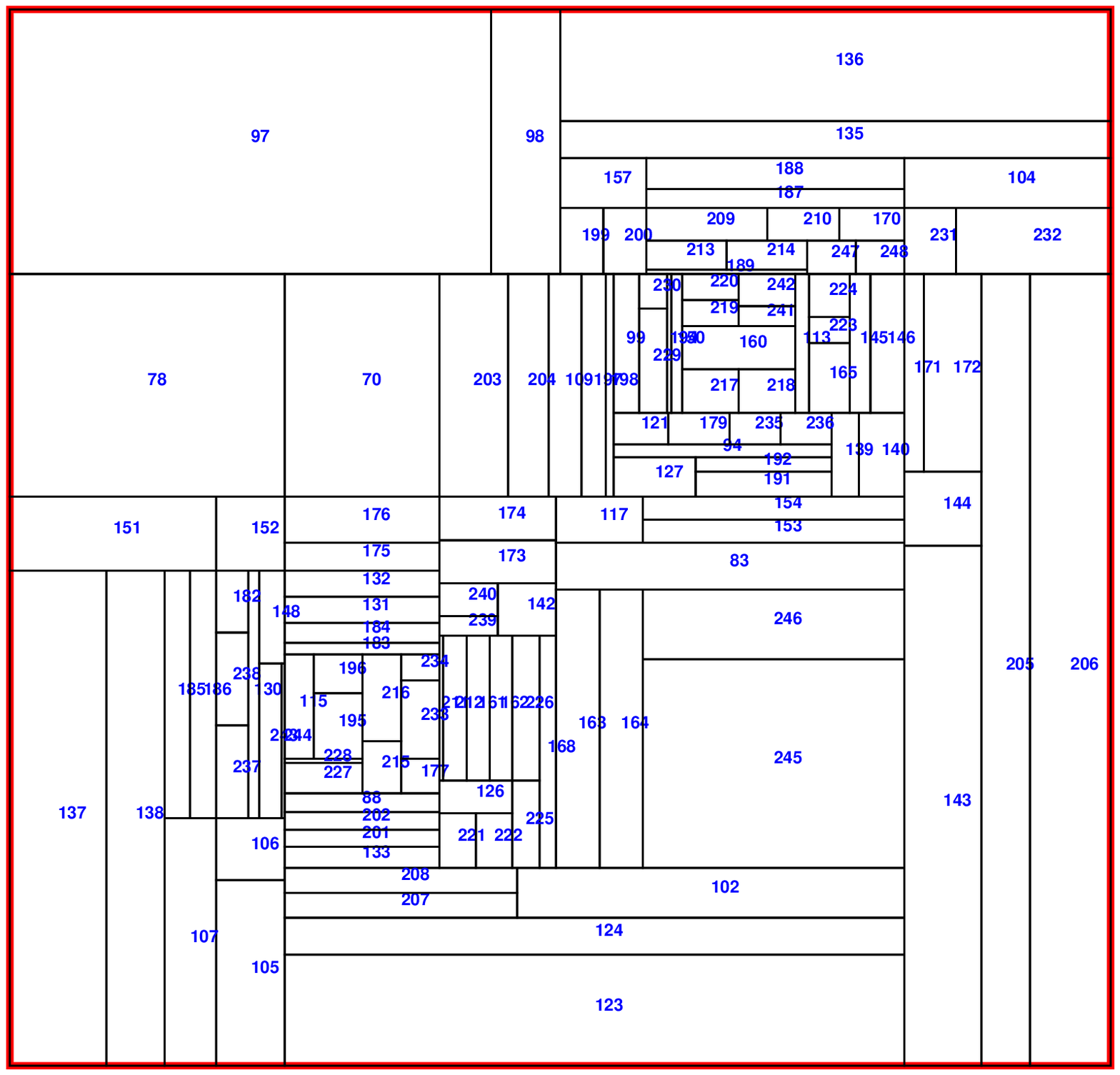,height=75mm}}
{\epsfig{file=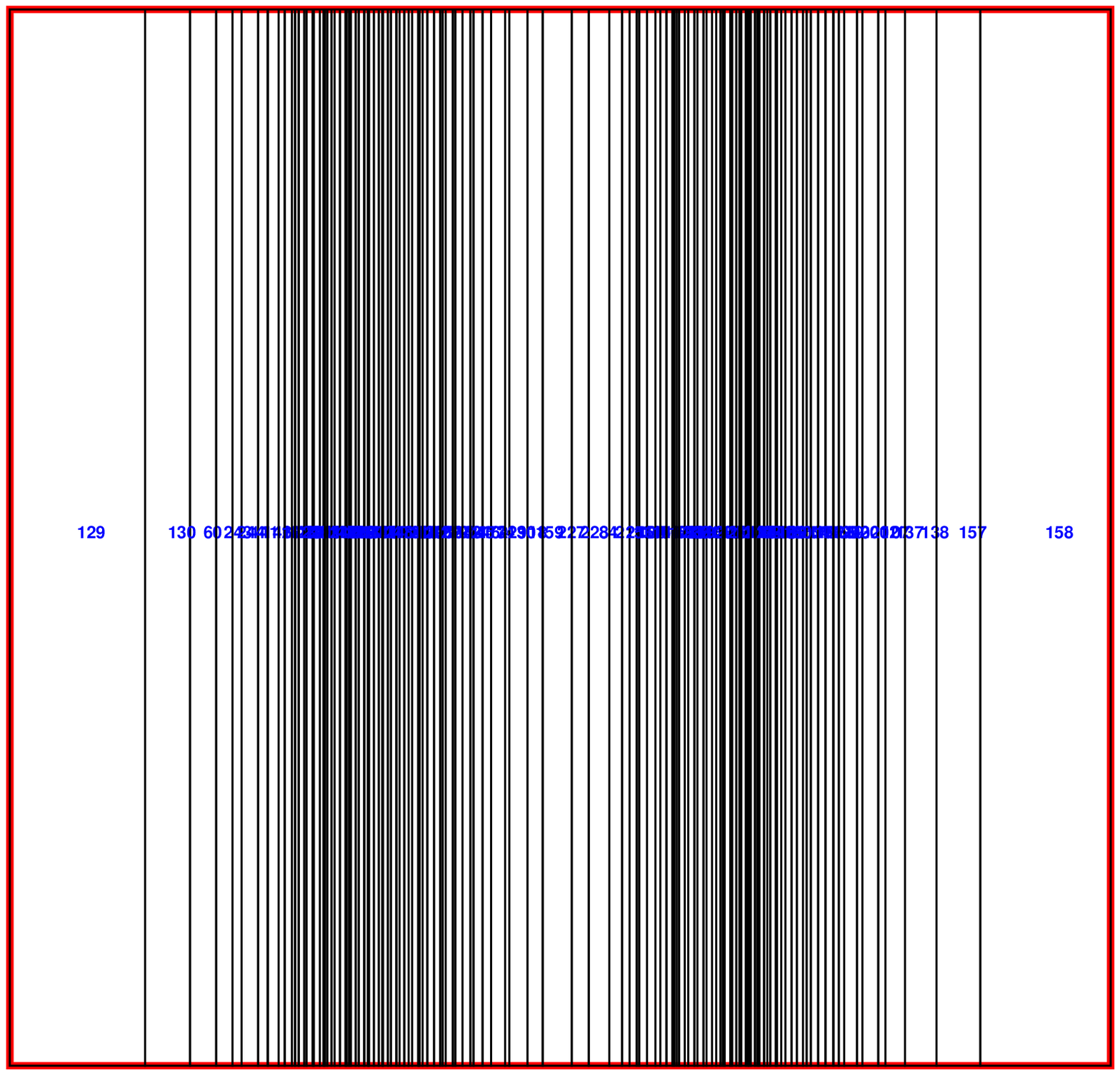,height=75mm}}
\end{center}
\caption{\small\sf
  Inhibited cell division for first variable, that is for $x_1$ (right). {\tt Foam} with 250 cells.
}
\label{fig:inhibi}
\end{figure}

\subsection{CPU time saving solution}
The final MC efficiency is improved mainly by increasing the number of cells $N_c$.
The CPU time of the cell build-up is $T \sim n\times N_c \times N_{samp}$, where $N_{samp}$
is the number of MC events used in the exploration of each newly created cell.
The important practical question is:
Can one somehow reduce $N_{samp}$ without much loss of final MC efficiency,
in order to be able to increase $N_c$, within the same CPU time budget?

A simple solution is the following:
during the MC exploration of a new cell we continuously monitor 
an accumulated ``number of effective events with $w=1$'' defined~\cite{James:1980} as
$N_{eff} = (\sum w_i)^2 / \sum w_i^2$,
and terminate cell exploration when%
\footnote{The actual limit of equivalent events per bin is the user-defined
  parameter, not necessarily equal to~25.}
$N_{eff}/n_{bin}>25$,
where $n_{bin}$ is the number of bins in each histogram, which is
used to estimate the best division direction/edge parameters.
This method helps to cut on the total CPU time, because
the increase of $N_{samp}$ is not wasted for cells in which
the distribution $\rho(x)$ is already varying very little.
At the later stage of the foam evolution this happens quite often.
In this method the user may set $N_{samp}$
to a very high value and the program will distribute the total CPU time 
(in terms of $N_{samp}$) among all cells economically, giving more CPU time to those cells
really need it, i.e. to cells with the stronger variation of $\rho(x)$.

\subsection{Inhibited variables -- flat dependence}
\label{sec:inhibit}
In some cases the user may not want {\tt Foam} to intervene into certain  variables in the
distribution $\rho(x)$, simply because there is little or no dependence
on them in $\rho(x)$.
The user may draw, of course, these variables directly from any uniform random number generator.
He may, however, find it more convenient to get them from the {\tt Foam} program.
This is easily implemented in {\tt Foam}:
any variable $x_i$ may be ``inhibited'' for the purpose of cell-splitting procedure.
In the {\tt Foam} code it is actually done in such a way that
{\tt Foam} excludes this variable (edge) from the procedure
of determining the best binary division of the cell.
This provision makes practical sense mainly for the
hyperrectangular part of the variable subspace.

In Fig.~\ref{fig:inhibi} we show two 2-dimensional foams (250 cells)
for the same testing distribution
$\rho(x)$ (two Gaussian peaks on the diagonal).
In one of them (right plot) we have inhibited a split in the first variable,
that is for $x_1$.

\begin{figure}[!ht]
\begin{center}
{\epsfig{file=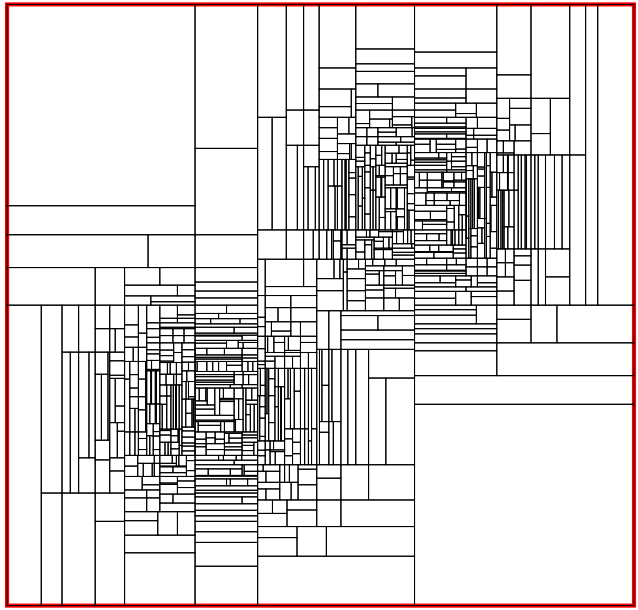,height=100mm}}
\end{center}
\caption{\small\sf
  Predefined division points at $x_1=0.30,0.40,$ and $0.65$, for 2000 cells.
}
\label{fig:xdivi}
\end{figure}

\subsection{Predefined split points -- provision for very narrow peaks}
In the practical applications, see Refs.~\cite{Boonekamp:2001wb,Jadach:1999vf},
one may encounter in certain variables extremely narrow spikes (narrow resonances).
The {\tt Foam} exploration algorithm may find it difficult to locate these spikes with the
usual method of MC sampling in the cells,
at the early stage of the foam build-up.
For very narrow spikes, or too low number of requested cells,
it may not find them at all!
The user usually knows in advance the position of these spikes
and the {\tt Foam} should have a built-in mechanism to exploit this knowledge.

The solution is simple.
(It applies for the hyperrectangular subspace of the parameter space only.)
The user has a possibility to provide {\tt Foam},
for each variable, with the list of a number of predefined values:
the first splitting positions of the root cell.
In the {\tt Foam} algorithm, it is checked if the list of predefined division points
is not empty. 
If it is the case, then instead of adopting the division parameter from the usual procedure
described in Section~\ref{sec:cell_split_geom}, {\tt Foam} takes the division
parameter from the list, and removes it from the list.
In this way the first few division points are taken from
the ``user-defined menu'', if available, and the next ones are chosen with the usual methods.
For narrow spikes this method helps {\tt Foam} to locate and surround them
with as a dense group of cells as necessary.

In Fig.~\ref{fig:xdivi} we show an example with two Gaussian peaks
in which we requested the {\tt Foam} program to use the three predefined division points
for the $x_1$ variable. They are clearly seen as three vertical division lines
dividing the entire root cell.
In the present case, peaks are not so narrow and there is no real need for a predefined division.
The example is just to illustrate the principle of the method.

\subsection{Mapping of variables}
If the structure of the singularities is known and/or {\tt Foam} is unable
to get a reasonable weight distribution for a reasonable number of cells,
it is then worth performing an additional change of variables,
so that the transformation Jacobian compensates for the singularities,
at least partly.
In such a case the user subprogram provides {\tt Foam} with the distribution
\begin{equation}
  \rho^\star(y)=
  \frac{d\rho}{dy_1\dots dy_n} =
  \rho(x_1(y),x_2(y),\dots,_n(y))
  \left| \frac{\partial x^{(j)}(y) }{ \partial y} \right|,
\end{equation}
instead of the original $\rho(x) = d^n \rho / dx^n$.
For each vector $y$ generated by {\tt Foam}, the image vector $x$ is well known
in the subprogram calculating $\rho^\star(y)$.
A mechanism for exporting $x$ to the outside world usually has to be provided by the user,
because {\tt Foam} cannot help -- it does not know anything about $x$; it only knows $y$.

Note that in the limiting case of the ``ideal mapping'' we have
\begin{equation}
  \left| \frac{\partial x(y) }{ \partial y} \right| \equiv \frac{R}{ \rho(x)};
\end{equation}
consequently, $\rho^\star(y)=R$ and in this case {\tt Foam} would play merely 
the role of a provider of the random numbers for $y$.

The user of {\tt Foam} may also need to apply mapping in the case of a ``weak'' integrable
singularity in the distribution $\rho$ like $\log(x)$ or $\sqrt{x}$.
{\tt Foam} can deal with them by brute force, at the expense of a larger number of cells.
However, a wiser approach is to apply mapping, in order to remove such
singularities from the distribution.

In the next section we shall describe how to combine
the mapping method with the multibranching.
Such a mixture is well known as the most powerful method of improving
the efficiency of the Monte Carlo method.

\subsection{Provisions for the multibranching}
\label{sec:mbranching}
In the following we elaborate on the various methods 
of implementing 
multibranching MC method~\cite{Berends:1983mi,Kleiss:1994qy,mcguide:1999,Skrzypek:1999td}
with the help of {\tt Foam}.
\subsubsection{Single discrete variable}
As a warm-up exercise, let us consider the question:
Is {\tt Foam} capable to generate (and sum up)
a discrete variable $i=1,2,\dots, N,$
according to the (unnormalized) distribution $r_1,\dots, r_N$?
Of course it can.
The simplest way is to define an auxiliary 1-dimensional  distribution
\begin{equation}
  \label{Eq.discrete1dim}
  \rho(x)= r_i,\quad \hbox{for}\;\;
   \frac{i-1}{N} \leq x \le \frac{i}{N},\;\;
   i=1,2,\dots,N.
\end{equation}
The user subprogram providing the above $\rho(x)$ is trivial.
If plotted, this $\rho(x)$  would look like a histogram with $N$ equal-width bins.
{\tt Foam} will build up its own grid of cells (intervals), and if we request
a high enough number of cells (that is $N_c>N$),
it will approximate the above $\rho(x)$ very well, with its own ``histogram-like''
distribution $\rho'(x)$.
However, the {\tt Foam} approximation will  never be ideal, because {\tt Foam}
is not able to detect the exact position of the discontinuities in $\rho(x)$.
(Nevertheless, this will be a workable solution with a very good weight distribution.)
The present {\tt Foam} algorithm provides for an essential improvement:
one may predefine the division points as $x^{(i)}=i/N$, $i=1,\dots,N$,
and set the number of cells to be $N_c\geq N$.
In such a case {\tt Foam} will define its cells matching exactly the shape of $\rho(x)$.
It will generate points with $w\equiv 1$ and provide the exact sum $R=R'=\sum r_i$,
already at the end of the foam build-up.
In the MC generation, {\tt Foam} will randomly generate variable $x\in(0,1)$,
which can be translated easily into discrete $i$ $1\leq i\leq N$.

\subsubsection{Discrete and continuous variables}
How does the above extend to the case of
the distribution $\rho$ depending on one discrete variable and
the usual $n$ continuous variables?
For such a distribution $\rho(y_1,\dots,y_n,i)$ we define
\begin{equation}
  \rho(x_1,x_2,\dots,x_{n+1})
  = \rho(x_1,\dots,x_n,i),\quad \hbox{for}\; 
   \frac{i-1}{N} \leq x_{n+1} \le \frac{i}{N},\;\;
  i=1,2,\dots,N,
\end{equation}
in completely analogy with Eq.~(\ref{Eq.discrete1dim}).
As previously, we provide for the variable $x_{n+1}$
a list of predefined division points $x_{n+1}^{(i)}=i/N,\; i=1,2,\dots,N$,
and, of course, we request $N_c>> N$.
There is still one small problem:
{\tt Foam} may ``by mistake'' perform an unnecessary cell division for the variable $x_{n+1}$,
simply due to statistical errors in the ``projection histogram''
described in Section~\ref{sec:projections}.
This problem is solved in {\tt Foam} in an elegant way:
in addition to providing for $x_{n+1}$ predefined division points,
the user of {\tt Foam} may declare $x_{n+1}$
as an ``inhibited variable'' in the sense of Section~\ref{sec:inhibit}.
In this case {\tt Foam} will still split cells according to a list of predefined division points
for $x_{n+1}$, but will not perform any additional division in this variable!
(The translation of the continuous $x_{n+1}$ to the discrete $i$ is done as usual.)
The above method is the basic method of implementation of the ``multibranching''
(or ``multichannel'') MC method using {\tt Foam}.
Let us call it ``predefined and inhibited division'', for short PAID method.
We shall also describe below how to combine the PAID method with mapping, etc.

In order to appreciate more fully the advantages of PAID,
let us consider a more straightforward implementation of the multibranching.
In the object-oriented environment one may easily construct
$N$ instances of the {\tt Foam} object,
each of them for the $n$-dimensional function $\rho(x_1,\dots,x_n,i)$,
initialize them (creating a separate foam of cells) and generate event $(x,i)$
with the associated weight $w_i(x)$.
Index $i$ can be chosen according to probability $p_i=R'_i/\sum_j R'_j$,
where $R'_i$ are provided by the $i$-th object of the {\tt Foam} class
(at the end of its initialization).
The total weight of the event is $w(x,i)=w_i(x)/p_i$.
Let us call this scenario an ``externally organized multibranching'', for short EOM.

Both methods have certain advantages and disadvantages.
In PAID the user does not need to organize the  optimal/efficient generation
of the branching index $i$. 
The root cell is divided into $N$ equal-size {\em sub-root} cells, which then evolve
separately into an independent system of cells, adapting individually
to the singularities in the $i$-th component of $\rho$.
{\tt Foam} adjusts the relative importance of the sub-root cells and their descendants,
and finds automatically the optimal number of the division cells 
in the $N$ subfoams within the requested total memory limit.
In the EOM scheme these adjustments for the individual branches has to be done by
the user.
On the other hand, in some cases, the user may want to 
configure the {\tt Foam} objects for each branch individually.
In the EOM scheme, this can be done for each {\tt Foam} object separately.
In the PAID scheme this cannot be done,
because  all cells have the same properties, the cell-split algorithm is the same,
the cell geometry is common, etc.
In most cases, the PAID method will be preferred, because it is easier to organize.

In the following we shall concentrate on the PAID scheme.
In this case, the normalization integral is
provided by the {\tt Foam} at the end of the exploration phase,
and it includes the sum over a discrete variable:
\begin{equation}
  R'=\sum_{i=1}^N \int \rho_i'(x) dx^n.
\end{equation}
We also have the usual relation between the average weight and the integral
\begin{equation}
  R=\sum_{i=1}^N \int \rho(x)_i dx^n = R'\; \langle w \rangle.
\end{equation}
The above method extends trivially to the case of several discrete variables.
As already stressed, the relative probabilities of the discrete components 
$p_i \sim R'_i= \int \rho'_i(x) dx^n$
in the MC generation are automatically adjusted by the {\tt Foam} algorithm,
so that the maximum weight or the total invariance is minimized.
In the EOM scheme, arranging this in the user program would
require an extra programming effort, while in the PAID method this comes for free,
by exploiting standard {\tt Foam} functionality.

\subsubsection{Multilayer method}
There is an alternative PAID-type method of dealing with the problem of the discrete
variable, which generates points according to 
$\rho(x_1,\dots,x_n,i),\; i=1,2,\dots,N$.
It will produce the same distribution but will differ from PAID in the MC efficiency,
in terms of the maximum weight or variance.
One may simply generate, with the help of {\tt Foam},
the $n$-dimensional auxiliary distribution 
\begin{equation}
  \bar\rho(x_1,\dots,x_n) = \sum_{j=1}^N \rho(x_1,\dots,x_n,j)
\end{equation}
and next, for each generated $x$, choose randomly 
a discrete variable $i$ according to the probability
\begin{equation}
  p_i(x)= \rho(x,i)/ \sum_{j=1}^n \rho(x,j).
\end{equation}
Let us call it PAID$^*$, or the  multilayer method.
This method is slightly less convenient to implement, as is clearly seen
for $n=0$, where the user effectively has to generate by himself
the discrete variable $i=1,2,\dots,N$ according to the above probability,
by means of creating an inverse cumulative distribution,
mapping a random number into $i$, etc., while in the standard PAID scenario
all this job is done by the {\tt Foam} program%
\footnote{The mapping $x_{n+1}\to i$ is a simple arithmetic operation.}.
Furthermore, in the PAID method each component distribution $\rho(x,j)$ may have
a ``cleaner'' structure of the singularities than the sum.
Consequently, in the PAID method {\tt Foam} will probably find it 
easier to learn the shape of each component
distribution than that of the sum in PAID$^*$.
These two kinds of equivalent multibranching algorithms, PAID and PAID$^*$,
are described and analysed in Ref.~\cite{mcguide:1999}.
The PAID$^*$ method is used in the KKMC event generator of Ref.~\cite{Jadach:1999vf}
to generate index $i$ numbering the type (flavour) of the produced quark or lepton.

\subsubsection{Multibranching and mapping}
However, the most important reason for setting up {\tt Foam} according to the PAID scenario,
with the separate foam build-up for each component distributions $\rho(x,j)$,
is that for each component one may apply {\em individually adjusted mapping} of variables,
which makes every component distribution much less singular.
The combination of the mapping and multibranching is probably the most
powerful known methods of improving MC efficiency~\cite{Kleiss:1994qy,mcguide:1999}.
How it can be actually realized with the help of {\tt Foam}
depends on the properties of the distribution $\rho(x)$ to be generated.
In the case where we have an explicit sum over many components
\begin{equation}
  \rho(x_1,\dots,x_n) = \sum_{j=1}^N \rho(x_1,\dots,x_n,j),
\end{equation}
each of the components being positive, with a distinctly different
and well known structure of the singularities, we would recommend
the use of {\tt Foam} in the PAID scheme.
Knowing the structure of singularities,
we may be able to introduce mapping in each component separately, which compensates
for these singularities with the Jacobian factor.
In such a case {\tt Foam} is provided with the following distribution:
\begin{equation}
  \label{eq:rhoj}
  \rho(y_1,\dots,y_n,j) =
  \rho(x^{(j)}_1(y),\dots,x^{(j)}_n(y),j)\;
  \left| \frac{\partial x^{(j)}(y) }{ \partial y} \right|,\;\; j=1,2,\dots,N,
\end{equation}
understanding that the translation of the discrete index $j$ 
into a continuous variable $y_{n+1}$, is done in the usual way.
The foam of cells is, of course, build-up in the $y$-variables,
different for each $j$-th branch.
The user is fully responsible for the proper mapping
$x^{(j)}(y),\; j=1,2,\dots,N$,
and the calculation of the Jacobian factor in every component (branch).
In the user subprogram providing the $\rho$-distribution,
the variable $y_{n+1}$ will first be translated into index $j$
and then,  depending on the value of $j$, a given type of mapping will be applied.
For the outside part of the code, the index $j$ can be made available,
or it may be hidden (erased from the record),
depending on the needs of a specific application.

In some cases, however, one does not know any unique 
way of splitting the $\rho(x)$ into
well defined positive components as Eq.~(\ref{eq:rhoj}),
but one may have a rough idea about the leading singularities.
This means, that one is able to construct the distribution
\begin{equation}
  \rho(x_1,\dots,x_n) \sim
  \bar\rho(x_1,\dots,x_n) = \sum_{j=1}^N \bar\rho(x_1,\dots,x_n,j),
\end{equation}
where $\bar\rho(x,j)$ have the same type of the leading singularities as $\rho(x)$,
and one controls the normalization of singularities in $\rho(x)$
up to a constant factor;
that is for $x$ in the neighbourhood of the $j$-th singular point, a line or a (hyper)plane,
only $\bar\rho(x,j)$ really matters, that is $\rho(x) \simeq C_j \bar\rho(x,j)$,
where $C_j$ is not known a priori%
\footnote{This definition is not very precise; it roughly means that
  each component is approximately a product of the singular factors
  and cannot be reduced into the sum of these.}.

In addition, let us assume
that we are able to compensate for the singularities in 
each $\bar\rho(x,j)$ {\em exactly}
by dedicated mapping specific to singularities in the $j$-th branch.
In other words, the mapping is {\em ideal} in each branch:
\begin{equation}
\label{eq:ideal_mapping}
  \left| \frac{\partial x^{(j)}(y) }{ \partial y} \right| = \frac{\bar R_j}{ \bar\rho(x,j)}.
\end{equation}
The above also means, that we know analytically
the exact values of the integrals%
\footnote{In fact, we could normalize $\bar\rho(x,j)$ to unity, $\bar R_j=1$, if we wanted.}:
$\bar R_j = \int \bar\rho(x,j) dx^n$.

In such a case we may employ the algorithm of {\tt Foam} successfully by means of
defining the ``branching ratio''
\begin{equation}
  \label{eq:bfun}
  b_j (y_1,\dots,y_n) 
  = \bar\rho(y_1,\dots,y_n,j)/ \bar\rho(y_1,\dots,y_n),\quad \sum b_i(y)=1,
\end{equation}
constructing the distribution to be digested by {\tt Foam} as
\begin{equation}
  \label{eq:PAID_map}
  \rho(y_1,\dots,y_n,j) =
  b_j(x) \rho(x^{(j)}_1(y),\dots,x^{(j)}_n(y))\;
  \left| \frac{\partial x^{(j)}(y) }{ \partial y} \right|
  =  \frac{\bar R_j\;  \rho\big(x(y)\big) }{\sum_l \bar\rho\big(x(y),l\big)},
\end{equation}
and proceeding as in the PAID scheme described previously.

The role of the function $b_j(x)$ is to isolate out from $\rho(x)$ ``a layer''
including just one known type of singularity.
In order to see how this method works,
let us consider a $\delta^{(n)}(x-a)$ shape singularity in the $j$-th component
(narrow Gaussian peak etc.) of size $\epsilon$.
Then, in the vicinity of that singularity we have
$b_j(x)=1$ and $\rho(x) \simeq C_j \bar\rho(x,j)$,
while further away we have $\rho(x) \sim {\cal O}(\epsilon^n)$.
The {\tt Foam} program will, of course, include the $C_j$ factor properly
in the normalization, and build up the foam of cells everywhere, close to a singularity
and far away.
It will do it really  not in the $x$ variables but in the $y$ variables.
Now, the mapping $x\to y$ (specific to the $j$-th branch)
will expand the singularity neighbourhood of ${\cal O}(\epsilon)$
into a region of size of ${\cal O}(1)$, while the $y$-image of the
remaining $x$-space will be of the size ${\cal O}(\epsilon)$.
This can be the source of the following drawback:
in the small $y$-domain of ${\cal O}(\epsilon)$ described above,
at the positions of the singularities from the other components $i\neq j$,
$\rho(y,j)$ may get narrow spikes or dips of height of ${\cal O}(1)$,
such that their integral contribution will be negligible, of ${\cal O}(\epsilon^n)$,
and this may not look as a problem.
However, the {\tt Foam} algorithm may find it difficult to locate these structures,
and this may lead to a small but finite bias of
the generated distributions and calculated integrals.
This should be kept in mind, and special tests
(MC runs with a maximum number of cells, and high MC statistics)
should be performed, in order to check that this effect is absent.

The above method is quite similar to that of Ref.~\cite{Kleiss:1994qy}.
One difference is that in the latter
there are several iterations with an aim af adjusting the relative
normalization of the components $\bar\rho(x_1,\dots,x_n,j)$ to $\rho(x)$.
Our scheme could effectively be regarded as the method of Ref.~\cite{Kleiss:1994qy}
with just one iteration; that is the first step being the foam build-up,
and the second step (1st iteration) being the MC simulation.
One iteration is sufficient in the limit of vanishing overlap of 
the components $\bar\rho(x_1,\dots,x_n,j)$ in the entire $\bar\rho(x)$.
While in the method of Ref.~\cite{Kleiss:1994qy} a better adjustment is provided
by the next iterations, in {\tt Foam} the cellular adaptive method provides an extra mileage.
One cannot therefore say which one is better in general --
it depends on the distribution $\rho(x)$.

In fact, in the PAID scheme with the mapping, extra iterations are also possible.
It can be done as follows:
(a) read $R_j$ from all $N$ ``leading cells'' after the foam build-up,
(b) rescale 
$\bar\rho(y_1,\dots,y_n,j) \to (R_j/ \bar R_j)  \bar\rho(y_1,\dots,y_n,j)$, and 
(c) repeat the foam build-up%
\footnote{When programming with {\tt Foam}
  it is possible to erase all cells from memory and rebuild them.}
for the new branching ratios $b_j(x)$ in Eq.~(\ref{eq:PAID_map}).
The above procedure can be repeated.
Whether such an iteration is profitable depends on the particular distribution --
we expect that in most cases it is not necessary,
thanks to the adaptive capabilities of {\tt Foam}.

Last, but not least, let us also consider the case of a sum of integrals
with different dimensionality or, in other words,
the distribution in which the number $n_i$ of the continuous variables $x_1,...,x_{n_i}$
depends on a certain discrete ``master variable'' $i=1,...,N$
(for example $n_i=i$):
\begin{equation}
  R=\sum_{i=1}^{N} \int \rho_i(x_1,...,x_{n_i}).
\end{equation}
{\tt Foam} can deal with this case too.
The simplest solution is to find the maximum dimension $n_{\max}$ and
add extra dummy variables on which some of $\rho_i$ does not depend,
such that formally all sub-distributions have the same dimension $n_{\max}$.
In this way one is back to a situation described earlier, and may apply the PAID
method, with or without the additional mapping.
The slight drawback of this solution is that in the present implementation of {\tt Foam}
we cannot inhibit the unnecessary
cell divisions across the directions of the newly introduced dummy variables --
simply because they are not the same in all branches%
\footnote{A more sophisticated procedure of inhibiting the division could be implemented
  in {\tt Foam}, if there is a strong demand.}.
This is the reason why this kind of problem
can, in some cases, be dealt with more efficiently using
the EOM scenario, with a separate {\tt Foam} object for each branch.

For additional practical examples 
on how to realize multibranching with {\tt Foam},  see Section~\ref{sec:NumeResults}.

\section{Cell split algorithm and geometry}
\label{sec:cell_split_geom}
As already indicated, our algorithm of the cell split covers two strategies:
(A) minimization of the maximum weight $w_{\max}$ and 
(B) minimization of the variance $\sigma$, where
both $w_{\max}$ and $\sigma$ are calculated in the Monte Carlo generation,
using the MC weight $w=\rho/\rho'$.
The distribution $\rho'$ is the result of the exploration (it is constant over each cell)
and is frozen at the end of exploration.
The subsequent MC event generation of events is done according to $\rho'(x)$.
Its integral $R'=\int \rho'(x) d^n x$
has to be known {\em exactly} before the start of the MC generation.
On the other hand, the best estimate of the integral $R=\int \rho(x) d^n x$ is obtained,
up to a statistical error, at the end of the MC event generation with the usual relation:
$R= R'\langle w \rangle_{\rho'}$.
The average $\langle ... \rangle_{\rho'}$ is over events generated according to $\rho'$.

There is another important ingredient in the algorithm of the cell split:
in addition to the auxiliary distributions $\rho'(x)$ we also define
another distribution $\rho_{loss}(x)$ related to the integrand $\rho(x)$.
The important role of the distribution $\rho_{loss}(x)$
is to guide the build-up of the foam of cells;
the function $R_{loss}=\int \rho_{loss} d^n x$ is minimized in the process --
its value is decreasing step by step, at each cell split.
Obviously, both $\rho_{loss}(x)$ and $\rho'(x)$
are evolving step by step during the foam build-up.
Once the division process is finished, the distribution $\rho_{loss}(x)$
is not used anymore; MC events are generated with $\rho'(x)$.
Nevertheless, $\rho_{loss}(x)$ is strongly related to the properties
of the weight distribution in the MC generation phase.

(A) In the case when our ultimate aim is to minimize $w_{\max}$, we define
\begin{equation}
  \begin{split}
    &\rho'(x) \equiv \max_{y\in Cell_I} \rho(y),\quad \hbox{for}\quad x\in Cell_I,\\
    &R_{loss}= \int d^nx\; [\rho'(x) -\rho(x) ]=\int d^nx\;\rho_{loss}(x).
  \end{split}
\end{equation}
The distribution $\rho_{loss}$ is the difference between 
the ``ceiling distribution'' $\rho'$ and the actual distribution $\rho$
from which it is derived.
The rejection rate in the final MC run is just proportional
to the integral over the loss distribution $\rho_{loss}(x)$ by construction, i.e.
the rejection rate $= R_{loss}/R$.
(This justifies the name ``loss''.)
The distribution $\rho_{loss}(x)$ also has a clear geometrical meaning, see below.

(B) In the case when we do not care so much about the maximum weight and the rejection rate,
but we want rather to
minimize the ratio of the variance to the average of the weight, $\sigma/\langle w \rangle$,
in the final MC generation, we are led to the following definition:
\begin{equation}
  \begin{split}
    &\rho'(x) \equiv \sqrt{ \langle \rho^2 \rangle_I},\quad \hbox{for}\quad x\in Cell_I,\\
    &\rho_{loss}(x) \equiv \sqrt{ \langle \rho^2 \rangle_I} -\langle \rho \rangle_I,
    \quad \hbox{for}\quad x\in Cell_I.
  \end{split}
\end{equation}
The average $\langle ... \rangle_I$ is over the $I$-th cell;
see Appendix A for a detailed derivation of the above prescription.
The ratio $\sigma/\langle w \rangle$ in the final MC generation
is a monotonous ascending function of $R_{loss}=\int \rho_{loss}(x) dx^n$, see Appendix A.
Consequently, minimization of $R_{loss}$ is equivalent to minimization of
$\sigma/\langle w \rangle$.

\subsection{Rules governing the binary split of a cell}

The basic rules governing the development of the foam of cells are the following:
\begin{itemize}
\item[{(a)}]
  For the next cell to be split we chose a cell with the largest%
  \footnote{In the {\tt Foam} code there is also an option of choosing the next
    cell to be split randomly, according to probability proportional to $R_{loss}$,
    instead of a cell with the largest $R_{loss}$.}
  $R_{loss}$.
\item[{(b)}]
  The position/direction of a plane dividing a parent cell into two daughter cells
  $\omega \to \omega'+\omega''$
  is chosen to get the largest possible decrease
  $\Delta R_{loss}= R_{loss}^{\omega} -R_{loss}^{\omega'}-R_{loss}^{\omega''}$.
\end{itemize}
How is the split of a given cell into two daughter cells in step (b) done in practice?
The method relies upon a small MC exercise within a cell,
in which a few hundreds of events are generated with a flat distribution.
They are weighted with $\rho(x)$ and projected onto $k$ (hyperrectangular case) or
$n(n+1)/2$ (simplical case) {\em edges} of the cells.
In the mixed case of a cell being the Cartesian product
of a $k$-dimensional hyperrectangle and an $n$-dimensional simplex,
there are $k+n(n+1)/2$ projections/edges.
Resulting histograms are analysed and
the best ``division edge'' and ``division hyperplane position'' are found -- the one
for which the estimate (forecast) of the $\Delta R_{loss}$ is the largest.
In the actual {\tt Foam} algorithm, each new-born cell is immediately explored,
its $R_{loss}$, $R$ and $R'$ are calculated, and
the best candidate as to the direction and position of the dividing plane 
are established and memorized, as the attributes of the cell; see below for details.
In this way, every newly created cell is ready for an immediate binary division.

\subsection{Geometry of the binary split of a cell}

\begin{figure}[!ht]
\begin{center}
{\epsfig{file=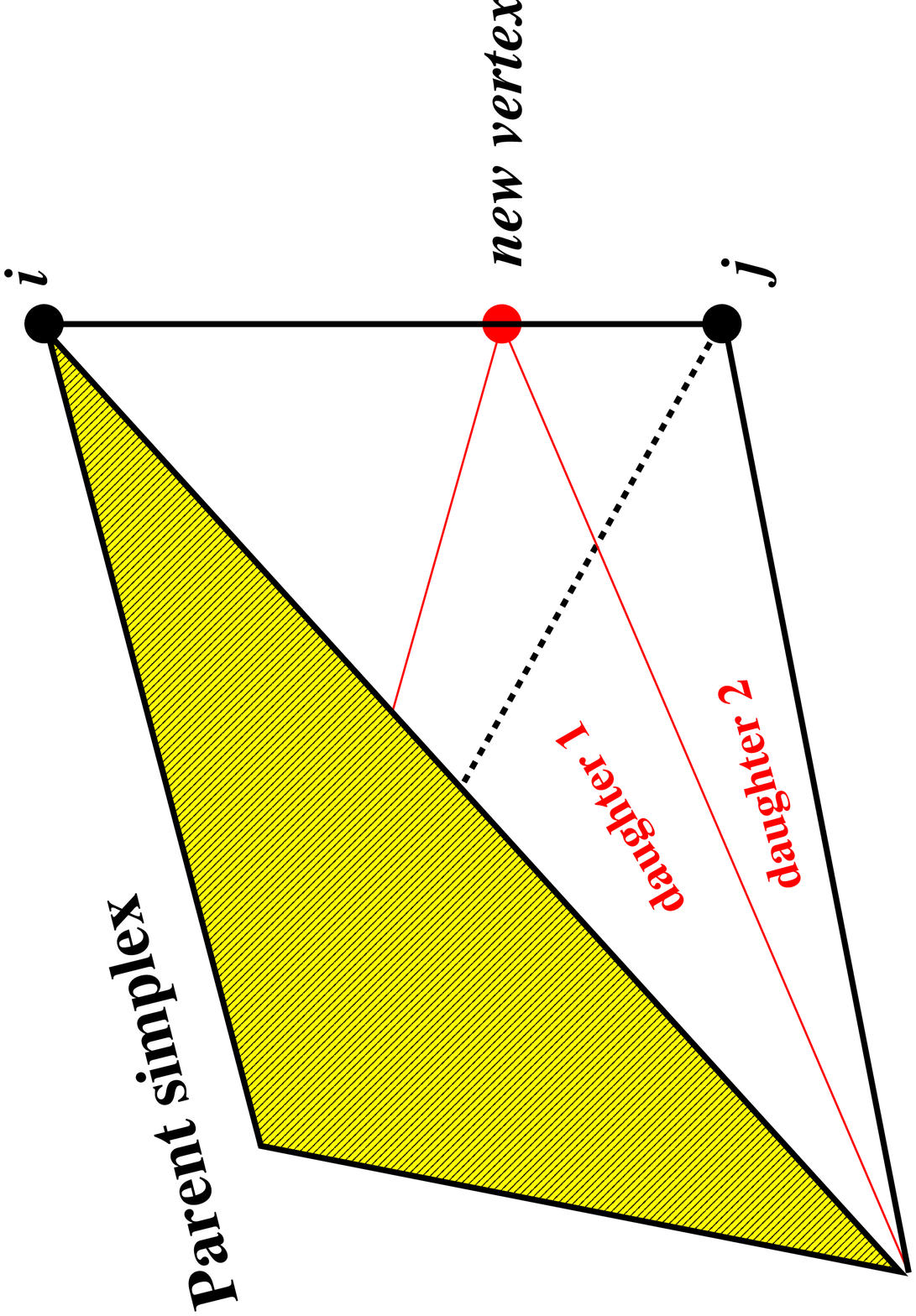,height=75mm,angle=270}}
{\epsfig{file=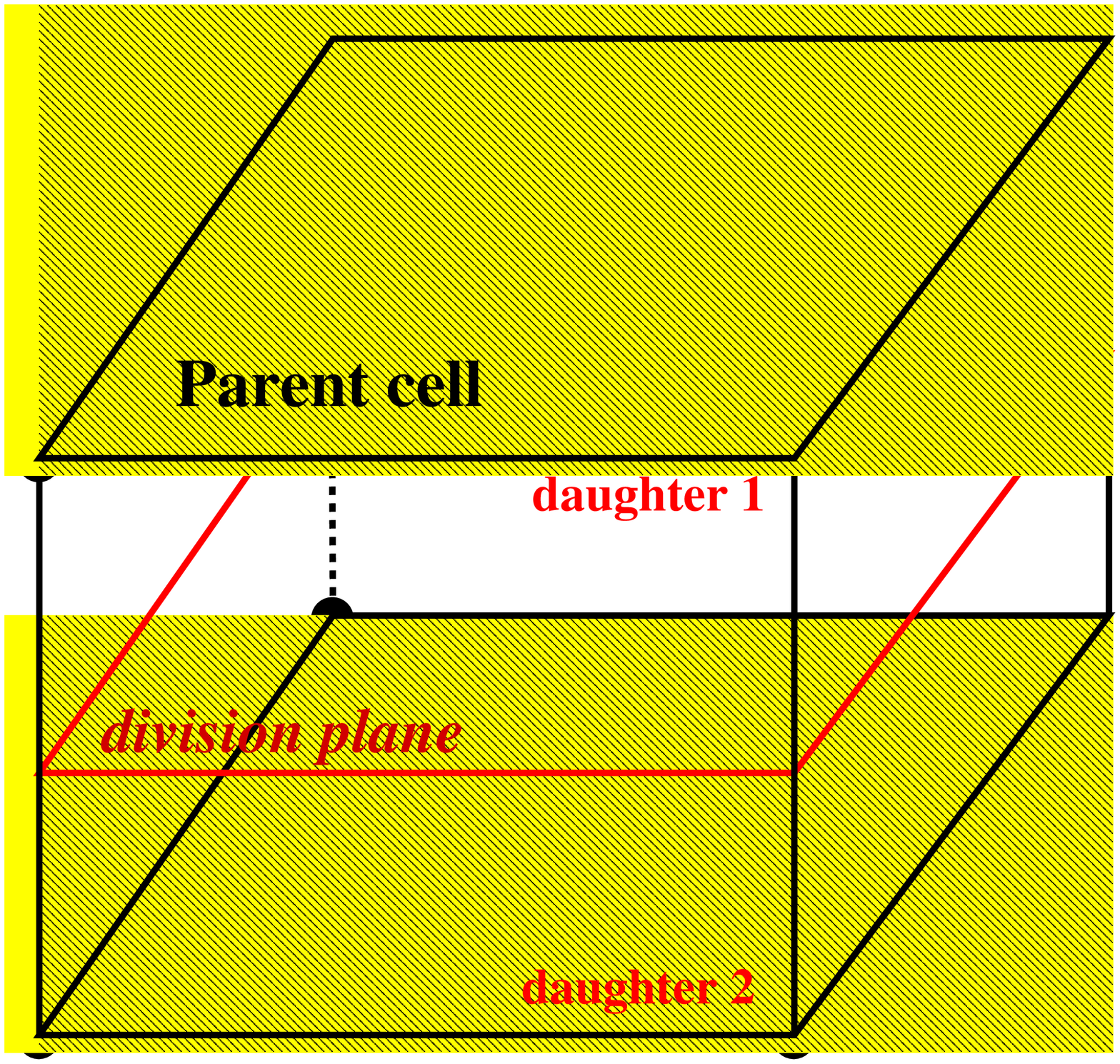,height=75mm,angle=270}}
\end{center}
\caption{\small\sf
  Geometry of the split of a 3-dimensional cell, simplex or hyperrectangle.
}
\label{fig:split_geometry}
\end{figure}

In Fig.~\ref{fig:split_geometry}
we show a 3-dimensional cell being a simplex or a hyperrectangle
and we visualize the geometry of their split.

Let us first describe the split of the $n$-dimensional
simplical parent cell into two daughter cells.
In this case~\cite{foam1:2000}
we need to know which of $n(n+1)/2$ edges defined
by any pair of vertices of a given simplex is the best for the split.
Suppose that it is an edge defined by a pair
of indices $(i,j),\; i\neq j,$ where $i,j=1,2,...,N_V$, of the vertices
$(\vec V_{K_i},\vec V_{K_j})$, see Sect.~\ref{sec:data_org}
for the method of numbering the vertices.
A new vertex $V_{N_V+1}$ is put somewhere on the line (edge) in between
the two vertices:
\begin{equation}
  V_{N_V+1}=\lambda \vec V_{K_i}  +(1-\lambda_{div}) \vec V_{K_j},\;\;   0<\lambda_{div}<1,
\end{equation}
where the division parameter $\lambda_{div}$ is determined by using
an elaborate procedure described later in this section,
and the number of vertices is updated $N_V \to N_V+1$.
With the new vertex
two daughter simplices are formed with the following two lists of vertices (their pointers):
\begin{equation}
  \begin{split}
    (K_1,K_2,...,K_{i-1}, (N_V+1), K_{i+1},...,K_{j-1},K_j,K_{j+1},...,K_n,K_{n+1}),\\
    (K_1,K_2,...,K_{i-1},K_i,K_{i+1},...,K_{j-1}, (N_V+1), K_{j+1},...,K_n,K_{n+1}).
  \end{split}
\end{equation}

For the $k$-dimensional
hyperrectangular cell defined with a pair of vectors $(\vec q, \vec h)$ we
first decide on the direction of the division hyperplane.
Assuming that this hyperplane is perpendicular to the $i$-th direction,
the two daughter cells (a) and (b) are defined
with the two pairs of new vectors as follows:
\begin{equation}
  \begin{split}
    &\vec q^{(a)} = (q_1,q_2,                      \dots,q_k),\quad
     \vec h^{(a)} = (h_1,h_2,\dots,h_{i-1},h_i\lambda_{div},h_{i+1},\dots,h_k),\\
    &\vec q^{(b)} = (q_1,\dots,q_{i-1},q_i+h_i\lambda_{div},h_{i+1},\dots,q_k),\quad
     \vec h^{(b)} = (h_1,\dots,h_{i-1},h_i(1-\lambda_{div}),h_{i+1},\dots,h_k).
  \end{split}
\end{equation}

The 3-dimensional case of the simplical and hyperrectangular cell split
made in this way is illustrated
in Fig.~\ref{fig:split_geometry}.

\begin{figure}[!ht]
\begin{center}
{\epsfig{file=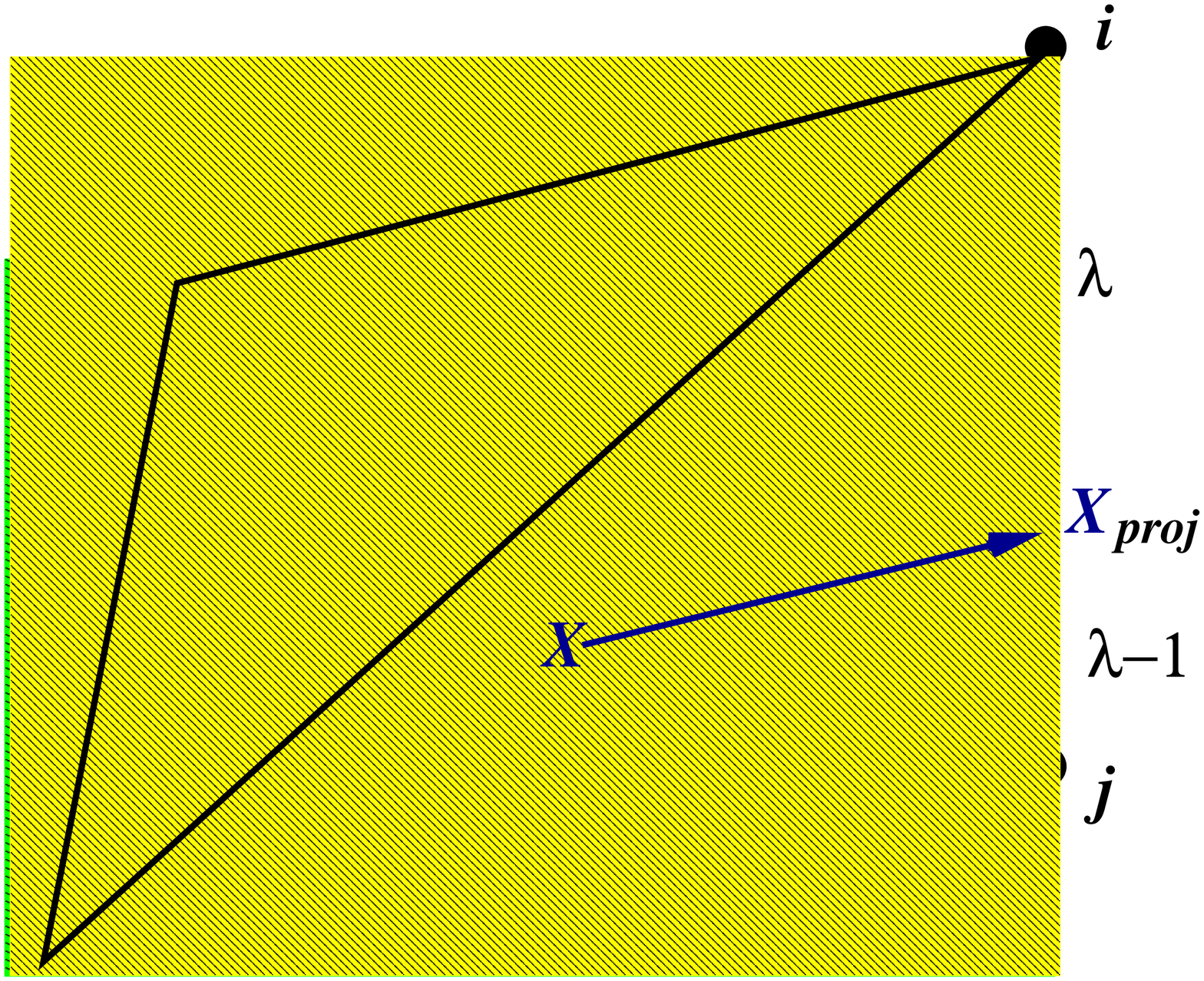,height=75mm,angle=270}}
\vspace{-5mm}
{\epsfig{file=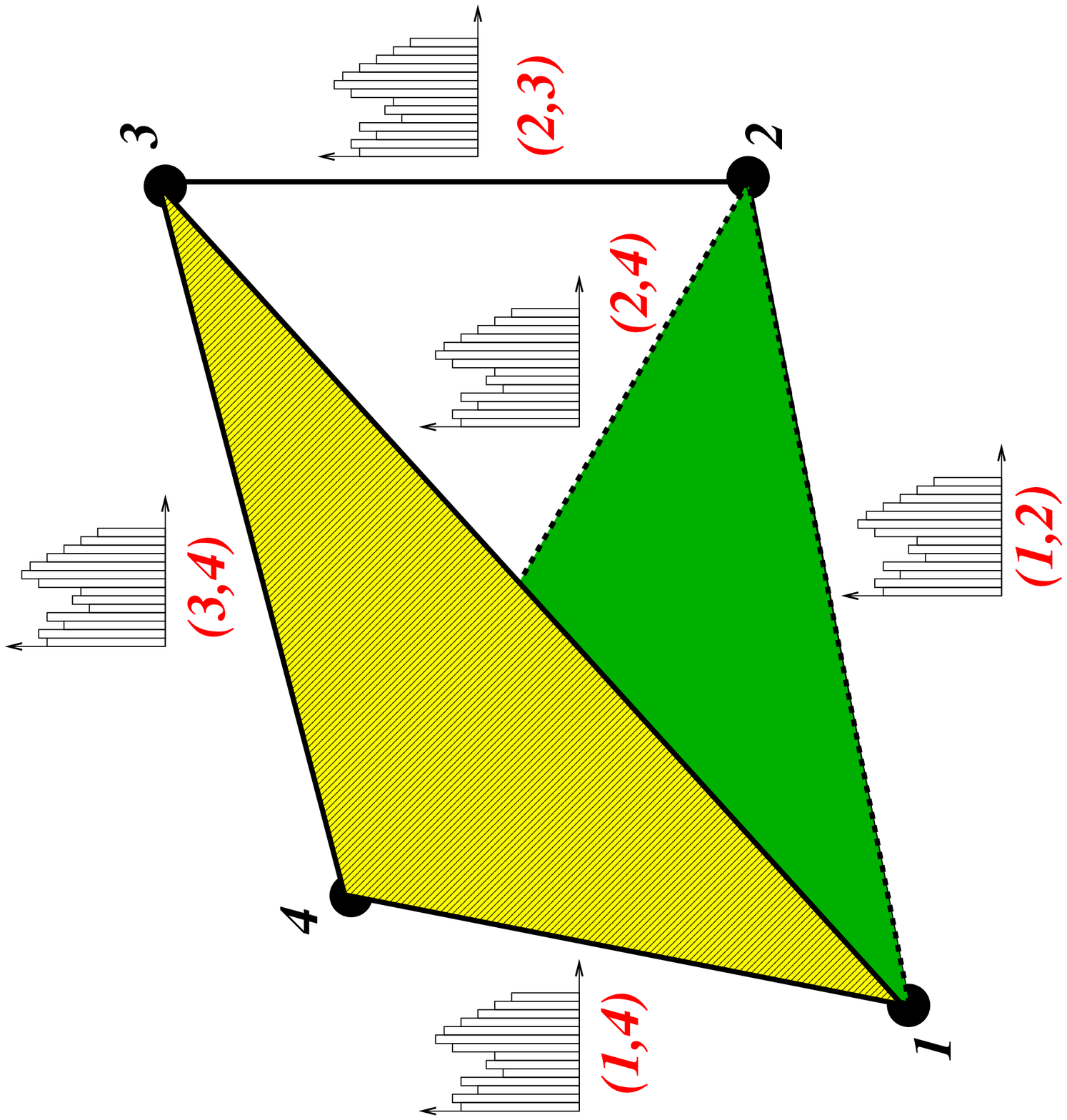,height=80mm,angle=270}}
\end{center}
\caption{\small\sf
  Geometry of the split of the 3-dimensional simplex cell.
}
\label{fig:split_simplex}
\end{figure}
\begin{figure}[!ht]
\begin{center}
{\epsfig{file=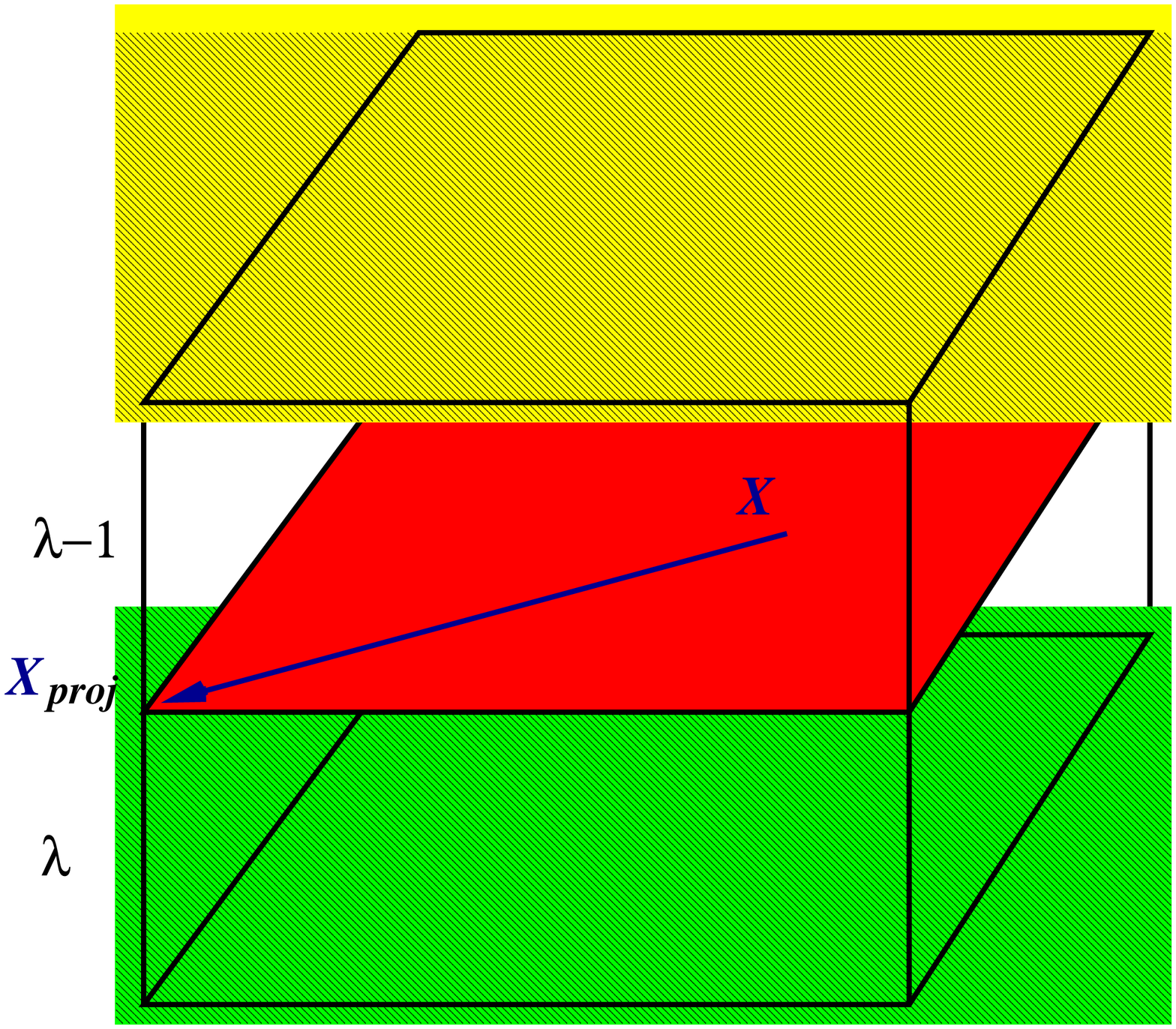,height=75mm,angle=270}}
\vspace{-5mm}
{\epsfig{file=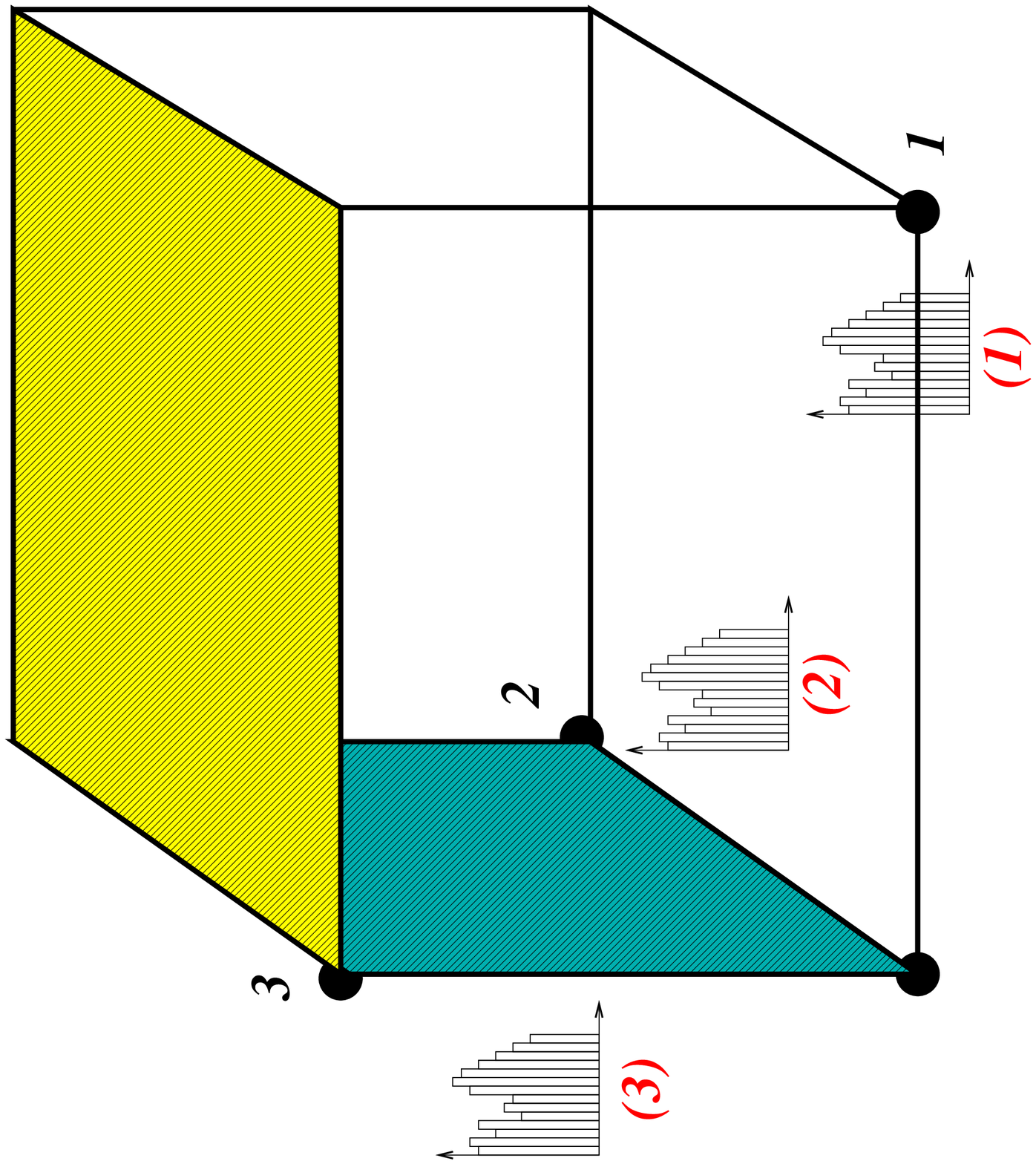,height=80mm,angle=270}}
\end{center}
\caption{\small\sf
  Geometry of the split of the 3-dimensional hyperrectangular cell.
}
\label{fig:split_hcub}
\end{figure}

\subsection{Projecting points into an edge}
Before we describe the determination of the division edge and of
the division parameter $\lambda_{div}$,
let us still discuss certain geometric aspects of the {\tt Foam} algorithm -- that is
how we project a point $\vec x$ inside a cell onto one of the edges of the cell.
In the case of a simplex the edges are numbered by the pair of indices $(i,j)$, $i>j$,
which number edges spanned by a pair of vertices%
\footnote{As explained in Sect.~\ref{sec:data_org}, the numbering of vertices
  is done using pointers $K_i$ to the elements of the array of vertices.}
$(\vec V_{K_i}, \vec V_{K_j})$,
while in the case of the hyperrectangle the $i$-th edge is spanned by
the pair of vectors $\vec q$ and 
$\vec{q}_+=(q_1,\dots,q_{i-1},q_i+h_i,\dots,q_n)$.
The point $x$ inside a cell is projected onto the edge and parametrized
using the parameter $\lambda\in(0,1)$.
The parameter $\lambda$ will be used to define an auxiliary projection $d \rho /d\lambda$
for each edge in the following subsection.
In particular we have to know how to evaluate $\lambda$ in an efficient way.
In the case of a simplex cell we have:
\begin{equation}
  \vec x_{proj}=\lambda \vec V_{K_i}  +(1-\lambda_{ij}) \vec V_{K_j},\;\;   0<\lambda_{ij}<1,
\end{equation}
where
\begin{equation}
  \begin{split}
    &\lambda_{ij}(\vec x) = \frac{ |{\rm Det}_i| }{ |{\rm Det}_i| +|{\rm Det}_j| },\\
    &{\rm Det}_i = {\rm Det}(\vec r_1,\dots,\vec r_{i-1},\vec r_{i+1},\dots,\vec r_n,\vec r_{n+1}),\\
    &{\rm Det}_j = {\rm Det}(\vec r_1,\dots,\vec r_{j-1},\vec r_{j+1},\dots,\vec r_n,\vec r_{n+1}),\qquad
    \vec r_j = \vec V_{K_j} -\vec x ,
  \end{split}
\end{equation}
and ${\rm Det}(x_1,x_2,...,x_n)$ is determinant.
The case of a hyperrectangular cell is much simpler:
\begin{equation}
  \lambda_i = (x_i-q_i)/h_i.
\end{equation}
Obviously, owing to the time-consuming evaluation of the determinants
at higher dimensions, the above projection
procedure will be much slower for simplices than for hyperrectangles.

In Fig.~\ref{fig:split_simplex} we illustrate a projection procedure into six edges
for 3-dimensional simplex and in Fig.~\ref{fig:split_hcub} the case of the
three edges of the 3-dimensional hyperrectangular cell.

\begin{figure}[!ht]
\begin{center}
  \epsfig{file=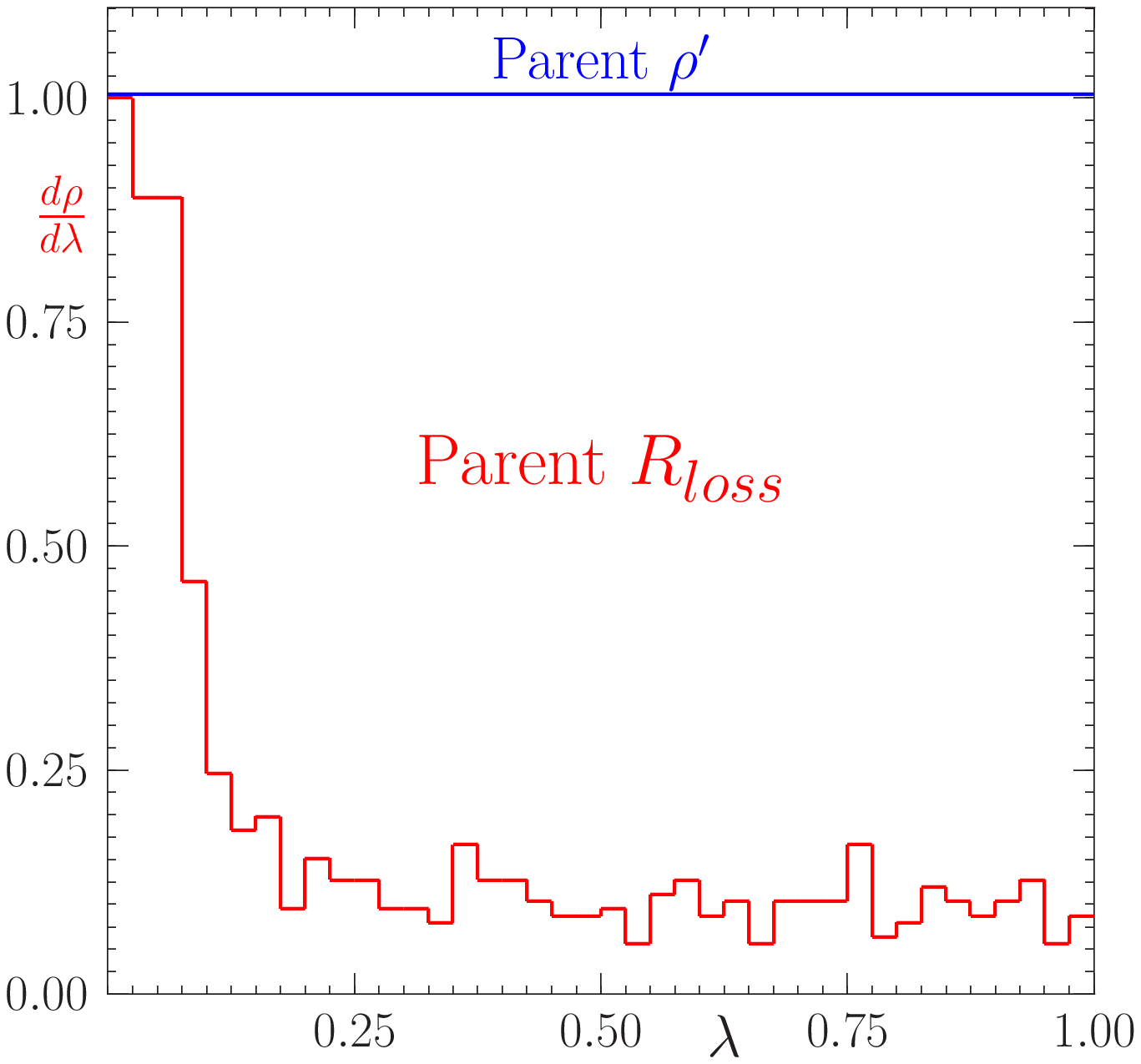,width=75mm}
  \epsfig{file=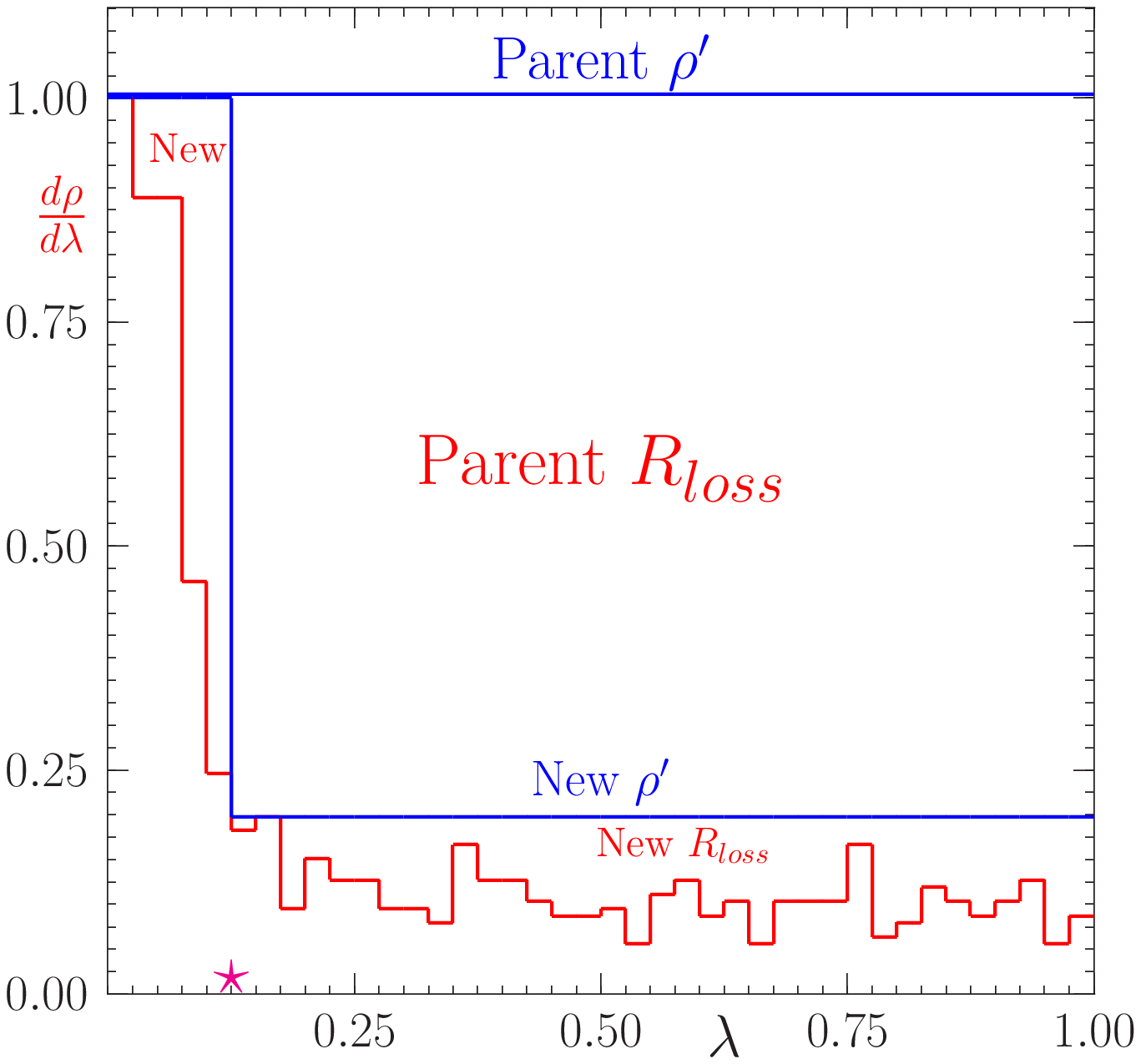,width=75mm}
\end{center}
\caption{\small\sf
  Projection histogram. The case of optimizing the maximum weight.
}
\label{fig:split_hist1}
\end{figure}
\subsection{Determination of an optimal division edge and of $\lambda_{div}$}

Our aim is to find out which division hyperplane, cutting through which edge,
provides the best gain of the total integral $R_{loss}$, summed over two daughters,
as compared to the parent cell.
In order to do that, we first analyse all possible division planes,
for all edges, and find out the best one, in terms of the gain in $R_{loss}$.
In other words, we go through all edges 
($k$ edges for a hyperrectangle and/or $n(n+1)/2$ for a simplex) and find,
for each edge, the optimal parameter $\lambda_{div}$ and the corresponding
best gain in $R_{loss}$.
We then compare between the gains in $R_{loss}$ for all edges and define the optimal edge
as the one with the best gain in $R_{loss}$.
The procedure of finding the best $\lambda_{div}$ is essentially the same for
simplical and hyperrectangular cells;
on the other hand, in the algorithm for
finding the best $\lambda_{div}$ there is a difference between the cases
of optimization of the maximum weight and of the variance; see the following discussion.

\subsubsection{Optimization of the maximum weight -- choosing $\lambda_{div}$}
\label{sec:projections}
Let us consider first the case of finding the best $\lambda_{div}$
for $R_{loss}$ corresponding to the {\em optimization of the maximum weight}.
The 1-dimensional case is a good starting point.
The cell in this case is just an interval $(q, q+h)$ and $\lambda=(x-q)/h$.
In the left-hand part of Fig.~\ref{fig:split_hist1},
we see a histogram with $N_b$ bins,
made of 1000 events generated inside a cell (interval)
using the weight $w=\rho(x)$,
so that the histogram represents approximately the distribution 
$d \rho/d\lambda$, $\lambda\in(0,1)$.
This distribution (histogram) peaks close to the lower edge.
The function $\rho'(x)= \max_{x\in Cell}\rho(x)$ is constant over the cell
and is depicted as an upper horizontal line marked ``Parent $\rho'$''.
The contribution of this particular (parent) cell to 
$R_{loss}=\int_{cell} \rho_{loss} dx =\int_{cell} (\rho'(x)-\rho(x)) dx$
is easily recognized as an area between the line marked ``Parent $\rho'$''
and the histogram line.
If we have stopped the exploration at this stage, with this parent cell,
then in the MC run points would be generated with the flat ``Parent $\rho'$''
and the weight would be $w=\rho(x)/\rho$.
Turning weighted events into unweighted ones by accepting $r<w$
events and rejecting $r\geq w$, where $0\leq r \leq 1$ is a uniform random number,
would correspond to generating points $(\lambda,r)$ within the rectangle
below the ``Parent $\rho'$'' line, accepting all points below the histogram line
and rejecting all those above, in the area marked ``Parent $R_{loss}$''.
This justifies the subscript ``loss''.

The best cell division is found by examining all $N_b-1$
end-points $\lambda= q+ih/N_b$, $i=1,2,...,N_b-1$
of the bins in the histogram, as a possible candidate for the division
point (plane in two and more dimensions) between the two daughter cells,
and choosing the best one.
In the right-hand part of Fig.~\ref{fig:split_hist1} 
we have marked such a candidate division point with a star.
For a given division point, we determine for two daughter cells
the new ``ceiling function'' $\rho'$; in Fig.~\ref{fig:split_hist1} it is 
the line marked ``New $\rho'$''. For each daughter cell we evaluate $R_{loss}$.
The summary $R_{loss}$ for both daughter cells is easily recognized
as an area between the line marked ``New $\rho'$'' and the histogram.
Of course, we automatically get the new total $R_{loss}$ smaller than for
the original parent cell!
This procedure is repeated for all possible $j=1,2,...,N_b-1$ division points
and each time we record the net gain in 
$\Delta_j R_{loss} = R_{loss,parent}-R_{loss,daughter1}-R_{loss,daughter2}$.
For the actual best division point we chose the division point
with the largest gain $\Delta_j R_{loss}$.
In Fig.~\ref{fig:split_hist1} the star marks the best division point.

\begin{figure}[!ht]
\begin{center}
  {\epsfig{file=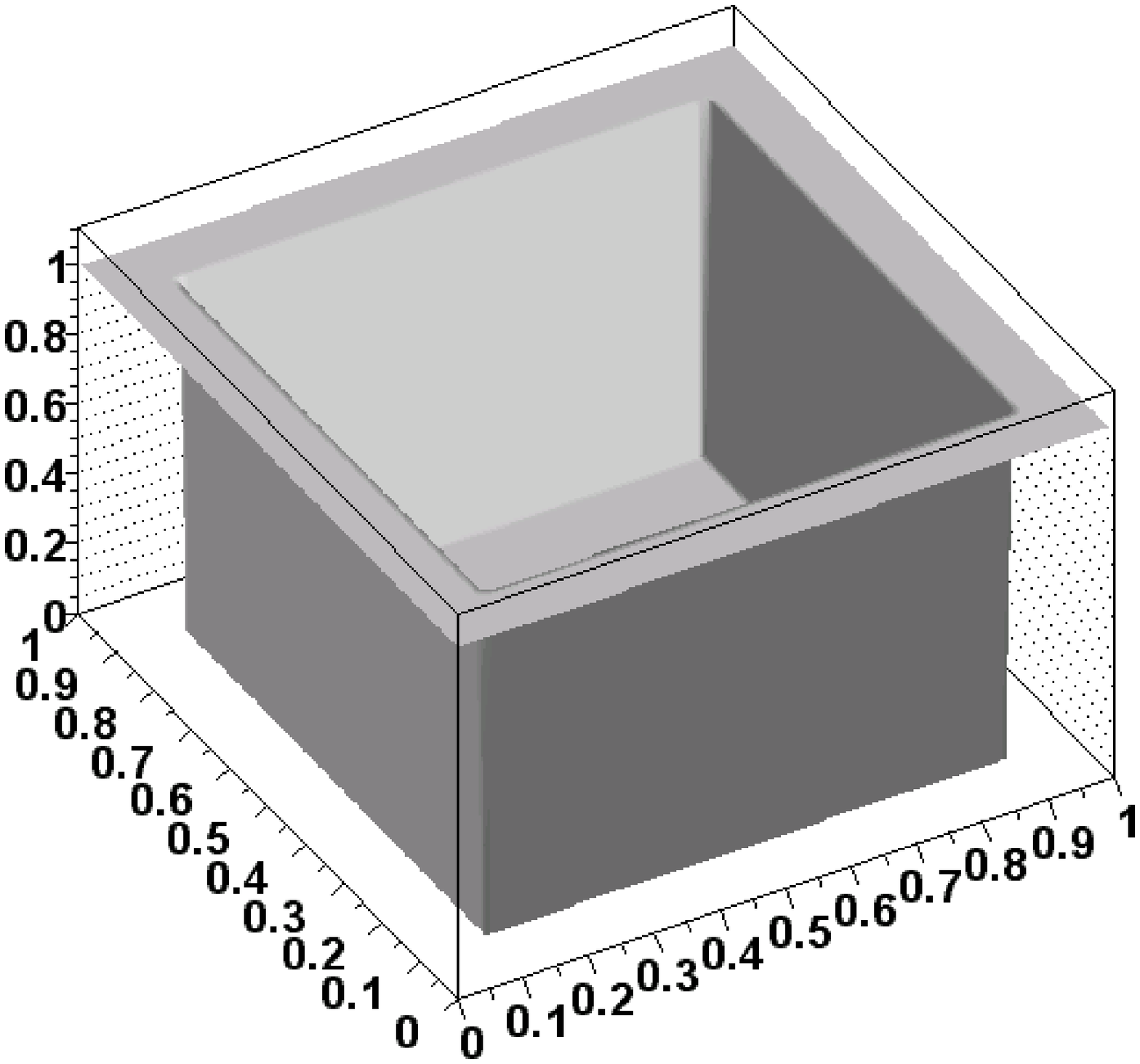,width=80mm}}
  {\epsfig{file=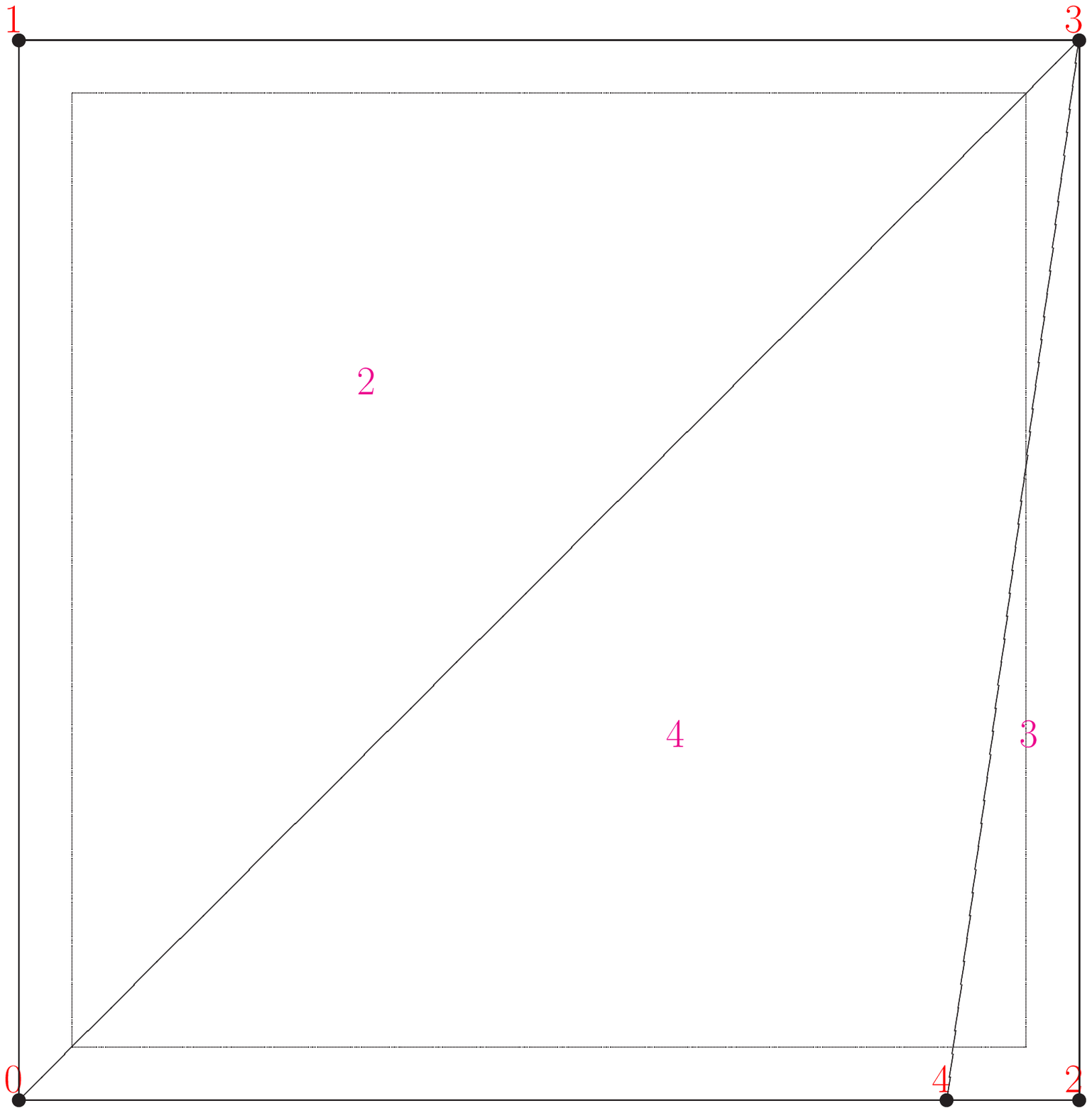,width=70mm}}
\end{center}
\caption{\small\sf
  Two-dimensional $\rho(\vec x)$ (left)
  and the geometry of the first three simplical cells (right).
  Inside the area marked by the inset rectangle $\rho(\vec x)=0$.
}
\label{fig:ashtray1}
\end{figure}

\begin{figure}[!ht]
\begin{center}
\epsfig{file=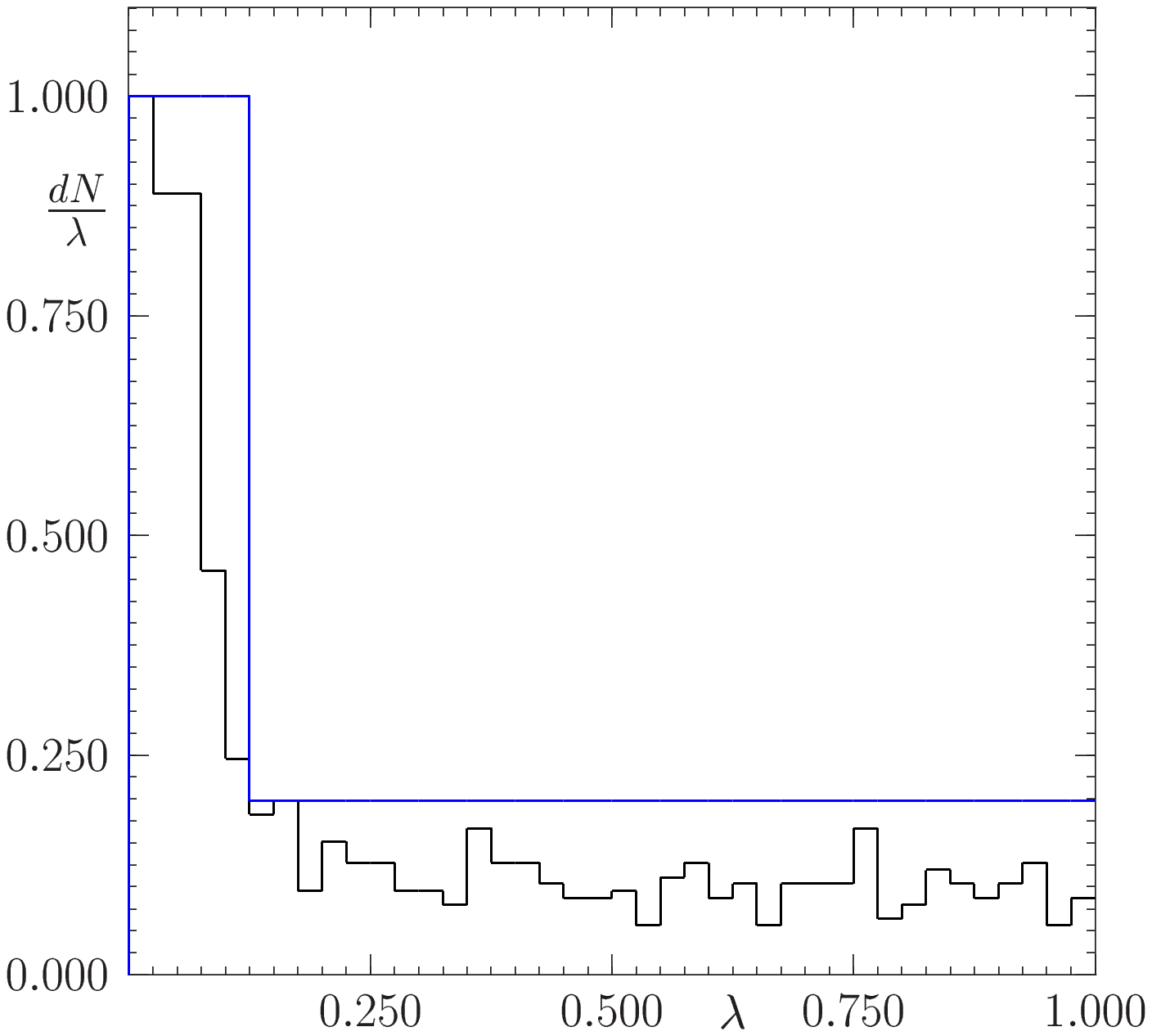,width=50mm}
\epsfig{file=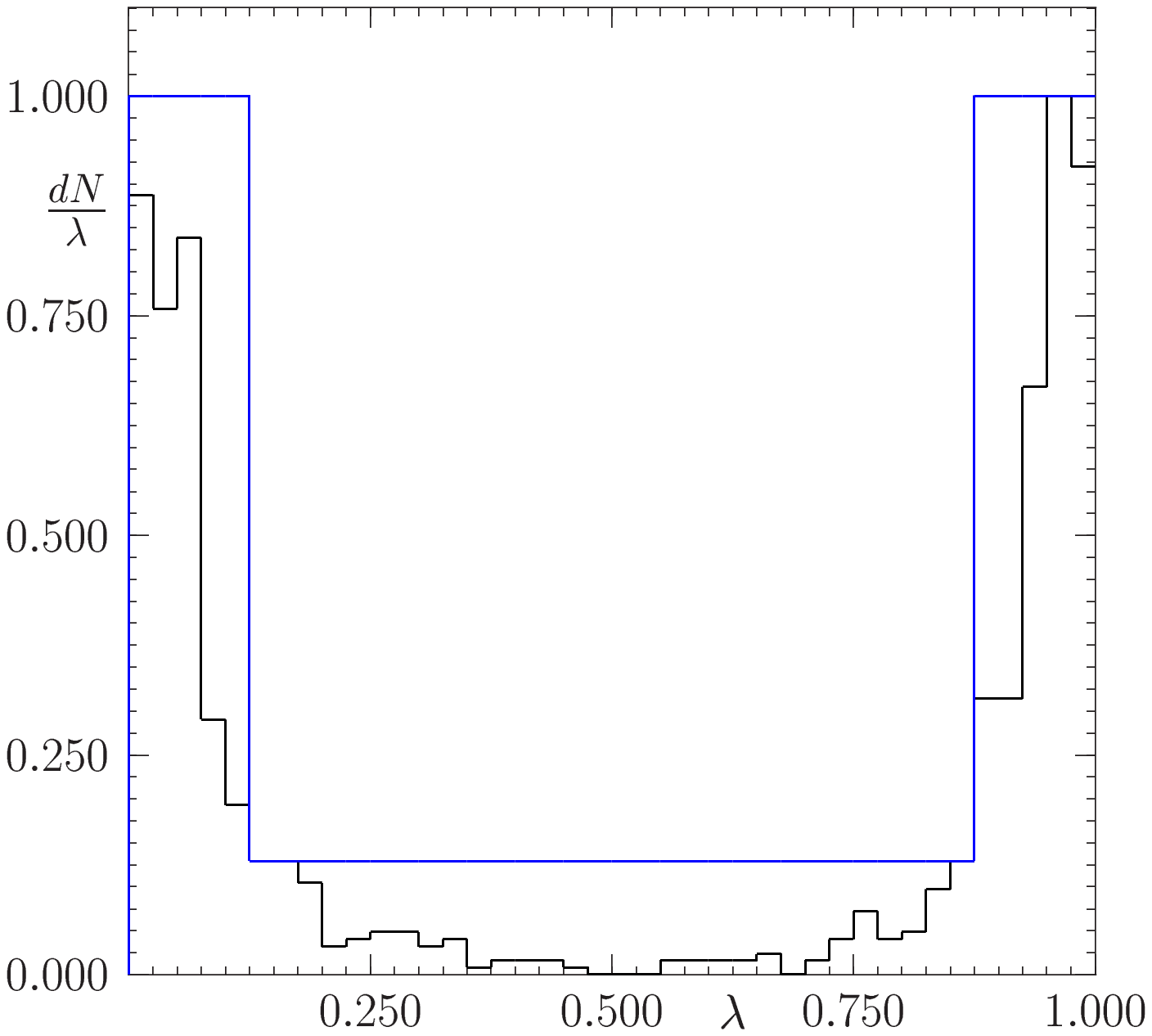,width=50mm}
\epsfig{file=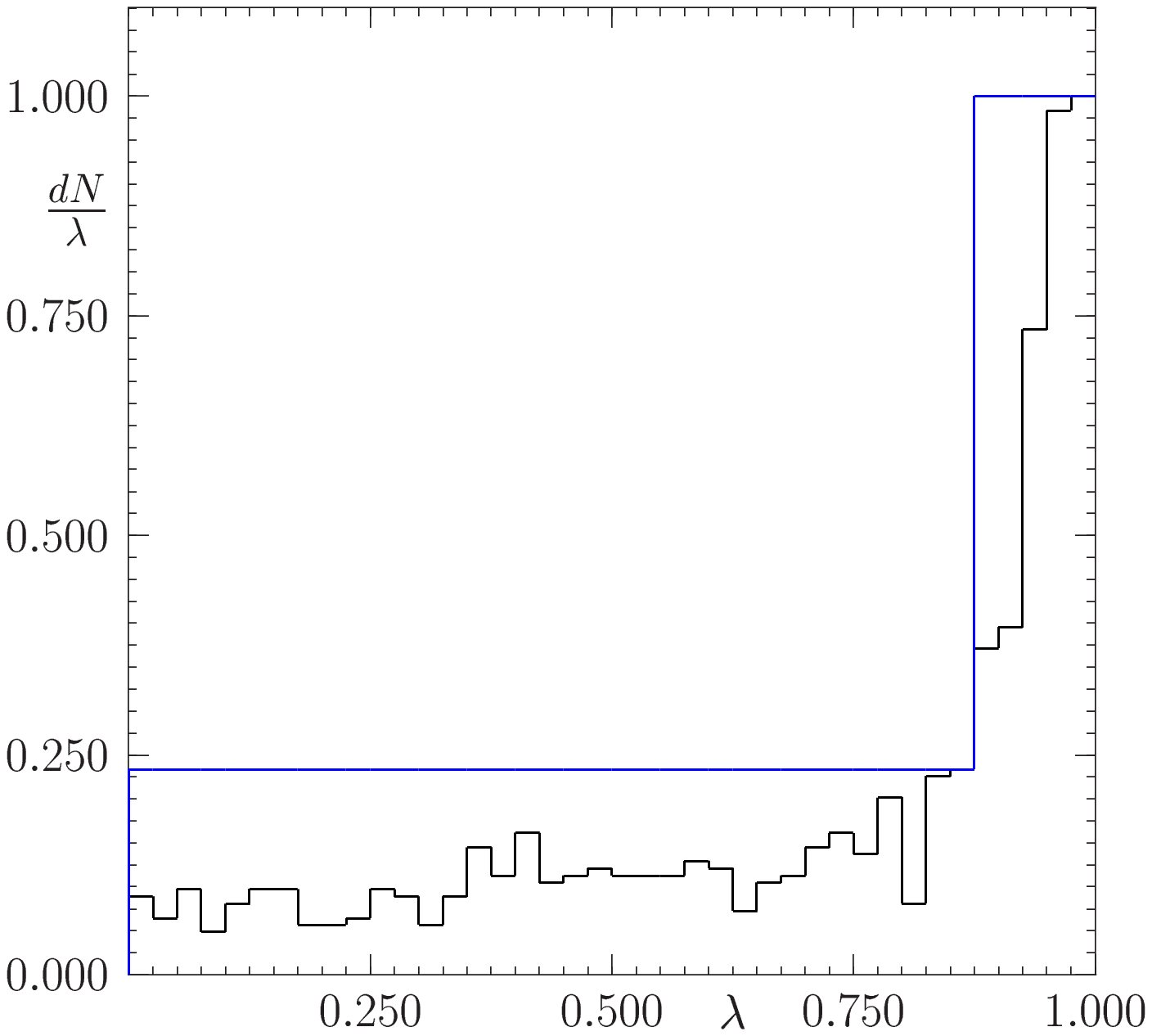,width=50mm}
\end{center}
\caption{\small\sf
  Distributions used in the construction of $\lambda_{div}$ 
  in the case of optimizing the maximum weight.
}
\label{fig:projs}
\end{figure}

Let us now consider the 2-dimensional distribution $\rho(x)$
depicted in the left part of Fig.~\ref{fig:ashtray1},
which is non-zero within the narrow strip along
the four edges of the rectangle.
In the simplical mode {\tt Foam} divides starting $n$-dimensional hyperrectangle
into $n!$ simplices -- in this case into two triangles.
We concentrate on the division procedure of the lower triangle;
see the right part of Fig.~\ref{fig:ashtray1}.
In the exploration of this triangular cell we use 1000 MC points
and project them onto three edges.
The corresponding three histograms are shown in Fig.~\ref{fig:projs},
where the middle histogram represents the projection onto the diagonal --
this is why it features two distinct peaks at the ends.
In all three plots we have also drawn the curve for $\rho'(x)$ for
the best hypothetical split.
The most promising split in terms of the gain in $R_{loss}$
turn outs to be related to the middle plot and is marked
in the right part of  Fig.~\ref{fig:ashtray1}.

The reader may notice that the $\rho'(x)$ in the middle
plot of Fig.~\ref{fig:projs} is not of the type
discussed above, because it has two discontinuities instead of one.
This is because in {\tt Foam} we have introduced an important refinement
of the algorithm to find an optimal $\lambda_{div}$.
One may easily notice that the algorithm described above
could not correctly locate the drop in the
distribution $d\rho/d\lambda$ of the middle plot, because there are
two equally strong peaks at the ends of this distribution%
\footnote{ The algorithm would pick up $\lambda_{div}$ at a random point
between the two peaks.}.
In the improved version of the algorithm the search of the
optimal $\lambda_{div}$ uses all pairs of the bin edges
$(\lambda_i,\lambda_j)=(q+ih/N_b,\;q+jh/N_b)$, $0\leq i<j\leq N_b$.
For every pair $(i,j)$ a new ceiling function $\rho'(x)$
is determined, such that it is unchanged outside the subinterval $(\lambda_i,\lambda_j)$
and is ``majorizing'' the histogram bins inside this subinterval.
Once we find out the best pair $(i,j)$ in terms of $R_{loss}$, then
we take either $\lambda_i$ or $\lambda_j$ as a division point $\lambda_{div}$
(at least one of them is not equal to 0 or 1).
In the case of two or more peaks in $d\rho/d\lambda$
the resulting division point $\lambda_{div}$ happens to be
close to one of the edges of the gap between the two peaks.
This feature prevents the {\tt Foam} algorithm (at least partly)
from placing a new division plane across a void 
in the multidimensional distribution $\rho(x)$.
In other words such a void will ``repel'' the division hyperplanes.
In the case of the double peak structure in 
the middle plot of Fig.~\ref{fig:projs}, the improved algorithm
will of course allocate the big value of $R_{loss}$ to a new cell (interval),
which includes the gap.
In the next step of the foam build-up this cell (interval) will have
a big chance to be split, and for this split the position of the split point
will be located at the second edge of the gap.
This is exactly what we need for an efficient foam evolution.

\begin{figure}[!ht]
\begin{center}
\epsfig{file=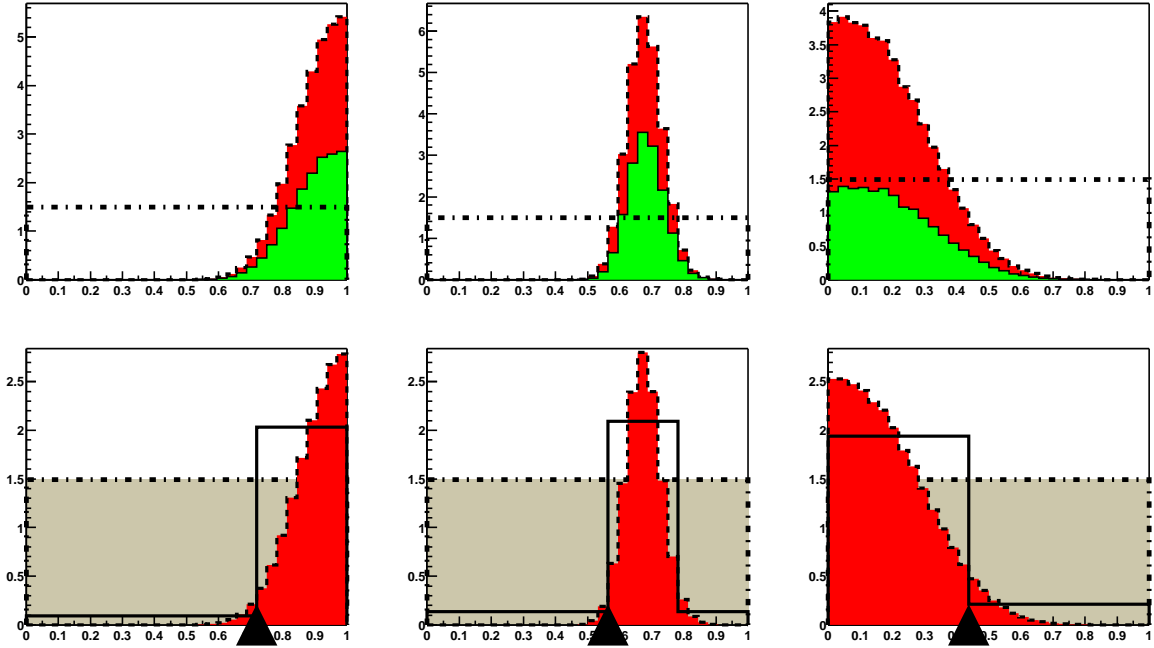,width=160mm}
\end{center}
\caption{\small\sf
  Distributions used in the construction of $\lambda_{div}$
  in the case of optimizing the variance.
}
\label{fig:projv}
\end{figure}

\subsubsection{Optimization of the variance -- choosing $\lambda_{div}$ }
Let us consider now the case of finding out the best $\lambda_{div}$
for $R_{loss}$ corresponding to {\em optimization of the variance}.
The strategy is again to choose  $\lambda_{div}$ by minimizing $R_{loss}$.
In Fig.~\ref{fig:projv} we illustrate our algorithm on the example
of the three projections of the triangular cell.
The three projections correspond to a triangular cell in two dimensions.
(We do not specify $\rho(x)$, as it is irrelevant for the purpose of our explanation.)
In the upper row of the three plots in Fig.~\ref{fig:projv}
we show as a horizontal line the value of the 
distribution $\rho_{loss}= \sqrt{\langle \rho^2 \rangle}$
(it is the same for all three projections).
The solid histogram is the distribution $d\rho/d\lambda$
and the dashed histogram is the distribution $d\rho_{loss}/d\lambda$
calculated bin by bin using $\sqrt{\langle \rho^2 \rangle}$,
treating every bin as a separate cell.
The properly normalized difference of the above two distributions
$\sqrt{\langle \rho^2 \rangle}-\langle \rho \rangle$
is plotted separately as the dashed histograms in the lower row of
the three plots in  Fig.~\ref{fig:projv}.
The horizontal line for the total $\rho_{loss}$ is shown once again there.
The histogram of $d\rho_{loss}/d\lambda$ gives us an idea, where
the biggest source of the variance is located and our aim is to
``trap'' it properly with an intelligent choice of $\lambda_{div}$.
We follow an algorithm similar to that of the maximum weight minimization,
namely we loop over pairs of the bin edges
$(\lambda_i,\lambda_j)=(q+ih/N_b,q+jh/N_b)$, $0\leq i<j\leq N_b$,
and for every pair we calculate $R_{loss}$
inside the interval $(\lambda_i,\lambda_j)$ and outside it.
We find out which $(i,j)$ provides the biggest
gain $\Delta_{ij} R_{loss} = R_{loss,parent} -R_{loss,Inside}-R_{loss,Outside}$.
In the lower row of the plots in Fig.~\ref{fig:projv} we show as a solid line
the distribution of $R_{loss}$ for the best pair  $(i,j)$.
Depending on the peak structure, at least one of 
the division points of the optimal pair $(\lambda_i,\lambda_j)$ is different from
0 or 1, and we take this one as $\lambda_{div}$.
In Fig.~\ref{fig:projv} the chosen $\lambda_{div}$
are marked with black triangles.
The above procedure is done for each edge, and the best $\Delta_{ij} R_{loss}$
is used as a guide to define an edge for which the next cell division will be executed.
The information about the best edge and the best division point $\lambda_{div}$
is recorded in the cell object.
As seen in Fig.~\ref{fig:projv}, $\lambda_{div}$
tends to fall at the position where $d\rho_{loss}/d\lambda$ drops or increases sharply.
Note that since the division point is always at the edge of the bin,
it is therefore a rational number, $\lambda_{div}=j/N_b$.
This has interesting consequences,
since the number of the bins $N_b$ is fixed,
it is therefore sufficient to memorize this integer index $j$ (2~Bytes) together
with the integer index of the division cell edge (also 2~Bytes) as an attribute of the cell,
in order to define fully and uniquely the geometry of the division of the cell!
See Section \ref{sec:save_memory} for more details on how this is exploited to
save the computer memory needed to encode the entire foam of cells.

\begin{figure}[!ht]
\begin{center}
\epsfig{file=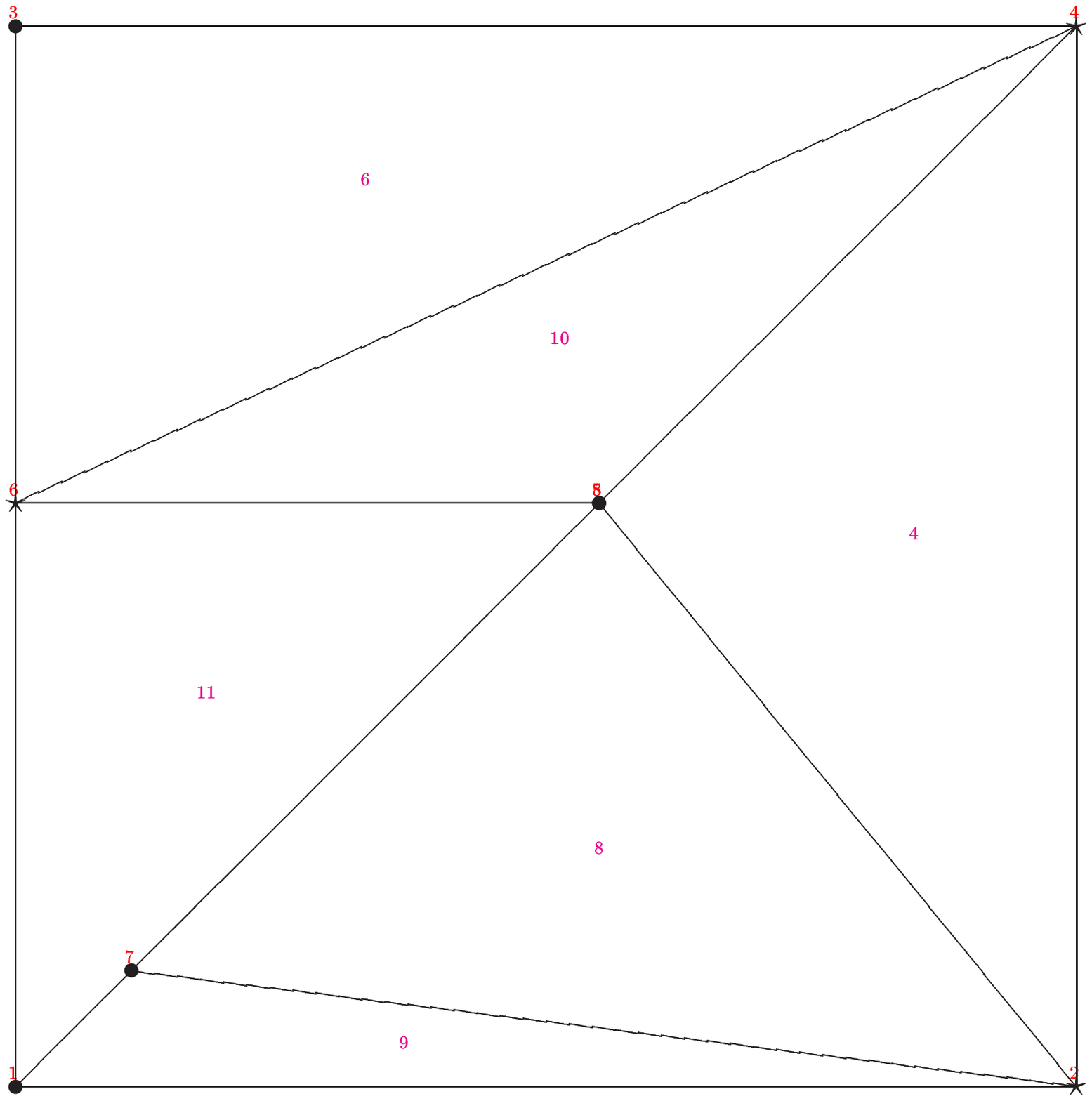,width=50mm}
\epsfig{file=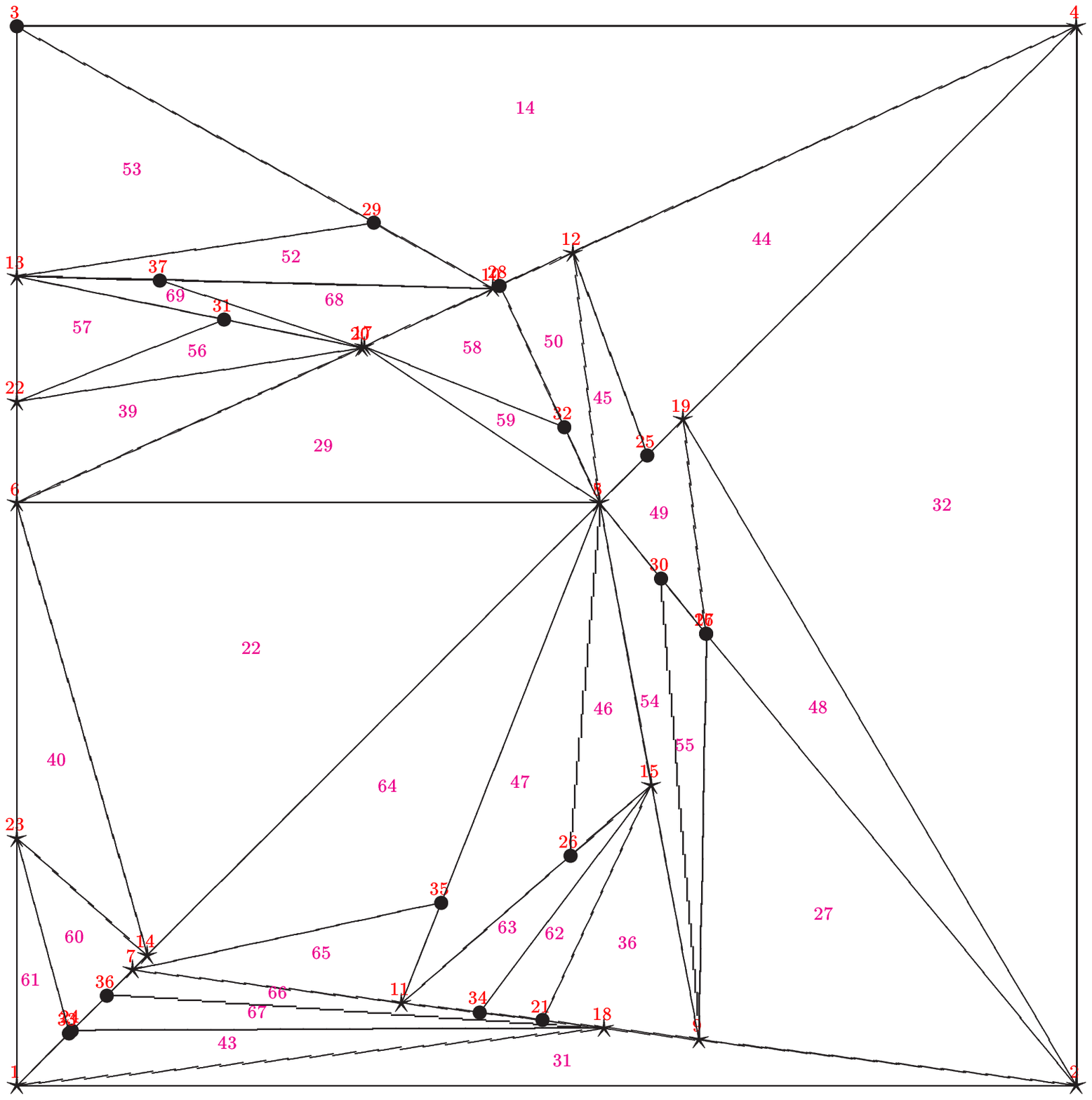,width=50mm}
\epsfig{file=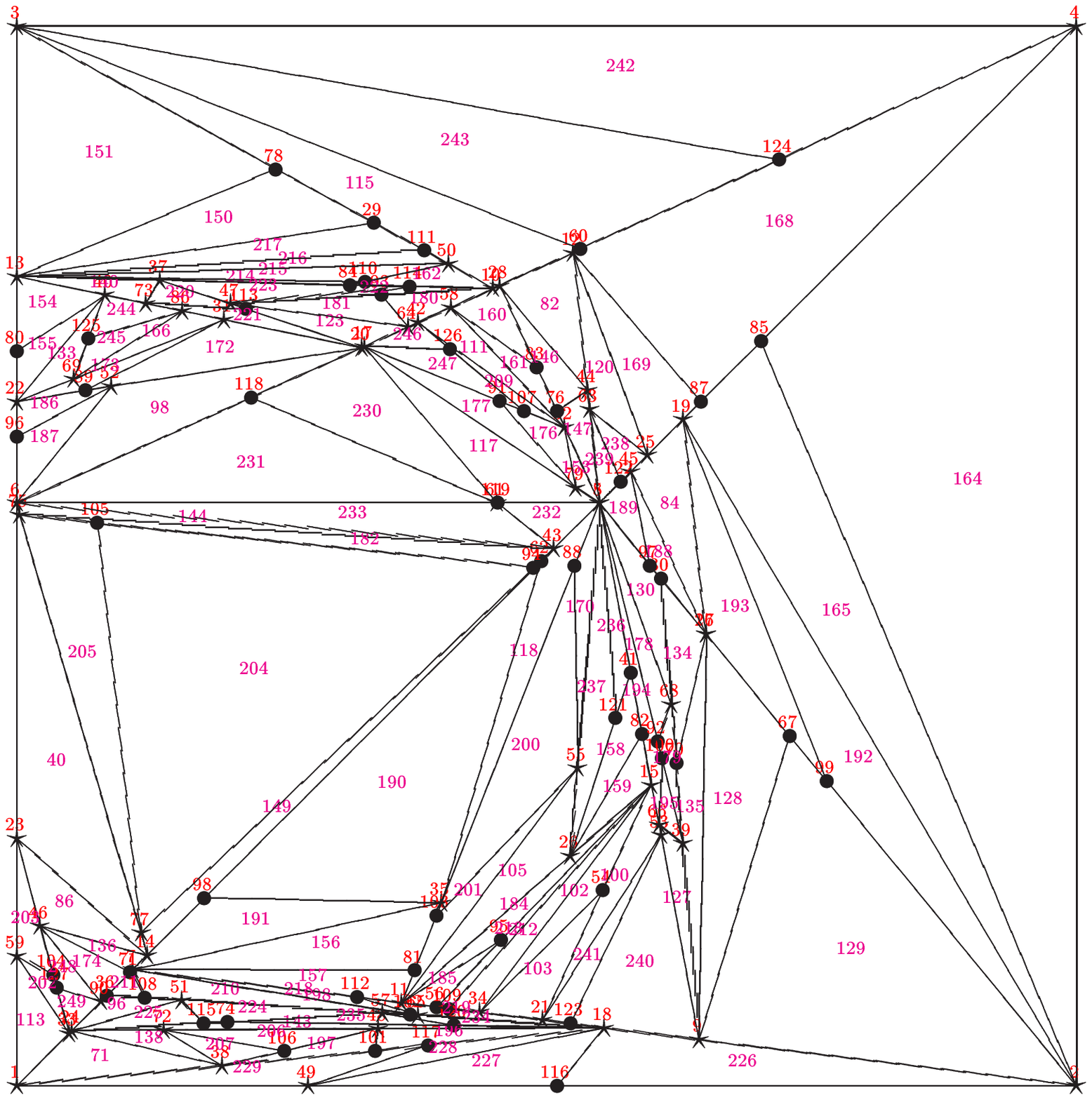,width=50mm}\\
\epsfig{file=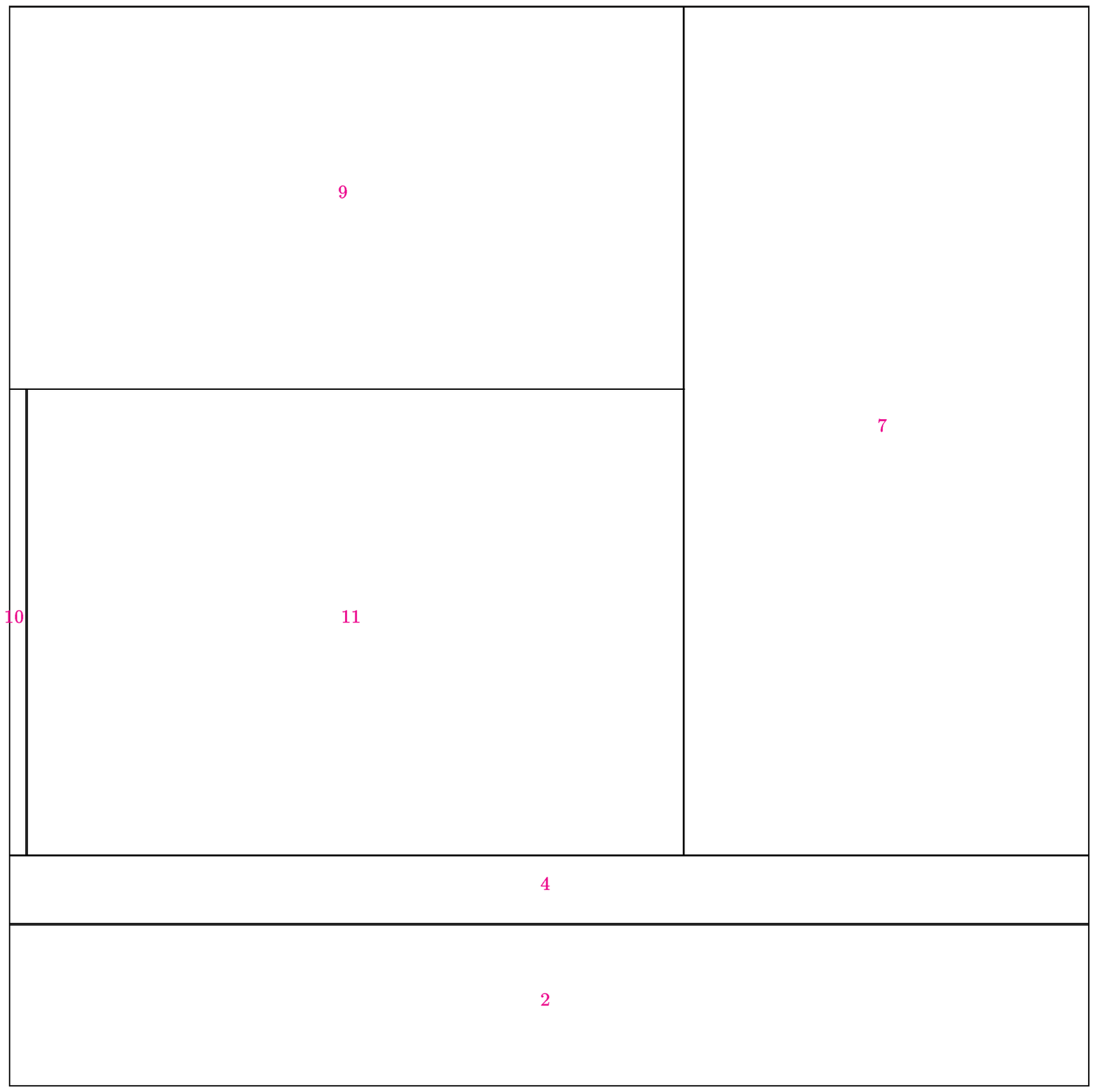,width=50mm}
\epsfig{file=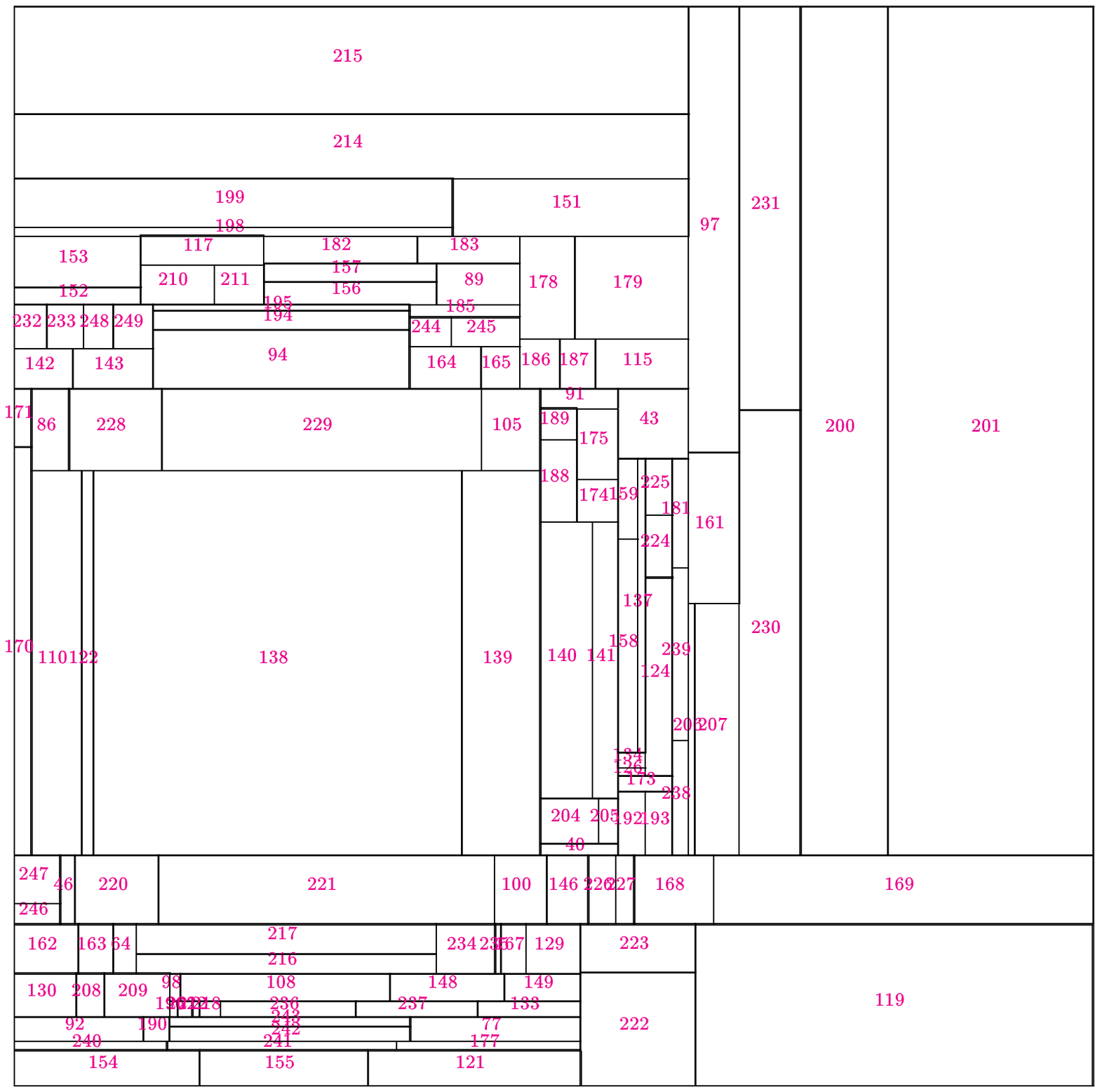,width=50mm}
\epsfig{file=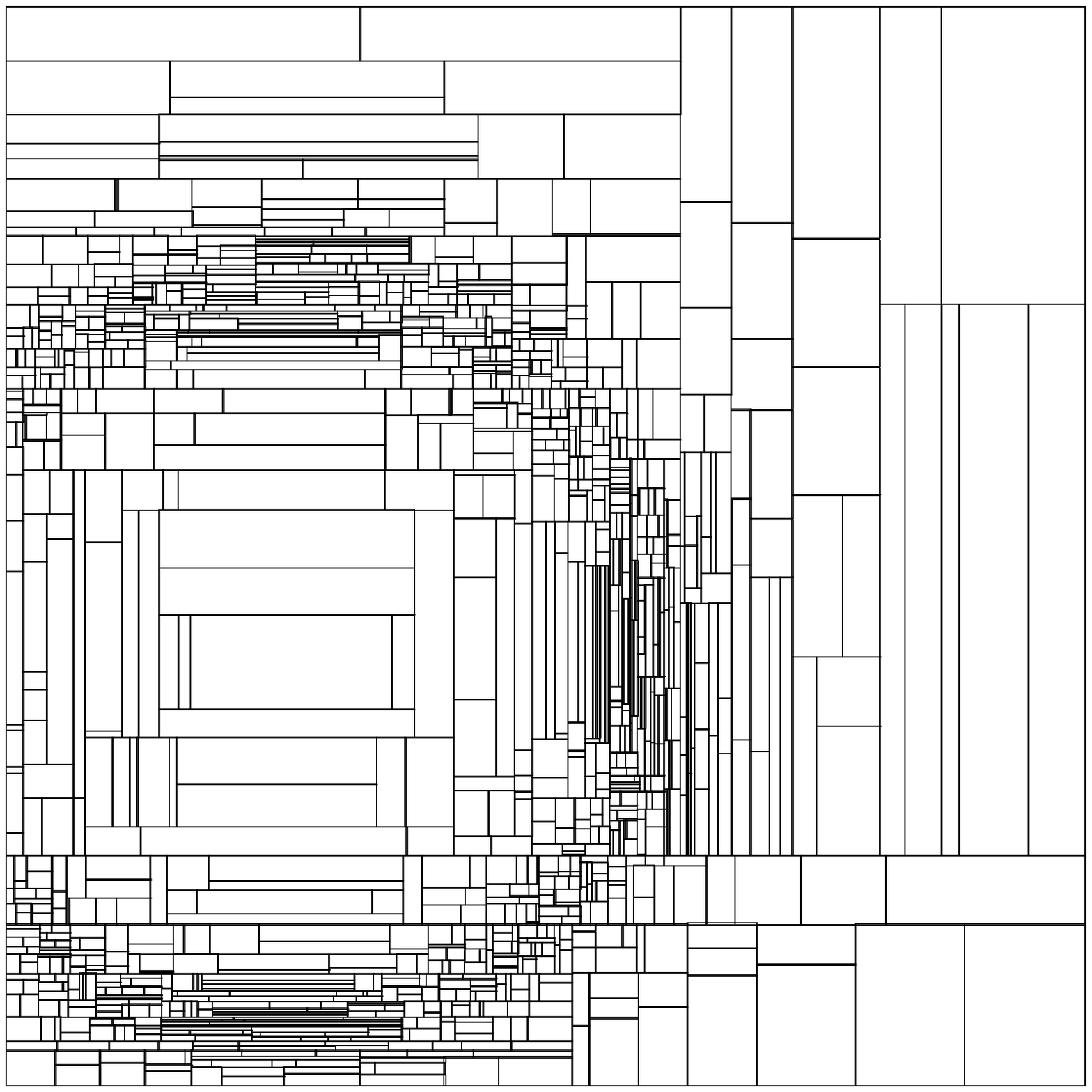,width=50mm}\\
\epsfig{file=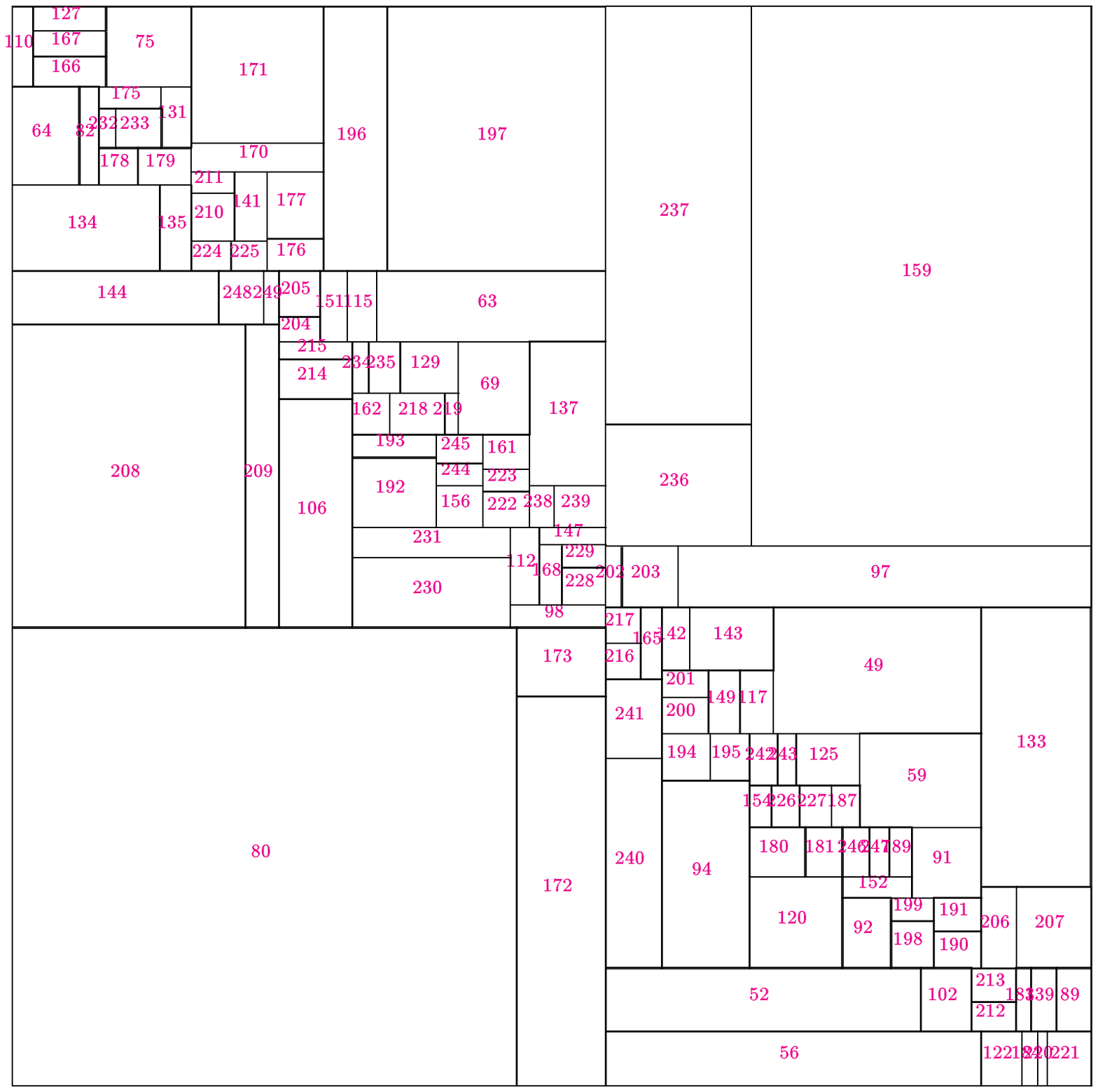,width=50mm}
\epsfig{file=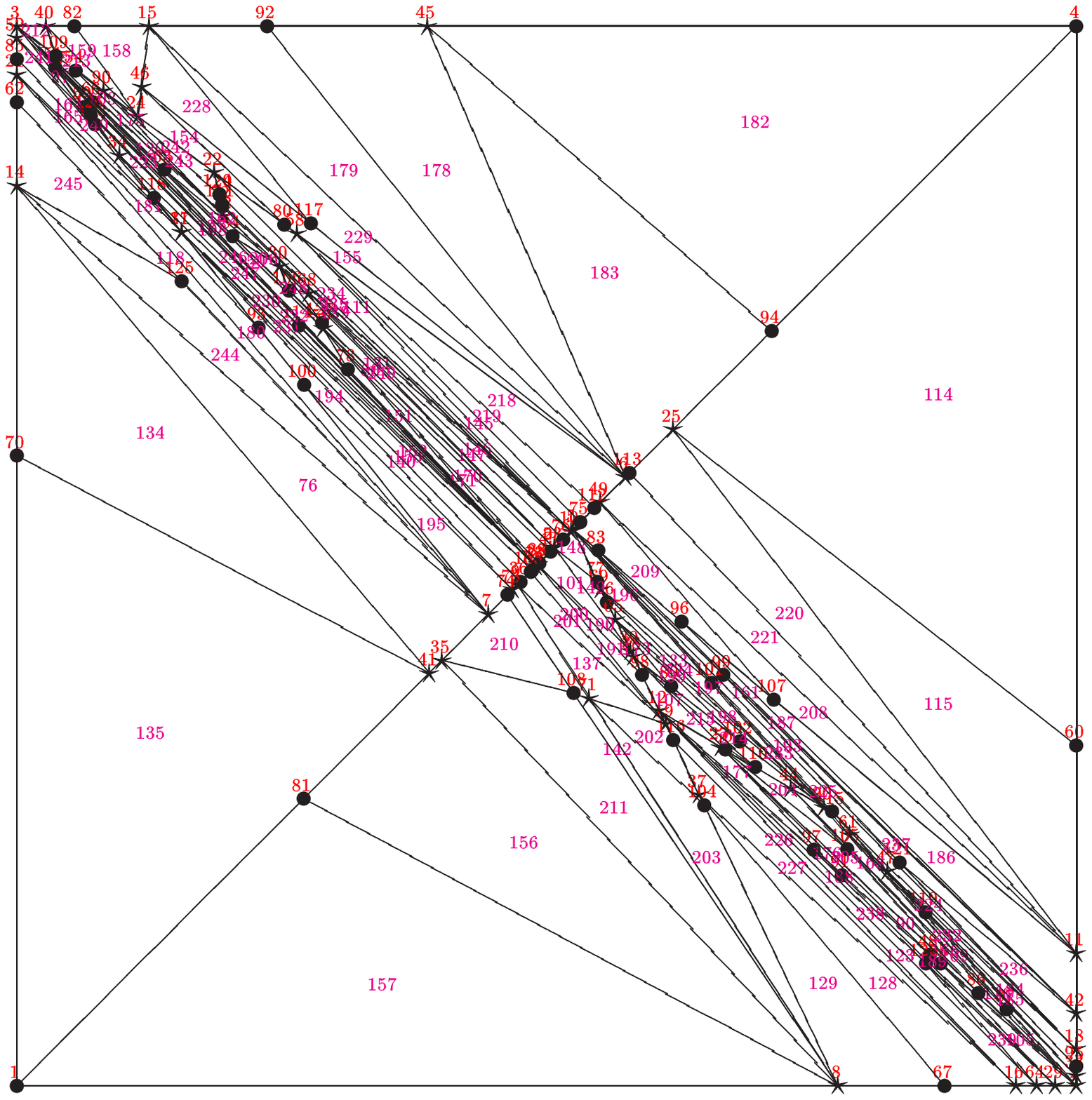,width=50mm}
\epsfig{file=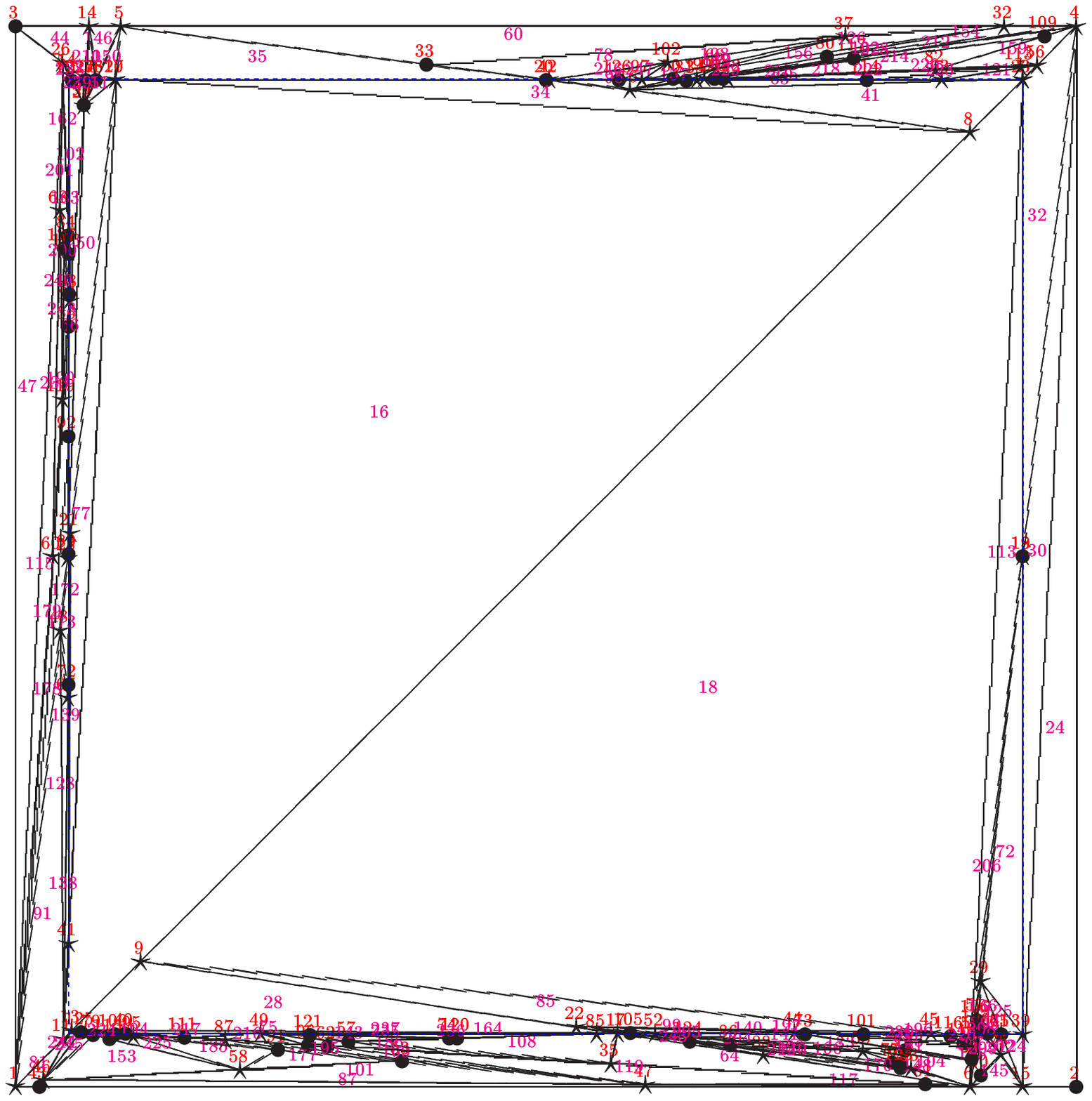,width=50mm}
\end{center}
\caption{\small\sf
  Examples of the 2-dimensional foam. Number of cells from 10 to 2500.
}
\label{fig:evolution}
\end{figure}
\begin{figure}[!ht]
\begin{center}
\epsfig{file=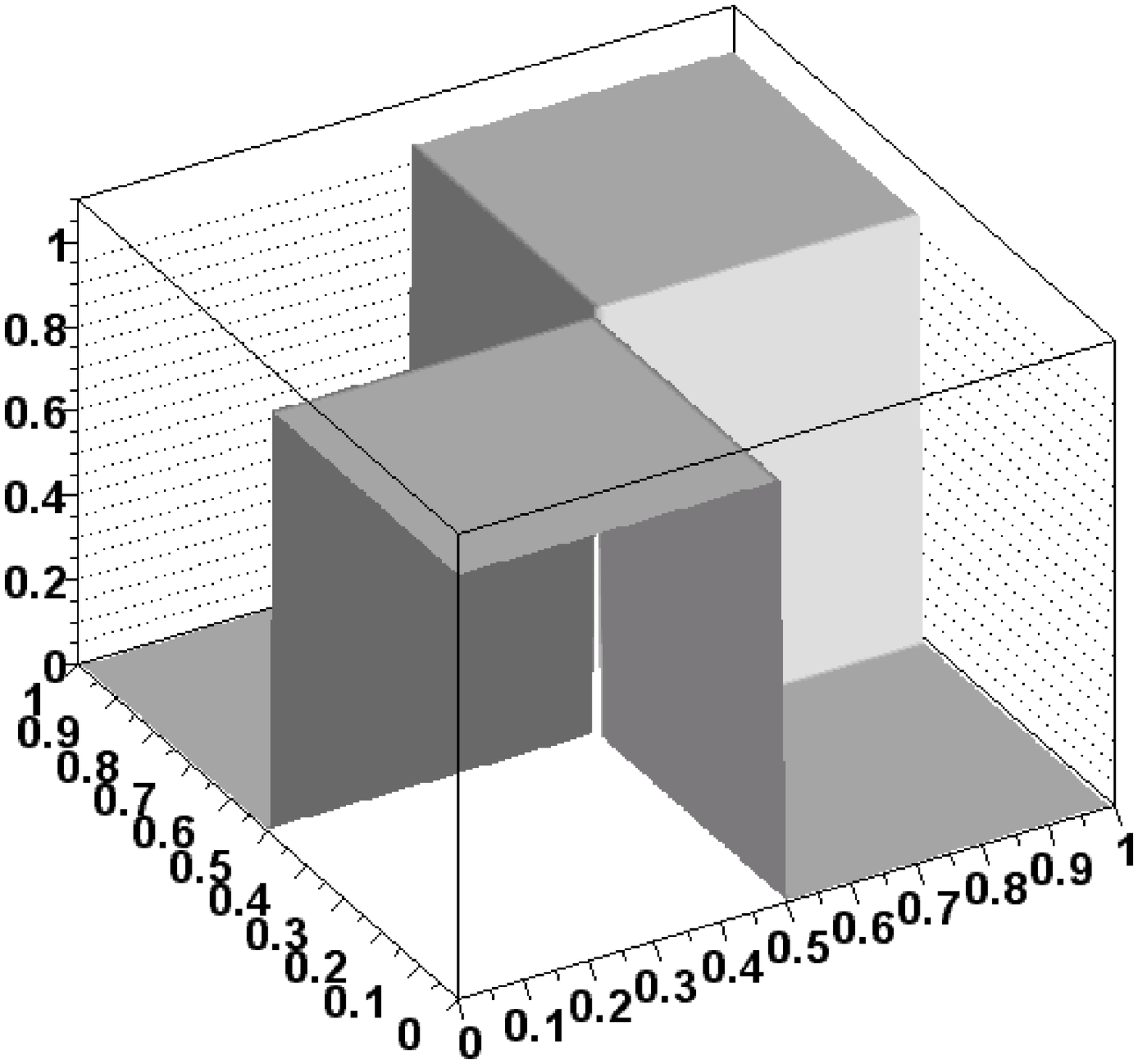,width=80mm}
\end{center}
\caption{\small\sf
  Example of $\rho(x)$ for which the {\tt Foam} algorithm of cell division fails.
}
\label{fig:failure}
\end{figure}

\subsubsection{Concluding remarks on the cell division algorithm}
The algorithm of the split of the cell is the important and most sophisticated part
of the new {\tt Foam}.
Let us therefore add a couple of final remarks:
\begin{itemize}
\item
  The new, much improved procedure of the choice of the division plane
  is the most significant difference%
  \footnote{ I would like to thank A. Para for a discussion which ignited this new development.}
  with respect to {\tt Foam} 1.x of Ref.~\cite{foam1:2000}.
\item
  The choice of the edge based on the histograms for each edge makes sense
  if we use histograms with at least 2--4 bins and at least 100 MC events
  per cell. This might be a serious limitation for these $\rho(x)$, which
  require a lot of CPU time per function call.
\end{itemize}
Finally, in Fig.~\ref{fig:evolution} we
show examples of the evolution of the foam of cells
as the number of the cells grows gradually.
The 2-dimensional case is easily visualized and we show it in 
Fig.~\ref{fig:evolution} for triangular and rectangular cells.
In the upper six plots, $\rho(x)$ features a circular ridge, in the bottom two plots
it is concentrated along the antidiagonal $x_1+x_2=1$,
the last one corresponding to the $\rho(x)$ of Fig~\ref{fig:ashtray1}.

\subsection{Limitations}
We are aware that the present procedure of selecting the next cell for the split and the
procedure of defining the division plane, although quite sophisticated, is not a perfect
one and has certain shortcomings.
Some of them can probably be corrected for, but some of them are inherent to the method.
In Fig.~\ref{fig:failure} we show a surprisingly simple example of a function for which
our method of finding a good division point $\lambda_{div}$ fails.
It fails simply because both projections of $\rho(x)$ onto two edges of the rectangle
are just flat and our procedure will pick up some $\lambda_{div}$ randomly within $(0,1)$,
while the most economic division point is in the middle $\lambda_{div}=1/2$.
On the other hand, although the {\tt Foam} algorithm gets disoriented for the first division,
it will recover and correct for the false start in the next divisions rather quickly.
It will eliminate the two voids from its area of interest.

Let us notice that the distribution of Fig.~\ref{fig:failure} violates maximally
the ``principle of factorizability''
$\rho(\vec x)=\prod_1^n \rho_i(x_i)$,
the principle on which the {\tt VEGAS} family of the programs is
built~\cite{Lepage:1978sw,Kawabata:1995th,Ohl:1998jn}.
Contrary to {\tt VEGAS} the problem with factorizability 
in {\tt Foam} is not a general one, but is limited to a single cell
and usually goes away after the cell split.
Nevertheless, the algorithm of {\tt Foam} analysing projections on all edges
in a single cell is relying on the ``principle of factorizability''.

\begin{table}[!ht]
\centering
\begin{small}
\begin{tabular}{|l|p{12.0cm}|}
\hline
Class & Short description\\ 
\hline\hline
{\tt TFOAM\_INTEGRAND} &
Abstract class (interface) for the {\tt Foam} integrand distribution $\rho(x)$\\
{\tt TFVECT} &
Class of vectors with dynamic allocation of the components.
Used in {\tt TFOAM} and {\tt TFCELL}\\
{\tt TFMATRIX} &
Square matrices, used for simplical geometry in the foam\\
{\tt TFPARTITION} &
Auxiliary small class for looping over partitions and permutations\\
{\tt TFCELL} &
Class of objects presenting single cell used in {\tt TFOAM}
(Cartesian product of the simplex and hyperrectangle)\\
{\tt TFOAM} &
Main class of {\tt Foam}. The entire MC simulator\\
\hline
{\tt TPSEMAR} &
Marsaglia etal. random number generator\cite{Marsaglia:1990ig}.\\
{\tt TFHST} &
Simple class of one-dimensional histograms.
Used only in the {\tt Foam} version without ROOT\\
{\tt TFMAXWT} &
Monitors MC weight, measures performance of the MC run\\
{\tt TFDISTR} &
Collections of distributions $\rho(x)$ for testing {\tt Foam}\\
\hline
\end{tabular}
\end{small}
\caption{\sf Description of C++ classes of {\tt Foam}.}
  \label{tab:classes}
\end{table}

\begin{table}[!ht]
\centering
\begin{small}
\begin{tabular}{|l|p{12.0cm}|}
\hline
{\tt TFOAM} member & Short description  \\ 
\hline\hline
  float m\_Version$^g$ & Actual version of the {\tt Foam} (like 2.34)\\
  char  m\_Date[40]    & Release date of the {\tt Foam}\\
  char  m\_Name[128]   & Name of a given instance of the {\tt TFOAM} class\\
  int m\_nDim$^s$      & Dimension of the simplical subspace\\
  int m\_kDim$^s$      & Dimension of the hyperrectangular subspace\\
  int m\_TotDim$^g$    & Total dimension = m\_nDim+m\_kDim\\
  int m\_nCells$^s$    & Maximum number of cells\\
  int m\_vMax          & Maximum number of vertices (calculated)\\
  int m\_LastVe        & Actual index of the last vertex\\
  int m\_RNmax         & Maximum number of random numbers generated at once\\
\hline
  int m\_OptDrive$^s$  & Type of optimization =1,2 for variance or maximum weight reduction\\
  int m\_OptEdge$^s$   & Decides whether vertices are included in the cell MC exploration\\
  int m\_OptPeek$^s$   & Type of cell peek =0,1,2 for maximum, random, random2\\
  int m\_OptOrd$^s$    & Root cell is simplex for OptOrd=1, hyperrectangle for OptOrd=2\\
  int m\_OptMCell$^s$  & =1 economic memory for hyperrectangles is on; =0 off\\
  int m\_Chat$^s$      & =0,1,2 chat level in output; =1 for normal output\\
  int m\_OptDebug      & =1, additional histogram (dip-switch)\\
  int m\_OptCu1st      & =1, numbering starts with hyperrectangle (dip-switch)\\
  int m\_OptRej        & =0 for weighted events; =1 for unweighted events in MC generation\\
\hline
  int  m\_nBin$^s$     & No. of bins in edge histogram for cell MC exploration\\
  int  m\_nSampl$^s$   & No. of MC events, when dividing (exploring) cell\\
  int  m\_EvPerBin$^s$ & Maximum number of effective ($w=1$) events per bin\\
  int  m\_nProj        & Number of projection edges (calculated)\\
\hline
\end{tabular}
\end{small}
\caption{\sf Data members of the {\tt TFOAM} class.
  Associated setters and getters marked as superscripts $s$ and $g$.}
  \label{tab:TFOAMmembers1}
\end{table}

\begin{table}[!ht]
\centering
\begin{small}
\begin{tabular}{|l|p{100mm}|}
\hline
{\tt TFOAM} member & Short description  \\ 
\hline
\multicolumn{ 2}{|c|}{{ Provision for the multibranching } }\\
\hline
  int    *m\_MaskDiv     &![m\_nProj] Dynamic mask for cell division\\
  int    *m\_InhiDiv     &![m\_kDim]  Flags inhibiting cell division, h-rectang. subspace\\
  int     m\_OptPRD      &Option switch for predefined division, for quick check\\
  TFVECT **m\_XdivPRD    &!Lists of division values encoded in one vector per direction\\
\hline
\multicolumn{ 2}{|c|}{{ Geometry of cells } }\\
\hline
  int m\_NoAct           &Number of active cells\\
  int m\_LastCe          &Index of the last cell\\
  TFCELL **m\_Cells      &[m\_nCells] Array of ALL cells\\
  TFVECT **m\_VerX       &[m\_vMax] Array of pointers to vertex vectors\\
\hline
\multicolumn{ 2}{|c|}{{ Monte Carlo generation } }\\
\hline
  double m\_MaxWtRej;    &Maximum weight in rejection for getting $w=1$ events\\
  TFMAXWT *m\_MCMonit;   &Monitor of the MC weight for measuring MC efficiency\\
  TFCELL **m\_CellsAct   &!Array of pointers to active cells, 
                          constructed at the end of foam build-up\\
  double *m\_PrimAcu     &!Array of cumulative $\sum_{i=1}^k R'_i$, for cell index generation\\
  TObjArray *m\_HistEdg  &Histograms of $w$, one for each edge, with ROOT\\
  TObjArray *m\_HistDbg  &Histograms for debug (m\_OptDebug=1), with ROOT\\
  TH1D      *m\_HistWt;  &Histograms of MC weight,  with ROOT\\
  TFHST    **m\_HistEdg  &Array of pointers to histograms, without ROOT\\
  TFHST    *m\_HistWt;   &Histograms of MC weight,  without ROOT\\
\hline
  double *m\_MCvect$^g$  &[m\_TotDim] Generated MC vector for the outside user\\
  double  m\_MCwt$^g$    &MC weight\\
  double *m\_Rvec        &[m\_RNmax] Random number vector from r.n. generator,
                             up to m\_TotDim+1 maximum elements\\
\hline
\multicolumn{ 2}{|c|}{{ Externals } }\\
\hline
  {\small TFOAM\_INTEGRAND} *m\_Rho$^gs$ &The distribution $\rho$ to be generated/integrated\\
  TPSEMAR         *m\_PseRan$^gs$        &Generator of the uniform pseudo-random numbers\\
\hline
\multicolumn{ 2}{|c|}{{ Statistics and MC results } }\\
\hline
  long m\_nCalls$^g$      &Number of function calls\\
  long m\_nEffev          &Total no. of effective $w=1$ events in build-up\\
  double m\_SumWt, m\_SumWt2&Sum of weight $w$ and squares $w^2$\\
  double m\_NevGen        &No. of MC events\\
  double m\_WtMax, m\_WtMin &Maximum/Minimum weight (absolute)\\
  double m\_Prime$^g$     &Primary integral $R'$, ($R=R\langle w \rangle$)\\
  double m\_MCresult      &True integral $R$ from the cell exploration MC\\
  double m\_MCerror       &and its error\\
\hline
\multicolumn{ 2}{|c|}{{ Working space for cell exploration } }\\
\hline
  double *m\_Lambda      &[m\_nDim] Internal parameters of the simplex:
                            $\sum \lambda_i <1$\\
  double *m\_Alpha       &[m\_kDim] Internal parameters of the h-rectang.: $0<\alpha_i<1$\\
\hline
\end{tabular}
\end{small}
\caption{\sf Data members of the class {\tt TFOAM}. Cont.}
  \label{tab:TFOAMmembers2}
\end{table}

\begin{table}[hp]
\centering
\begin{small}
\begin{tabular}{|l|p{70mm}|}
\hline
{\tt TFOAM} method & Short description  \\ 
\hline
\multicolumn{ 2}{|c|}{ Constructors and destructors }\\
\hline
  TFOAM()                          & Default constructor (for ROOT streamer)\\
  TFOAM(const char*)               & User constructor\\
  \~TFOAM()                        & Explicit destructor\\
  TFOAM(const TFOAM\&)             & Copy Constructor  NOT USED\\
  TFOAM\& operator=(const TFOAM\& )& Substitution      NOT USED \\
\hline
\multicolumn{ 2}{|c|}{ Initialization, foam build-up }\\
\hline
  void Initialize(TPSEMAR*,        & Initialization, allocation of memory\\
  ~~~~~~~~~~~~TFOAM\_INTEGRAND*)   & and the foam build up\\
  void InitVertices(void)          & Initializes first vertices of the root cell\\
  void InitCells(void)             & Initializes first $n!$ cells in h-rect. root cell\\
  void Grow(void)                  & Adds new cells to foam, until buffer is full\\
  int  Divide(TFCELL *)            & Divides cell into two daughters\\
  void Explore(TFCELL *Cell)       & MC exploration of cell main subprogram\\
  void Carver(int\&,double\&,double\&)& Determines the best edge, $w_{\max}$-reduction\\
  void Varedu(double[~],int\&,double\&,double\&)& Determines the best edge, $\sigma$-reduction\\
  long PeekMax(void)               & Chooses one active cell, used in {\tt Grow}\\
  TFCELL* PeekRan(void)            & Chooses randomly one active cell, in {\tt Grow}\\
  void MakeLambda(void)            & Generates random point inside simplex\\
  void MakeAlpha(void)             & Generates rand. point inside h-rectangle\\
  int  CellFill(int, TFCELL*,      & \\
  ~~~~~~~~~~ int*,TFVECT*,TFVECT*) & Fills next cell and return its index\\
  void MakeActiveList(void)        & Creates table of all active cells\\
\hline
\multicolumn{ 2}{|c|}{ Generation }\\
\hline
  void   MakeEvent(void)           & Makes (generates) single MC event\\
  void   GetMCvect(double *)       & Provides generated random MC vector\\
  void   GetMCwt(double \&)        & Provides MC weight\\
  double MCgenerate(double *MCvect)& All the above in single method\\
  void GenerCell(TFCELL *\&)       & Chooses one cell with probability $\sim R'_j$\\
  void GenerCel2(TFCELL *\&)       & Chooses one cell with probability $\sim R'_j$\\
\hline
\multicolumn{ 2}{|c|}{ Finalization, reinitialization }\\
\hline
  void Finalize(double\&, double\&)  & Prints summary of MC integration\\
  void GetIntegMC(double\&, double\&)& Provides MC integral\\
  void GetIntNorm(double\&, double\&)& Provides normalization\\
  void GetWtParams(const double, & \\
  ~~~~double\&, double\&, double\&)  & Provides MC weight parameters\\
  void LinkCells(void)               & Restores pointers after restoring from disk\\
\hline
\multicolumn{ 2}{|c|}{ Debug }\\
\hline
  void CheckAll(const int)   & Checks correctness of the data structure\\
  void PrintCells(void)      & Prints all cells\\
  void PrintVertices(void)   & Prints all vertices\\
  void LaTexPlot2dim(char*)  & Makes LaTeX file for drawing 2-dim. foam\\
  void RootPlot2dim(char*)   & Makes C++ code for drawing 2-dim. foam\\
\hline
\end{tabular}
\end{small}
\caption{\sf Methods of {\tt TFOAM} class.}
  \label{tab:TFOAMmethods1}
\end{table}

\section{The {\tt Foam} code}
At present, the C++ version of the {\tt Foam} code
is more advanced than the Fortran77 version.
(We do not plan to develop F77 code any further.)
In this section we shall therefore describe mainly the C++ code.

The code of the {\tt Foam} version 1.x was originally written in Fortran77 
with popular language extensions, such as long variable names etc.
It was already written in an object-oriented style, as much as it had been possible.
In particular the main classes {\tt TFOAM} and {\tt TFCELL} of the present C++ version
were already present under a certain form.
The important shortcoming of the F77 version is the lack of dynamic allocation of the memory.
Otherwise, it has most of the functionality of the C++ version;
see later in this section for a list of limitations.

\subsection{Description of C++ classes}

In Table~\ref{tab:classes} we list all classes of the {\tt Foam} package.
The main class is {\tt TFOAM}, which is the MC simulator itself.
It is served by the class {\tt TFCELL} of the cell objects,
and three auxiliary classes {\tt TFVECT}, {\tt TFMATRIX} and {\tt TFPARTITION}.
The other classes are not related directly to the {\tt Foam} algorithm --
they are utilities used by {\tt Foam}: 
random-number generator class {\tt TPSEMAR}~\cite{Marsaglia:1990ig}
and the histogramming class {\tt TFHIST}.
The class {\tt TFDIST} provides a menu of the distributions for testing {\tt Foam}.
In the following we shall describe in more detail
the key classes {\tt TFOAM} and {\tt TFCELL}.

\subsection{ {\tt TFOAM} class}
{\tt TFOAM} is the main class.
Every new instance of this class (properly initialized) is another
independent {\tt Foam} event generator.
In Tables~\ref{tab:TFOAMmembers1} and \ref{tab:TFOAMmembers2}
we provide a full list of the data members of the class {\tt TFOAM}
and their short description.
As seen in these tables, we have added the prefix ``{\tt m\_}''
to all names of the data members, so that they are visually different, in the code,
from the other variables 
(which is recommended practice in C++ coding).

Generally, one may notice that many data members
could be declared (and allocated) as the local variables in the procedures,
instead of being data members.
For example, vector {\tt m\_MCvect}, which transpors
random numbers out from the random number generator {\tt m\_PseMar}
could be declared locally at every place where {\tt m\_PseMar} is called.
We opted for a more ``static'' structure of the data,
with more than necessary of the data members in the class,
at the expense of the human readability of the code,
in order to:
(a) facilitate the implementation persistency with ROOT,
(b) gain in execution speed and
(c) facilitate the translation to other languages.

Most of the methods (procedures) of the class {\tt TFOAM} are listed in
Table~\ref{tab:TFOAMmethods1}.
We omitted in this table ``setters'' and ``getters'', which provide access
to some data members, and simple inline functions,
such as {\tt sqr} for squaring a {\tt double} variable.
Data members which are served by the setters and getters are marked
in Tables~\ref{tab:TFOAMmembers1} and \ref{tab:TFOAMmembers2}
by the superscripts ``$s$'' or/and ``$g$''.

Let us now characterize briefly the role of most important methods
of the class {\tt TFOAM} in the {\tt Foam} algorithm.

\begin{figure}[!ht]
\begin{center}
\epsfig{file=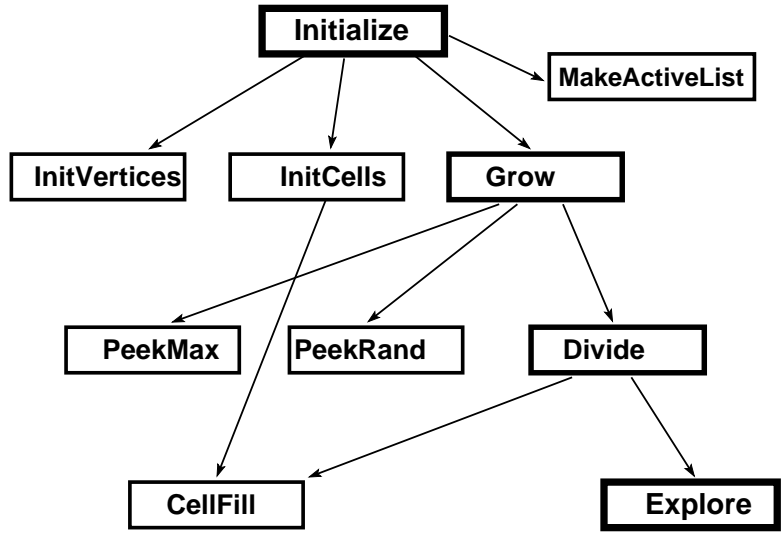,width=80mm,angle=270}
\end{center}
\caption{\small\sf
  Calling sequence of the {\tt Foam} procedures during the foam build-up (initialization).
}
\label{fig:initialize}
\end{figure}

\subsubsection{Procedures for {\tt Foam} initialization and foam build-up}
The constructor {\tt TFOAM(const char*)} is for creating
an object of the class {\tt TFOAM}.
Its parameter is the name given by the user to an object.
The principal role of this constructor
is to initialize data members to their default values --
no memory allocation is done at this stage.
After resetting all kinds of steering parameters of the {\tt Foam} to preferred values
(using setters) the user calls the {\tt Initialize} method, which builds up the foam of cells.
The two methods {\tt InitVertices} and {\tt InitCells} allocate arrays
of vertices and cells (pointers) with empty cells.
The empty cells are allocated/filled using {\tt CellFill}.
Next comes the procedure {\tt Grow} which loops over cells,
picking up the most promising cell for the split,
either randomly using {\tt PeekRand} or deterministically using {\tt Peekmax}.
The chosen cell is split using {\tt Divide}.
It is, however, the procedure {\tt Explore} called by {\tt Divide}
(and by {\tt InitCells} for the root cell)
which does the most important job in the {\tt Foam}
build-up: it performs a small MC run for each newly allocated daughter cell.
It calculates how profitable the future split of the cell will be
and defines the optimal cell division geometry with the help
of {\tt Carver} or {\tt Varedu} procedures,
for maximum weight or variance optimization respectively.
All essential results of the exploration are written into the explored cell object.
At the very end of the foam build-up, {\tt MakeActiveList}
is invoked to create a list of pointers to all active cells,
for the purpose of the quick access during the MC generation.
The procedure {\tt Explore} uses two procedures
{\tt MakeLambda} and {\tt MakeAlpha}, which generate randomly (uniformly)
coordinates of the MC points inside a given cell.
The above sequence of the procedure calls is depicted in
Fig.~\ref{fig:initialize}.

\subsubsection{Procedures for MC generation}
The MC generation of a single MC event is done by invoking
{\tt MakeEvent}, which chooses randomly a cell
with the help of procedure%
\footnote{Method {\tt GenerCell1} exists, but it is not used.}
{\tt GenerCell2} and, next, the internal coordinates
of the point within the cell using {\tt MakeLambda} and/or {\tt MakeAlpha}.
The absolute coordinates of the MC event
are calculated and stored in the data member double-precision vector {\tt m\_MCvect}.
The MC weight is calculated using an external procedure
that provides the density distribution $\rho(x)$,
which is represented by the pointer {\tt m\_Rho}.
A class to which the object {\tt m\_Rho} belongs must inherit
from the abstract class {\tt TFOAM\_INTEGRAND}.
The MC event (double-precision vector) and its weight are available through getters
{\tt GetMCvect} and {\tt GetMCwt}.
Note that the variables of the hyperrectangular subspace come first in the {\tt m\_MCvect},
before variables of the simplical subspace.

The user may alternatively call {\tt MCgenerate}, which invokes {\tt MakeEvent}
and provides a MC event and its weight, all at the same time.

\subsubsection{Procedures for finalization and debugging}
The use of the method {\tt Finalize} is not mandatory.
It prints statistics and calculates the estimate of the
integral using the average weight from the MC run.
The amount of printed information depends on the values of {\tt m\_chat}.
For the normalization of the plots and integrals,
the user needs to know the exact value of $R'=\int \rho'(x) dx$, which is
provided by the method {\tt GetIntNorm} or {\tt Finalize}.
The actual value of the integrand from the MC series is provided
by {\tt GetIntegMC}.
Note that for the convenience of the user
{\tt GetIntNorm} provides $R'$ or MC estimate of $R=\int \rho(x) dx$,
depending on whether the MC run was with weighted or $w=1$ events.

Another useful finalization frocedure,
\begin{verbatim}
GetWtParams(const double eps, double &AveWt, double &WtMax, double &Sigma)
\end{verbatim}
provides three parameters that characterize the MC weight distribution:
the average weight {\tt AveWt},
the ``intelligent'' maximum weight {\tt WtMax}~$=w^\varepsilon_{\max}$
for a given value of {\tt eps}~$=\varepsilon$
(see Sect.~\ref{sec:NumeResults} for its definition)
and the variance {\tt sigma}~$=\sigma$.
In particular, in the case of unweighted events, $w^\varepsilon_{\max}$ can be used
as an input for the next MC run.

The {\tt Foam} program includes procedure
{\tt CheckAll}, which checks the correctness of the pointers in the
doubly linked tree of cells (this can take time for large $N_c$).
It can sometimes be useful for debugging purposes.
Another two methods {\tt PrintVertices} and {\tt PrintCells}
can be used at any stage of the calculation in order to print
the list of all cells and vertices.
In the case of the two-dimensions there is a possibility
to view the geometry of the cells with a 2-dimensional plot,
which is either a LaTeX file produced by {\tt LaTexPlot2dim},
or a ROOT file produced by {\tt RootPlot2dim}.

\begin{table}[!ht]
\centering
\begin{small}
\begin{tabular}{|l|p{11.0cm}|}
\hline
TCELL member & Short description  \\ 
\hline\hline
\hline
\multicolumn{ 2}{|c|}{ ``Static'' members, the same for all cells! }\\
\hline
  short int   m\_kDim     & Dimension of hyperrectangular subspace\\
  short int   m\_nDim     & Dimension of simplical subspace\\
  short int   m\_OptMCell & Option of economic memory for usage (hyperrectangular subspace) \\
  short int   m\_OptCu1st & =1, Numbering of dims starts with hyperrectangles; =0 simplices\\
  int         m\_nVert    & No. of vertices in the simplex = m\_nDim+1\\
  TFCELL **m\_Cell0       &! Pointer of the root cell\\
  TFVECT **m\_Vert0       &! Pointer of the vertex list\\
\hline
\multicolumn{ 2}{|c|}{ Linked tree organization}\\
\hline
  int    m\_Serial        & Serial number (index in m\_Cell0)\\
  int    m\_Status        & Status (active or inactive)\\
  int    m\_Parent        & Pointer to parent cell\\
  int    m\_Daught0       & Pointer to daughter 1\\
  int    m\_Daught1       & Pointer to daughter 2\\
\hline
\multicolumn{ 2}{|c|}{The best split geometry from the MC exploration}\\
\hline
  double m\_Xdiv          & Factor $\lambda$ of the cell split\\
  int    m\_Best          & The best edge candidate for the cell split\\
\hline
\multicolumn{ 2}{|c|}{Integrals of all kinds}\\
\hline
  double m\_Volume        & Cartesian volume of this cell\\
  double m\_Integral      & Integral over cell (estimate from exploration)\\
  double m\_Drive         & Driver  integral $R_{loss}$ for cell build-up\\
  double m\_Primary       & Primary integral $R'$ for MC generation\\
\hline
\multicolumn{ 2}{|c|}{Geometry of the cell}\\
\hline
  int   *m\_Verts         & [m\_nVert] Pointer to array of vertices in simplical subspace\\
  TFVECT *m\_Posi         & Pointer to position vector,  hyperrectangular subspace\\
  TFVECT *m\_Size         & Pointer to size vector,      hyperrectangular subspace\\
\hline
\end{tabular}
\end{small}
\caption{\sf Data members of the class {\tt TFCELL}.}
  \label{tab:TFCELLmembers}
\end{table}

\subsection{{\tt TFCELL} class}
{\tt TFCELL} is an important class of objects representing a single cell
of the foam.
Data members of the class are listed in Table~\ref{tab:TFCELLmembers}.

Most of the methods of the {\tt TFCELL} class are setters and getters.
The non-trivial methods are {\tt GetHcub} and {\tt GetHSize},
which calculate the absolute position and size of hyperrectangles
in the algorithm of Section~\ref{sec:save_memory},
and {\tt MakeVolume}, which calculates the Cartesian volume of the cell.
In the simplical subspace, the volume is a determinant
of a square matrix of the class {\tt TFMATRIX}.

\subsection{Persistency with the help of ROOT}
The C++ language does not provide any built-in mechanism of the persistency of the classes.
For this purpose we use the ROOT package~\cite{root}, with the help of its
``automatic streamers''.
ROOT is a useful C++ library for histogramming,
organizing large databases of identical objects of the type used
in high energy physics experiments.
It also provides an efficient input/output, with compressing capabilities.

Providing full persistency for C++ classes 
preserving all the structures of the pointers is not a trivial task in general.
ROOT can do it, even for pointers.
For this, our code had to be  reorganized slightly:
puting additional directives for ROOT in the comment lines,
remooving static variable, explicit integer indices instead of pointers in some places.
As a whole, this solution is relatively simple, and works correctly.
In Tables~\ref{tab:TFOAMmembers1} and \ref{tab:TFOAMmembers2}
we have at the beginning of the description
certain characteristic marks which are directives for the persistency mechanism of ROOT,
see the ROOT manual~\cite{root} for more details.

One has to remember, when reading a {\tt TFOAM} class object from the disk,
that the method {\tt LinkCells()} has to be invoked in order
to fully reconstruct all pointers in the doubly linked tree of cells.
Moreover, any object of the class {\tt TFOAM} restored from the disk file
will have its internal object for the random number generator
and distribution function.
There is a method that provides access (pointer)
to these objects, if necessary.
The relevant fragment of the code may look as follows:
\begin{verbatim}
 TPSEMAR *RNGen= FoamX->GetPseRan();        //get pointer of RN generator
 TFDISTR *RHO  = (TFDISTR*)FoamX->GetRho(); //get pointer of distribution
\end{verbatim}
It might be useful if, for instance, one wants to reinitialize
the random number generator used by the {\tt TFOAM} class object, which has been
read from the disk file, with a new ``random seed''.

{\tt Foam} can be used with or without ROOT.
In the code all parts of the code dependent on ROOT enclosed
in the pair of preprocessor commands {\tt \#ifdef ROOT\_DEF ... \#endif},
where the {\tt  ROOT\_DEF} environmental variable is defined centrally
in the header file {\tt ROOT\_DEF.h}.
Eliminating ROOT requires removing this variable and modifying
{\tt makefile} accordingly (the {\tt TFHST} class has to be linked).
The version without ROOT does not feature persistency,
and is employing its own simple histogramming class {\tt TFHST}
instead of the ROOT class {\tt  TH1D}.
ROOT also helps to create documentation of the {\tt Foam} in the html format.
We recommend a version tied up with ROOT.

\subsection{Fortran77 version and its limitations}
We also provide users with the Fortran77 versions of {\tt Foam}.
There are two of them at the development level 2.02 (May 2001)
of the algorithm.
In the first one cells can be simplical, hyperrectangular
and the Cartesian product of the two.
This version is limited to dimension five for the simplical subspace
and not very useful for large dimensions ($n\geq 5$) in the hyperrectangular space, because
it does not feature the memory-saving algorithm of Section~\ref{sec:save_memory}.
Another version, named {\tt MCell} (standing for Mega-Cell),
features only hyperrectangular cells; on the other hand,
it includes the memory-saving algorithm described in Section~\ref{sec:save_memory}.
We recommend the reader to use the version {\tt MCell}.

Neither of these versions can have dynamic memory allocation; they
have a maximum dimensions of the integration/simulation subspaces
(simplical and hyperrectangular) hard-coded in the source code.
Any change of these maximum dimensions requires recompilation of the code.

Present versions in Fortran77
are substantially improved with respect to the original version of Ref.~\cite{foam1:2000}.
For the option of minimizing the maximum weight, they have exactly the same
algorithm (of the cell split) as the C++ version.
They feature, however, an older, more primitive version of the algorithm of finding
the best cell division for the variance reduction.

The structure of the programs, naming of procedures and variables,
configuration parameters and their meaning are very similar in the F77 and C++ versions.
Some differences in the usage will be indicated in the next subsections.

\subsection{Future development}
In the following we indicate some of the possible future developments 
of the  {\tt Foam} package.
As already indicated we do not plan to develop the Fortran77 version any further.
On the other hand, it would be interesting to upgrade the existing {\tt Foam}
version 1.x in the JAVA language\cite{CieslaSlusarczyk}
to the level of the present version 2.x.

As for the C++ version,
it would be a logical development to inherit the class of pseudo-random number generators
{\tt TPSEMAR}
from a unique abstract class, and in this way to define a universal interface
for a library of the number generators.
We intend to collect a library of a few random number generators
with a universal interface (or find one)
for the use in {\tt Foam} and applications based on it.

Concerning the version of {\tt Foam} adapted to parallel processing,
as in Refs.~\cite{Ohl:1998jn,Doncker:1998,Doncker:1999,Doncker:parint1},
we do not have plans in this direction in the immediate future.
Here, we have of course in mind the use of the true CPU parallelism
in the foam of cells build-up.
One has to remember that in the high energy physics applications,
which are our main objective,
the foam of cell build-up will always be a tiny fraction
of the total CPU time. 
The main fraction will be the subsequent MC simulation
in which, as the vast experience with PC-farms in CERN and FNAL shows,
one may organize the MC simulation with a low level of parallelism,
with many simulators started with different random seeds running
in parallel but not communicating 
-- another specialized job combines all the results at the very end of the run.
However, the first practical examples of true parallelism in the massive
MC simulation for the purpose of the high energy physics experiments
has appeared recently~\cite{Jadach:2001mp}
and is used in the LEP2 data analysis.

The main algorithm of {\tt Foam} is quite stable at this stage,
and the main emphasis in its future development
will be on making it more user friendly
and better adapted to the use as a part of bigger MC projects.
In particular its provisions for multibranching will become more
sophisticated, as more feedback comes from real-life applications.

\section{Usage of {\tt Foam}}
In the following we provide basic information on how to use the {\tt Foam}
program in the user application.

\subsection{{\tt Foam} distribution directory of the C++ version}
The {\tt Foam} package is distributed together with the demonstration main programs
and some utilities in the form of about 20 files in a single UNIX directory
{\tt FOAM-export-v2.05}.
Demonstration runs can be executed using standard {\tt make } commands as follows:
\begin{verbatim}
       make Demo-run
       make DemoPers
       make Demo-map
       make DemoNR-run
\end{verbatim}
The essential fragments of the output from running command ``{\tt make Demo-run}''
are shown in Appendix B.
The compilation and linking procedure is encoded in the file {\tt Makefile},
and it has to be checked by the user whether it conforms to the local operating system.
In particular, if ROOT is used, then certain paths and environmental variables
in {\tt Makefile} have to be adjusted.
The use of ROOT is decided by the presence of the variable {\tt ROOT\_DEF}
in the file {\tt ROOT\_DEF.h}.
Without ROOT, the  user should execute command ``{\tt make DemoNR-run}''.

The essential part of {\tt Foam}, that is class {\tt TFOAM}, {\tt TFCELL}
and a few auxiliary classes, is located in the files
{\tt TFOAM.cxx} and {\tt TFOAM.h}.
This is the ``core'' of the {\tt Foam} source.
The source code of the other utility classes {\tt TFHST}, {\tt TFMAXWT}, {\tt TPSEMAR}
and {\tt TFDISTR} are in separate files.
The main programs are in files {\tt Demo.cxx} and {\tt DemoPers.cxx}.
They should serve as a useful template for the user's own application
based on {\tt Foam}.

There is also one Fortran77 source code, {\tt circe2.f}%
\footnote{We thank Thorsten Ohl for providing us with a preliminary version of this
  code~\cite{circe2}.},
which contains a testing density distribution 
(2-dimensional beamstrahlung spectrum~\cite{circe2} 
of the electron linear collider~\cite{Aguilar-Saavedra:2001rg})
used in {\tt TFDISTR}.
The {\tt Makefile} provides, therefore, also a useful example of linking C++
and F77 codes.

There are also two output files
{\tt output-Demo.linux} and {\tt output-DemoNR.linux},
which the reader may use to check whether he is able to reproduce
these benchmark output results.

\subsection{Simple example of an application}
\label{sec:application}
A very simple example of the use of {\tt Foam} may look as follows:
\begin{verbatim}
//   *** Initialization ***
double MCwt;
TFDISTR *Density1 = new TFDISTR(FunType); // Create integrand function
TPSEMAR *PseRan   = new TPSEMAR();        // Create random numb. generator
TFOAM   *FoamX    = new TFOAM("FoamX");   // Create Simulator
FoamX->SetkDim( 3);                       // Set dimension, h-rect.
FoamX->Initialize(PseRan,  Density1 );    // Initialize simulator
//   *** MC Generation ***
TFHST *hst_Wt = new TFHST(0.0,1.25, 25);  // Create weight histogram
double *MCvect =new double[3];            // Monte Carlo event
for(long loop=0; loop<1000000; loop++){
   MCwt = FoamX->MCgenerate(double *MCvect); // Generate MC event
   hst_Wt->Fill(MCwt,1.0);                   // Fill weight histogram
}
//   *** Finalization ***
double IntNorm, Errel;
FoamX->Finalize( IntNorm, Errel);  // Print statistics, get normalization
double MCresult, MCerror, AveWt, WtMax, Sigma;
FoamX->GetIntegMC( MCresult, MCerror);        // get MC integral
double eps = 0.0005;
FoamX->GetWtParams(eps, AveWt, WtMax, Sigma); // get MC wt parameters
hst_Wt->Print();                   // Print weight histogram
\end{verbatim}

The user normally provides his own density distribution function
belonging to the class that has to inherit from the
following abstract class:
\begin{verbatim}
class TFOAM_INTEGRAND{ // Abstract class of distributios for Foam
 public:
  TFOAM_INTEGRAND() { };
  virtual ~TFOAM_INTEGRAND() { };
  virtual double Density(int ndim, double*) = 0;
};
\end{verbatim}
In the above example the distribution {\tt *Density1}
belongs to the class {\tt TFDISTR}, which is provided
in the {\tt Foam} distribution directory.

\begin{table}[ht!]
\centering
\begin{small}
\begin{tabular}{|l|l|p{12.0cm}|}
\hline
Param. & Value & Meaning  \\ 
\hline\hline
  nDim     & 0$^*$    & Dimension of the simplical subspace\\
  kDim     & 0$^*$    & Dimension of the hyperrectangular subspace\\
  nCells   & 1000$^*$ & Maximum number of cells,\\
  nSampl   & 200$^*$  & No. of MC events in the cell MC exploration\\
  nBin     & 8$^*$    & No. of bins in edge-histogram in cell exploration\\
  OptRej   & 0$^*$    & OptRej = 0, weighted; =1, $w=1$ MC events\\
  OptDrive & 2$^*$    & Maximum weight reduction\\
           & 1        & or variance reduction\\
  OptPeek  & 0$^*$    & Next cell for split with maximum $R'_I$ (PeekMax) \\
           & 1        & or randomly with probability $\sim R'_I$ (PeekRan)\\
  OptEdge  & 0$^*$    & Vertices are NOT included in the cell MC exploration\\
           & 1        & or vertices are included in the cell MC exploration\\
  OptOrd   & 0$^*$    & Root cell is a hyperrectangle in simplical subspace\\
           & 1        & or root cell is  a simplex in simplical subspace \\
  OptMCell & 1$^*$    & Economic memory algorithm in hyperrectangular subspace is ON\\
           & 0        & or economic memory algorithm in hyperrectangular subspace is OFF\\
  EvPerBin & 25$^*$   & Maximum number of effective $w=1$ events/bin\\
           & 0        & or counting of number eff. events/bin is inactive\\
  Chat     & 1$^*$    & =0,1,2 is the ``chat level'' in the standard output\\
  MaxWtRej & 1.1$^*$  & Maximum weight used to get $w=1$ MC events\\
\hline
\end{tabular}
\end{small}
\caption{\sf Fourteen principal configuration parameters and switches of the {\tt Foam} program.
  The default values are marked with the star superscript.}
  \label{tab:TFCELLparams}
\end{table}

\subsection{Configuring {\tt Foam}}
{\tt Foam} has fourteen principal configuration parameters
plus parameters inhibiting and/or predefining the division geometry
in the cell split.
\subsubsection{Principal configuration parameters}
All of the principal parameters
listed in Table~\ref{tab:TFCELLparams}
are set to meaningful default values,
hence the beginner may stay ignorant of their role for some time,
and learn gradually how to exploit them in order to improve
the efficiency of {\tt Foam} in the actual application.
All these parameters are data members of the {\tt TFOAM} class,
see Table.~\ref{tab:TFOAMmembers1}.
If the user wants to redefine all of them, then the relevant piece of code
will look as follows:
\begin{verbatim}
  FoamX->SetnDim(     nDim);
  FoamX->SetkDim(     kDim);
  FoamX->SetnCells(   nCells);
  FoamX->SetnSampl(   nSampl);
  FoamX->SetnBin(     nBin);
  FoamX->SetOptRej(   OptRej);
  FoamX->SetOptDrive( OptDrive);
  FoamX->SetOptPeek(  OptPeek);
  FoamX->SetOptEdge(  OptEdge);
  FoamX->SetOptOrd(   OptOrd);
  FoamX->SetOptMCell( OptMCell);
  FoamX->SetEvPerBin( EvPerBin);
  FoamX->SetMaxWtRej( MaxWtRej);
  FoamX->SetChat(     Chat);
\end{verbatim}
In practical applications one will redefine only some of them.
The minimum requirement is that the user sets a non-zero value
of {\tt nDim} or {\tt kDim} such that the total dimension {\tt nDim+kDim}
is a non-zero positive integer.

\subsubsection{Inhibiting cell division in certain directions}
If the user of {\tt Foam}  decides to inhibit the division in certain variable 
in the hyperrectangular subspace, this can be done with the method 
{\tt SetInhiDiv(int iDim, int InhiDiv)} of the class {\tt TFOAM},
where {\tt iDim} is the dimension index for which inhibition is done
and {\tt InhiDiv} is the inhibition tag.
This method should be used before invoking {\tt Initialize},
after setting {\tt nDim} and/or {\tt kDim}.
The relevant code may look as follows:
\begin{verbatim}
  FoamX->SetInhiDiv(0, 1); //Inhibit division of x_1
  FoamX->SetInhiDiv(1, 1); //Inhibit division of x_2
\end{verbatim}
The allowed values are {\tt InhiDiv=0,1} and
the default value is {\tt InhiDiv=0}.
Note that the numbering of dimensions with {\tt iDim} starts from zero
and variables of the hyperrectangular subspace always come first,
before the simplical ones.

\subsubsection{Setting predefined cell division geometry}
The user may predefine divisions of the root cell
in certain variables in the hyperrectangular subspace using the method
{\tt SetXdivPRD(int iDim, int len, double xDiv[])}.
The relevant piece of the user code may look as follows:
\begin{verbatim}
  double xDiv[3];
  xDiv[0]=0.30; xDiv[1]=0.40; xDiv[2]=0.65;
  FoamX->SetXdivPRD(0, 3, xDiv);
\end{verbatim}
Again, this should be done before invoking {\tt Initialize},
after setting {\tt nDim} and/or {\tt kDim}.

\subsection{Persistency}
Persistency of the {\tt Foam} classes is arranged using ``default streamers''
of the ROOT~\cite{root} package.
Writing a {\tt TFOAM} class object into the disk file {\tt rmain.root} 
can be done with the single {\tt Write} as follows:
\begin{verbatim}
  TFile RootFile("rmain.root","RECREATE","histograms");
  ...
  FoamX->Write("FoamX"); //Writing Foam on the disk
  ...
  RootFile.Write();
  RootFile.Close();
\end{verbatim}
The instruction {\tt FoamX->Write("FoamX")} can be put
at any place of the code after the instruction {\tt FoamX->Initialize(...)},
see example of the user code shown in Section~\ref{sec:application}.

Next, in another program, the {\tt TFOAM} class object can be read from
the disk file {\tt rmain.root} as follows:
\begin{verbatim}
  TFile fileA("rmain.root");  // connect disk file
  fileA.cd();
  fileA.ls();                        // optional printout
  fileA.Map();                       // optional printout
  fileA.ShowStreamerInfo();          // optional printout
  fileA.GetListOfKeys()->Print();    // optional printout
  TFOAM  *FoamX = (TFOAM*)fileA.Get("FoamX"); // find object
  FoamX->LinkCells();  // restore pointers of the binary tree of cells
  FoamX->CheckAll(1);                // optional x-check of pointers
\end{verbatim}
and at this point the {\tt FoamX} object is ready to generate MC events,
as in the MC generation part of the code shown in Section~\ref{sec:application}.

\subsection{Fortran77 versions}
The distribution directory {\tt FoamF77-2.02-export} contains
the {\tt README} file, two demonstration main programs 
{\tt DemoFoam.f} and {\tt DemoMCell.f} to be compiled and run
with the help of commands
\begin{verbatim}
        make DemoFoam
        make DemoMCell
\end{verbatim}
encoded in the {\tt Makefile}.
The outputs from the above runs can be compared with the
benchmark outputs {\tt output-DemoFoam-linux}
and {\tt output-DemoMCell-linux}.

The basic {\tt Foam} source files are:
    {\tt FoamA.f}  with header file {\tt FoamA.h} and
    {\tt MCellA.f} with header file {\tt MCellA.h}.

For the description of the input (configuration) parameters, see comments
in {\tt FoamA.f} and {\tt MCellA.f} respectively.
The names of the configuration variables are the same as in the C++ version,
except {\tt nCells}, which is here renamed to {\tt nBuff}.
Their values and the meaning are the same.

Demonstration main programs {\tt DemoFoam.f} and {\tt DemoMCell.f} can serve as templates
for the user application programs.

The testing main program uses the histogramming package {\tt GLK} 
of the {\tt KKMC} program~\cite{Jadach:1999vf},
which the user may replace with any other histogramming package.

\begin{table}[ht!]
\centering
\begin{small}
\begin{tabular}{|r|r|r|r|r|r|r|r|r|}
\hline
 nDim& kDim& nCalls& nCells& nSampl& $w^{\varepsilon}_{\max}/\langle w \rangle$ 
          & $\sigma/\langle w \rangle$& $\Delta_{statist.} R$& $R$ \\
\hline\hline
 0&    1&      203719&     1000&      333&       0.99148&      0.014585&    1.031e-05&    0.99999782\\
 0&    1&      206192&     1000&     1000&       0.99147&      0.014752&    1.043e-05&    0.99999962\\
 0&    1&      206192&     1000&     3333&       0.99147&      0.014752&    1.043e-05&    0.99999962\\
 0&    1&      206192&     1000&    10000&       0.99147&      0.014752&    1.043e-05&    0.99999962\\
 0&    1&      206192&     1000&    33333&       0.99147&      0.014752&    1.043e-05&    0.99999962\\
\hline
 0&    3&      275421&     1000&      333&       0.50538&       0.54088&    0.000382&     0.99986832\\
 0&    3&      435112&     1000&     1000&       0.50886&       0.54033&    0.000382&     1.00017104\\
 0&    3&      663493&     1000&     3333&       0.49922&       0.55721&    0.000394&     0.99934710\\
 0&    3&      834094&     1000&    10000&       0.50359&       0.54674&    0.000386&     1.00056316\\
 0&    3&     1015157&     1000&    33333&       0.51091&       0.54035&    0.000382&     0.99983999\\
 0&    3&     2312759&    10000&      333&       0.72346&       0.27758&    0.000196&     0.99977045\\
 0&    3&     2675820&    10000&     1000&       0.72677&       0.27504&    0.000194&     0.99995080\\
 0&    3&     3054404&    10000&     3333&       0.72199&       0.27710&    0.000195&     1.00013270\\
 0&    3&     3333479&    10000&    10000&       0.72200&       0.27720&    0.000196&     1.00008994\\
 0&    3&     3575366&    10000&    33333&       0.72243&       0.27786&    0.000196&     0.99997875\\
\hline
 0&    4&     3825046&    10000&     1000&       0.50363&       0.51168&    0.000361&     1.00013082\\
 0&    4&     6559430&    10000&    10000&       0.50297&       0.51001&    0.000360&     0.99960319\\
 2&    2&     4493961&    10000&     1000&       0.43076&       0.63185&    0.000446&     1.00072564\\
 2&    2&     9374351&    10000&    10000&       0.44922&       0.60669&    0.000429&     1.00013171\\
 4&    0&     6642202&    10000&     1000&       0.21029&       1.19420&    0.000844&     1.00072248\\
 4&    0&    12337748&    10000&    10000&       0.20817&       1.20067&    0.000849&     1.00020405\\
\hline
 0&    6&     2311881&     1000&     3333&       0.04199&       2.12091&    0.001499&     0.99856206\\
 0&    6&     5542146&     1000&    10000&       0.03847&       2.38588&    0.001687&     0.99912901\\
 0&    6&    12844256&     1000&    33333&       0.03279&       2.61028&    0.001845&     0.99799089\\
 0&    6&    12737314&    10000&     3333&       0.15385&       1.15211&    0.000814&     1.00039754\\
 0&    6&    24134694&    10000&    10000&       0.15313&       1.19596&    0.000845&     0.99945766\\
 0&    6&    42827237&    10000&    33333&       0.14168&       1.22627&    0.000867&     0.99954178\\
 0&    6&    42808972&   100000&     1000&       0.30910&       0.71250&    0.000503&     0.99972833\\
 0&    6&    61803017&   100000&     3333&       0.30805&       0.71462&    0.000505&     1.00002674\\
 0&    6&    92531875&   100000&    10000&       0.30905&       0.71423&    0.000505&     0.99985093\\
\hline
 0&    9&    78325890&   100000&     1000&       0.03718&       1.64608&    0.001163&     0.99367339\\
 0&    9&   167710365&   100000&     3333&       0.05247&       1.73063&    0.001223&     1.00109792\\
 0&    9&   353943409&   100000&    10000&       0.05196&       1.80538&    0.001276&     1.00196909\\
 0&    9&   272162624&   400000&     1000&       0.08490&       1.30193&    0.000920&     1.00065580\\
 0&    9&   495260998&   400000&     3333&       0.09174&       1.35307&    0.000956&     0.99884358\\
 0&    9&   924011087&   400000&    10000&       0.08853&       1.38579&    0.000979&     1.00052122\\
\hline
 0&   12&   261911066&   100000&     3333&             0&       5.83954&    0.004129&     0.97304842\\
 0&   12&   671460574&   100000&    10000&       0.00640&       3.85823&    0.002728&     0.98878698\\
 0&   12&   913072065&   400000&     3333&       0.01285&       2.73991&    0.001937&     0.98688299\\
 0&   12&  2117963809&   400000&    10000&       0.01235&       2.92642&    0.002069&     0.99301117\\
\hline
\end{tabular}
\end{small}
\caption{\sf Numerical results of {\tt Foam} with the maximum weight reduction.
  Variable nCalls is the total number of the function calls in the foam build-up.}
  \label{tab:NumeResults}
\end{table}

\begin{table}[ht!]
\centering
\begin{small}
\begin{tabular}{|r|r|r|r|r|r|r|r|r|}
\hline
 nDim& kDim& nCalls& nCells& nSampl& $w^{\varepsilon}_{\max}/\langle w \rangle$ 
          & $\sigma/\langle w \rangle$& $\Delta_{statist.} R$& $R$ \\
\hline\hline
 0&    4&     3855289&    10000&     1000&       0.27659&       0.31944&    0.000225&     1.00025027\\
 0&    4&     7760907&    10000&    10000&       0.30313&       0.31483&    0.000222&     0.99978048\\
 2&    2&     4589024&    10000&     1000&       0.23086&       0.38050&    0.000269&     0.99967167\\
 2&    2&     8696153&    10000&    10000&       0.24696&       0.37153&    0.000262&     0.99959278\\
 4&    0&     6157799&    10000&     1000&       0.08498&       0.92314&    0.000652&     1.00006553\\
 4&    0&    10547749&    10000&    10000&       0.09881&       0.89859&    0.000635&     1.00024727\\
\hline
\end{tabular}
\end{small}
\caption{\sf Numerical results of {\tt Foam} with the variance reduction.}
  \label{tab:NumeVaredu}
\end{table}

\begin{figure}[!ht]
\begin{center}
\epsfig{file=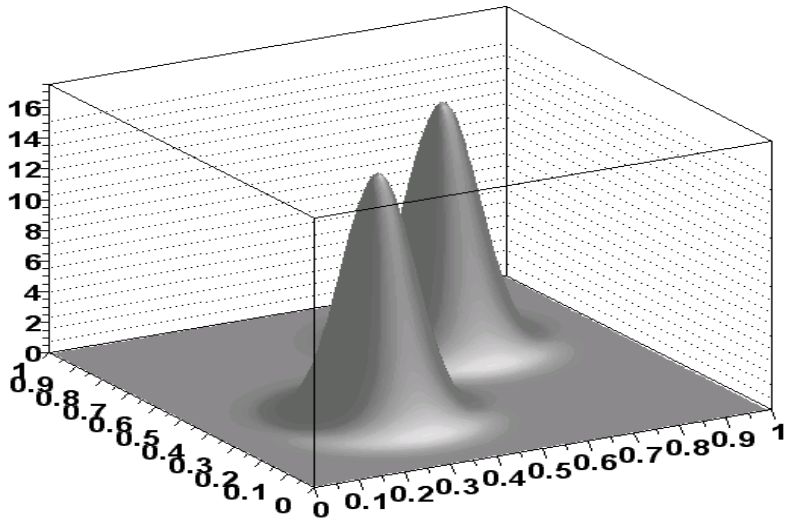,width=120mm,height=80mm}
\end{center}
\caption{\small\sf
  Testing distribution  $\rho_{camel}(x)$ of Ref.~\cite{Lepage:1978sw} in two dimensions.
}
\label{fig:camel}
\end{figure}

\section{Numerical studies and example applications}
\label{sec:NumeResults}
In the following subsection we examine MC efficiency of the {\tt Foam}
in a series of numerical exercises.
In some of them we shall also show examples of the  {\tt Foam} application 
with the distributions relevant to everyday practice in high energy physics.
\subsection{Dependence of the {\tt Foam} efficiency on the configuration parameters}
As a first numerical exercise, we examine the dependence of the {\tt Foam} efficiency
on the most important (input) configuration parameters,
including the dimension of the space.

In Table~\ref{tab:NumeResults}
we collect results from many MC runs for various dimensions, numbers of cells
and numbers of MC events in single cell exploration, varying also
the type of the cells.
We always use the same test distribution $\rho_{camel}(x)$ as in Ref.~\cite{Lepage:1978sw},
which features two relatively narrow gaussian peaks placed on the diagonal.
The 2-dimensional version of this distribution is shown in Fig.~\ref{fig:camel}.
We have used the non-default values {\tt nBin=4} and {\tt EvPerBin=50} and the default values
for the other configuration parameters.
In this table all tests were done for the default option 
of reduction of the maximum weight, {\tt OptDrive=2}
(for results with {\tt OptDrive=1}, see next table).
The efficiency $w^{\varepsilon}_{\max}/\langle w \rangle$ is calculated
using maximum weight $w^{\varepsilon}_{\max}$ defined as in%
\footnote{The $\varepsilon$-dependent maximum weight is defined such
  that events with  $w>w^{\varepsilon}_{\max}$ contribute an $\varepsilon$-fraction
  to the total integral.
  It is numerically more stable in the numerical evaluation
  than the one defined as the largest weight in the MC run.}
Ref.~\cite{foam1:2000} for $\varepsilon = 0.0005$.
The maximum weight $w^{\varepsilon}_{\max}$ is calculated
with the help of the small auxiliary class {\tt TFMAXWT}.

In Table~\ref{tab:NumeResults} the efficiency of the MC run
measured in terms of $w_{\max}/\langle w \rangle$  and $\sigma/\langle w \rangle$.
The value of the integral  $R \pm \Delta_{statist.} R$,
shown in the last four columns, was obtained from the
MC run in which the total number of MC events was $2\times 10^6$.
The value of the integral $R$ is well known, it is equal to 1, within $10^{-5}$.

The following observations based, on the results of Table~\ref{tab:NumeResults}, can be made:
\begin{itemize}
\item
  Looking at the results, for total dimension $n=4$ we see that the hyperrectangular cells
  clearly provide better MC efficiency than simplical ones.
  All other results are for hyperrectangular cells.
\item
  All of the results support the observation that the MC efficiency depends
  critically on the number of cells.
  In particular, see results for $n=6$, the increase of {\tt nSamp}
  (No. of MC events in cell exploration) beyond a certain value does not help
  at all.
\item
  In the case of the very inefficient {\tt Foam}, see  $n=12$
  with $\sigma/\langle w \rangle \sim 6$,
  the estimate of the MC statistical error can be misleading.
  We see an indication that one should not trust runs with $\sigma/\langle w \rangle > 3$.
\item
  For this particular testing function the dimension $n=12$ requires
  a minimum of 400k cells and resulting MC efficiency of order of 1\%
  is barely acceptable%
  \footnote{However, we still get the correct value of the integral within 0.2\%.}.
\end{itemize}

In Table~\ref{tab:NumeVaredu} we repeat the exercise of Table~\ref{tab:NumeResults}
for the option of the variance reduction {\tt OptDrive=1} at four dimensions.
As compared to Table~\ref{tab:NumeResults} we see a net improvement in the variance
and deterioration of the $w^{\varepsilon}_{\max}$.
This agrees with the expectations.

\begin{table}[ht!]
\centering
\begin{tabular}{|l|r||r|r|r|}
\hline
Functions at 2-dimens.       &{\tt Foam} 1.01&  Simpl.& H-Rect.& VEGAS  \\ 
\hline\hline
$\rho_a(x)$ (diagonal ridge)    &0.93 &   0.93  &   0.86 & 0.03 \\
$\rho_b(x)$ (circular ridge)    &0.82 &   0.82  &   0.82 & 0.16 \\
$\rho_c(x)$ (edge of square)    &0.57 &   1.00  &   1.00 & 0.53 \\
\hline
\hline
Functions at  3-dimens.      &{\tt Foam} 1.01& Simpl.&  H-Rect.& VEGAS  \\ 
\hline\hline
$\rho_a(x)$ (thin diagonal)    & 0.67 &   0.74  &   0.66 & 0.002 \\
$\rho_b(x)$ (thin sphere)      & 0.36 &   0.47  &   0.53 & 0.11 \\
$\rho_c(x)$ (surface of cube)  & 0.37 &   0.95  &   1.00 & 0.30 \\
\hline
\end{tabular}
\caption{\sf
  Efficiencies $\langle w \rangle/w^{\varepsilon}_{\max}$ for $\varepsilon=0.0005$.
  Functions $\rho_x(x)$ are the same as in Ref.~\cite{foam1:2000}.
  Results from {\tt Foam} are for 5000 cells (2500 active cells) 
  and cell exploration is done for a modest 200 MC events/cell.
  }
  \label{tab:NumOld}
\end{table}

\begin{table}[ht!]
\centering
\begin{small}
\begin{tabular}{|r|r|r|r|r|r|r|r|r|r|r|r|}
\hline
 {\footnotesize }&  $k$&   $n$&  $N_c$& $N_{s}$&  $N_{b}$&  $\frac{N_{eff}}{bin}$& 
 {\footnotesize nCall}&
 $\frac{\sigma}{\langle w \rangle}$& $\frac{\langle w \rangle}{w^{\varepsilon}_{\max}}$& 
 $R \pm \Delta R$& $\Delta R/R$ \\
\hline\hline
 1&   2&  0&  1K&  1K&  4&  25&  900K&   1168.8&     0.0&  5.40726$\pm$1.99871&  0.36963\\ 
\hline
 1&   0&  2&  1K&  1K&  4&  25&  169K&   0.2272&  0.6149&  3.14121$\pm$0.00022&  7.1$\cdot 10^{-5}$\\
 2&   0&  2&  1K&  1K&  4&  25&  215K&   0.2962&  0.7754&  3.14118$\pm$0.00029&  9.3$\cdot 10^{-5}$\\
\hline
 1&   0&  2&  5K&  1K&  4&  25&  656K&   0.0639&  0.8421&  3.14159$\pm$0.00006&  2.0$\cdot 10^{-5}$\\
 1&   0&  2& 10K&  1K&  4&  25& 1174K&   0.0487&  0.8877&  3.14156$\pm$0.00005&  1.5$\cdot 10^{-5}$\\
\hline
 1&   0&  2&  1K& 10K&  4&  25&  849K&   0.1479&  0.5920&  3.14118$\pm$0.00014&  4.6$\cdot 10^{-5}$\\
 1&   0&  2&  5K& 10K&  4&  25& 1457K&   0.0606&  0.8354&  3.14150$\pm$0.00006&  1.9$\cdot 10^{-5}$\\
\hline
 1&   0&  2&  1K&  2K&  8&  25&  621K&   0.0606&  0.8354&  3.14195$\pm$0.00026&  8.4$\cdot 10^{-5}$\\
 1&   0&  2&  1K&  8K&  8& 100& 1671K&   0.1048&  0.6652&  3.14168$\pm$0.00010&  3.3$\cdot 10^{-5}$\\
\hline
\end{tabular}
\end{small}
\caption{\sf Numerical results of {\tt Foam} for 2-dimensional distribution of Eq.~(\ref{eq:gallix})
  for $\mu=10^{-6}$.
  Variation of the configuration parameters:
  $k=${\tt kDim}, $n=${\tt nDim}, $N_c=${\tt nCells} (No. of function calls), 
  $N_b=${\tt nBin}, $\frac{N_{eff}}{bin}=${\tt EvPerBin}.
  In the first column we mark the type of the weight optimization {\tt OptDrive=1,2},
  for variance or maximum weight reduction.
  The value of the integral $R$ and its statistical error $\Delta R$ 
  are from MC run of $N_{MC}= 10^{7}$ events.
  $w^{\varepsilon}_{\max}$ is for $\varepsilon=0.0005$.
  {\tt nCalls} is the total number of the function calls in the foam build-up.
  }
  \label{tab:NumeResult2}
\end{table}

\subsection{Comparison with {\tt Foam} 1.x and classic {\tt VEGAS}}

In Table~\ref{tab:NumOld} we update the comparison of 
{\tt Foam} and {\tt VEGAS} of Ref.~\cite{foam1:2000},
adding results for the new hyperrectangular option.
The simplical results are now clearly improved with respect to Ref.~\cite{foam1:2000},
because of the better cell division algorithm.
Generally, the hyperrectangular cell mode provides as good an efficiency as the simplical one.
However, one should keep in mind that {\tt Foam} with hyperrectangular cells is a factor
of 2 or more faster in the execution.

\subsection{Example of sharply peaked distribution}
In Table~\ref{tab:NumeResult2} we examine the dependence of the MC efficiency/error
on the various input configuration parameters of {\tt Foam}.
All these numerical results are for the distribution
\begin{equation}
  \label{eq:gallix}
  \rho_g(x)= \frac{ \mu x_2}{ (x_1+x_2-1)^2 +\mu^2 },
\end{equation}
which, for $\mu=10^{-6}$, has a very sharp ridge across the diagonal $x_1+x_2=1$.
This distribution is taken from Refs.\cite{dice:1992,dice:1996} and is related to the photon
distribution at high energy electron--positron colliders.

\begin{figure}[!ht]
\begin{center}
{\epsfig{file=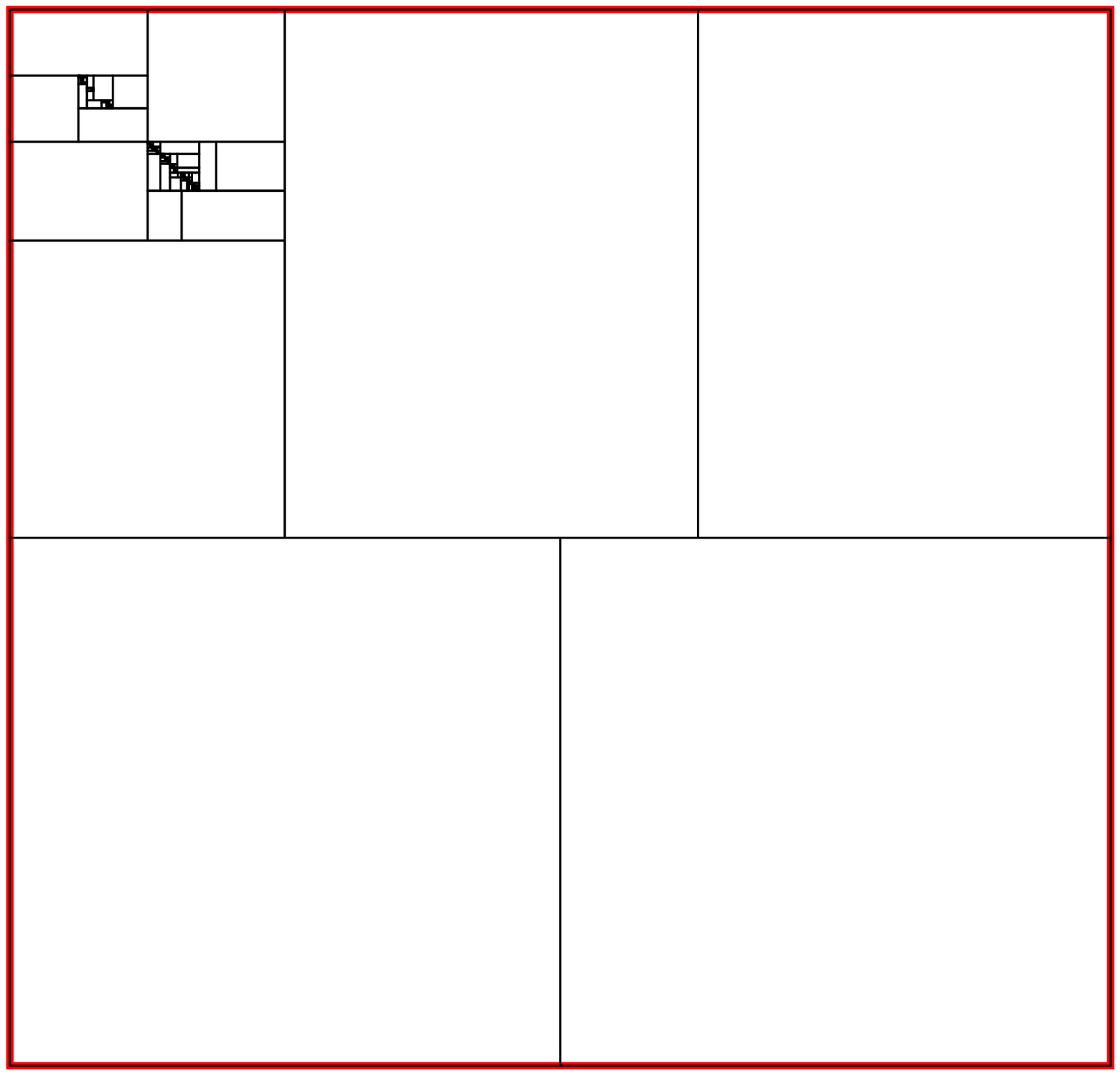,height=75mm}}
{\epsfig{file=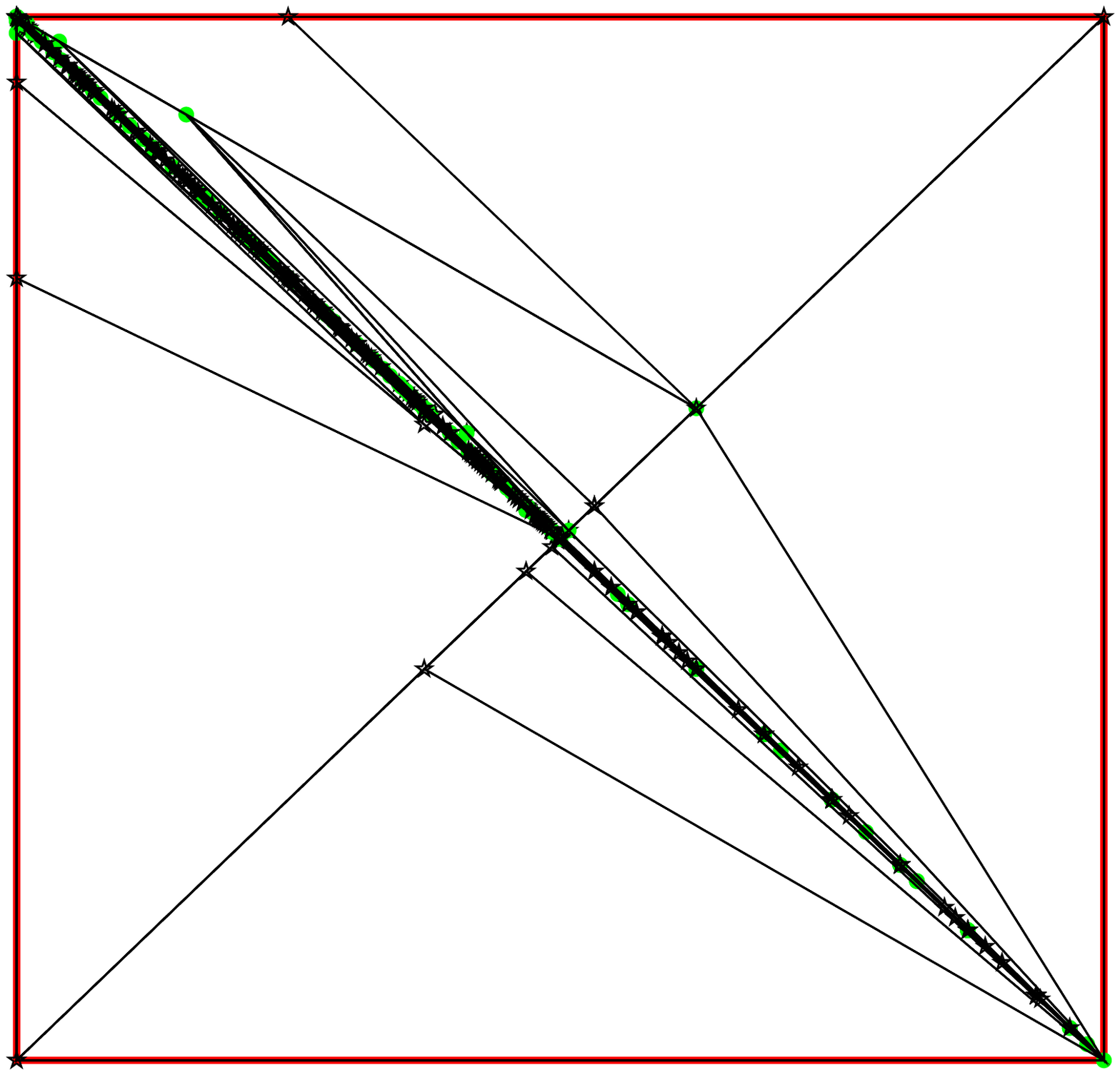,height=75mm}}
\end{center}
\caption{\small\sf
  Rectangular and triangular foam of 1000 cells for the distribution of  Eq.~(\ref{eq:gallix}).
}
\label{fig:asterix}
\end{figure}
What can we learn from the results in Table~\ref{tab:NumeResult2}?
First of all, in the first line,
we see a spectacular failure of {\tt Foam} with rectangular cells%
\footnote{Setting the number of the MC events
  in a single rectangle exploration to $10^4$ cures the problem partly.}.
The value of the integral is wrong by a factor of 2 and the statistical error is underestimated.
This illustrates the problem of the the lack of ``angular mobility'' of the rectangular cells
discussed in Section~\ref{sec:hr_vs_simp};
rectangles are unable to align with the singularity along the diagonal.

This we illustrate in the left plot of Fig.~\ref{fig:asterix}, for rectangular 1000 cells,
where we clearly see ``blind spots''.
In the right-hand plot the triangular cells in the foam are clearly aligned with the diagonal ridge.
In the rows 2 and 3 of Table~\ref{tab:NumeResult2} we see the reasonable numerical results
for the triangular foam, which
are for the maximum weight reduction and variance reduction options respectively;
the other configuration parameters are rather close to the default ones.
In rows 4 and 5 we are playing with the increase of the cell number and in the rows 6 and 7 with
the number of MC events used in the cell exploration.
Finally in rows 8 and 9 we change the binning of the histograms used in the MC cell exploration.
As we see, the most profitable in terms of MC efficiency/precision is the increase
of the number of cells however, adjusting other parameters can also help.
In all cases we display  variable {\tt nCalls}, the number
of calls of the density distribution function during the foam build-up.
In the best result of line 5 with 10000 triangular cells
we have obtained a 5-digit precision for about $10^7$ function calls%
\footnote{In Ref.~\cite{Doncker:2000} the same precision for the same function was attained
  for about $10^8$ function calls.}.

Summarizing, we see from the above exercise that the user of {\tt Foam} has a possibility
to adjust several configuration parameters, so that the MC efficiency for a given distribution
is improved quite significantly.

\begin{table}[ht!]
\centering
\begin{small}
\begin{tabular}{|rr|r|r|r|r|r|r|r|r|r|r|r|}
\hline
 \multicolumn{ 2}{|r|}{\scriptsize MAPPING}
 &  $k$&   $n$&  $N_c$& $N_{s}$&  $N_{b}$&  $\frac{N_{eff}}{bin}$& 
 {\footnotesize nCall}&
 $\frac{\sigma}{\langle w \rangle}$& $\frac{\langle w \rangle}{w^{\varepsilon}_{\max}}$& 
 $\Delta R/R$& {\footnotesize SIZE} \\
\hline\hline
(a1)& OFF& 8&    0&      1&    1K&     4&     25&     518&  2.1555&   0.038&   0.00481&  15KB\\ 
(a2)&  ON& 9&    0&      1&    1K&     4&     25&     218&  1.1391&   0.115&   0.00254&  15KB\\ 
\hline
(b1)& OFF& 8&    0&     20&    1K&     4&     25&    5767&  1.1847&   0.130&   0.00264&  15KB\\ 
(b2)&  ON& 9&    0&     20&    1K&     4&     25&    3548&  0.7626&   0.206&   0.00170&  15KB\\ 
\hline
(c0)& OFF& 0&    8&   1000&    1K&     4&     25&    487K&  1.6603&   0.085&   0.00371&  54KB\\ 
(c1)& OFF& 8&    0&   1000&    1K&     4&     25&    145K&  0.5298&   0.359&   0.00118&  53KB\\ 
(c2)&  ON& 9&    0&   1000&    1K&     4&     25&    125K&  0.7626&   0.394&   0.00104&  53KB\\ 
\hline
(d1)& OFF& 8&    0&   5000&    1K&     4&     25&    528K&  0.4330&   0.438&   0.00096& 209KB\\ 
(d2)&  ON& 9&    0&   5000&    1K&     4&     25&    596K&  0.4037&   0.467&   0.00090& 209KB\\
\hline\hline
\end{tabular}
\end{small}
\caption{\sf Numerical results from {\tt Foam} simulation/integration
  for the decay process $\tau\to \nu \pi^- \pi^+ \pi^-$, according to the matrix element squared
  of the {\tt TAUOLA} MC event generator~\cite{Jadach:1993hs,Jezabek:1992qp}.
  All MC averages are for 200K events generated after {\tt Foam} 
  initialization with the configuration parameters given in the table. 
  The notation and the meaning of the quantities are the same as in previous tables.
  The size of the ROOT disk file, in which the {\tt Foam} object was written is indicated
  in the last column.
  }
  \label{tab:NumeResult3}
\end{table}

\subsection{Decay of the $\tau$ lepton into 3 pions}
In Table~\ref{tab:NumeResult3} we collect numerical results
for an example of the application of {\tt Foam} to the very practical problem
of the MC simulation of the decay $\tau\to \nu \pi^- \pi^+ \pi^-$,
according to the distribution (matrix element squared) taken from 
the MC program {\tt TAUOLA}~\cite{Jadach:1993hs,Jezabek:1992qp}.

\begin{figure}[!ht]
\begin{center}
{\epsfig{file=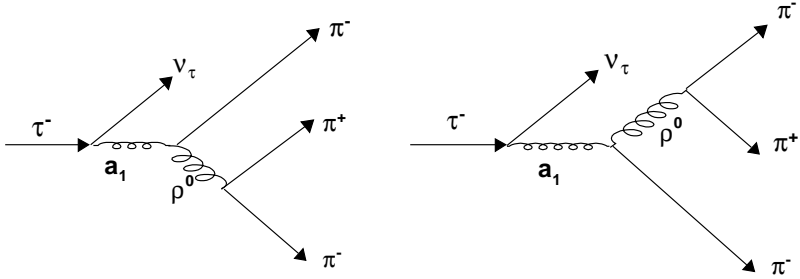,height=50mm}}
\end{center}
\caption{\small\sf
  Feynman diagrams for $\tau$ decay into 3 pions.}
\label{fig:taudec}
\end{figure}

\begin{figure}[!ht]
\begin{center}
{\epsfig{file=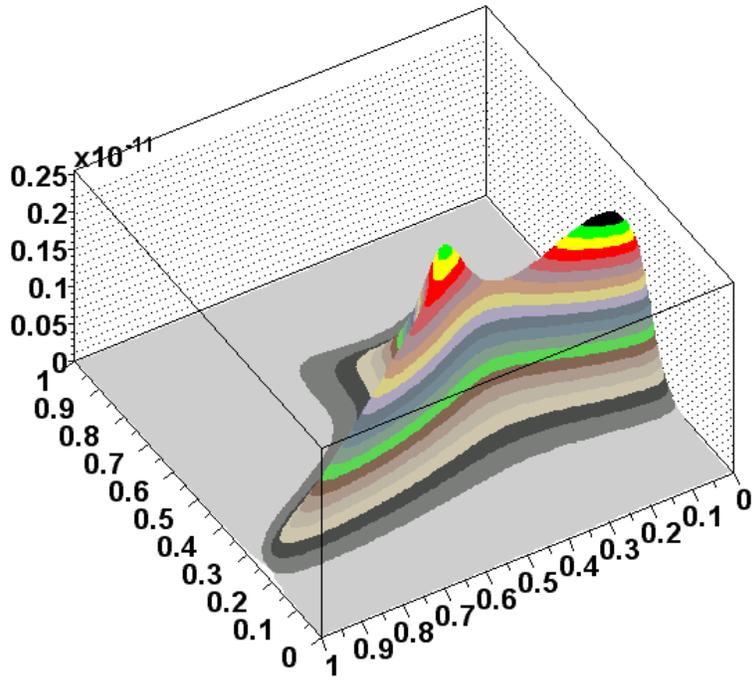,height=105mm}}
\end{center}
\caption{\small\sf
  The distribution of $\tau\to\nu 3\pi$ decay
  as a function of $x_1$ and $x_2$ (without mapping),
  fixing the other variables to some values.
  }
\label{fig:taudis}
\end{figure}

The amplitude of the decay process contains two distinct parts 
due to two Feynman diagrams, see Fig.~\ref{fig:taudec}, 
which have peaks due to the $a_1$ resonance and the $\rho$ resonance.
There are two peaks due to $\rho$ resonances that partly overlap in 
the Lorentz invariant phase-space,
such that the actual shape of the differential distribution is rather complicated.
We took for this exercise subroutine {\tt DPHTRE} of {\tt TAUOLA},
in which nine random numbers%
\footnote{Including two random numbers of the subroutine {\tt SPHERA}
  and two (Euler) angles corresponding to the overall rotation of the entire event.}
are replaced by the nine variables of {\tt Foam}.
The 4-particle phase-space is 8-dimensional.
The ninth variable is due to two branches
in the phase-space parametrization of {\tt TAUOLA}, and in case of {\tt Foam} as well 
(the method is similar to that of Section~\ref{sec:mbranching}).
In cases of ``no mapping and no multibranching'' we are back to eight dimensions.
The variables $x_1$ and $x_2$ of {\tt Foam} represent (up to a linear transformation)
the two effective masses of the $3\pi$ and $2\pi$ systems.
The next four variables $x_i,i=3,4,5,6$ are
the polar variables $\cos\theta$ and $\phi$ of the pions in the rest frame
of the $3\pi$ and $2\pi$ systems 
-- this is a completely standard phase-space parametrization,
see Ref.~\cite{Jezabek:1992qp}, and also Ref.~\cite{Skrzypek:1999td}.
The variables $x_7$ and $x_8$ are reserved for the overall rotation,
and the last one, $x_9$, is mapped into branch index $j=1,2$,
see Section~\ref{sec:mbranching}.
Variables $x_7$ and $x_8$  are inhibited (no cell-split in them),
because the distribution does not depend on them.
Variable $x_9$ (if present) has a predefined division value equal to $0.5$ and is
inhibited for the division (see Section~\ref{sec:mbranching}).
In Fig.~\ref{fig:taudis} we show the decay distribution 
as a function of $x_1$ and $x_2$ (without mapping),
fixing the other variables to some values.
The distribution is clearly a non-trivial one.

In Table~\ref{tab:NumeResult3} we show the MC efficiency of {\tt Foam}
for a gradually increasing number of hyperrectangular cells,
with the mapping compensating for the Breit--Wigner peaks
of the $a_1$ and $\rho$ resonances (as in Ref.~\cite{Jezabek:1992qp}) and without.
One example with simplical cells is also included.

The most striking result in Table~\ref{tab:NumeResult3} is the comparison of
lines (a2) and (b1): the {\tt Foam} algorithm {\em with only 20 cells} is performing
equally well as the doubly-branched mapping compensating for the resonance peaks
of the $a_1$ and $\rho$.
When going to higher number of cells,
the MC efficiency in the cases with and without mapping becomes almost the same.
This is expected, since {\tt Foam} also does the mapping compensating
for the resonances on its own.
From row (c0) we see also that the simplical mode of
{\tt Foam} is clearly underperforming.
We think that {\tt Foam} 
configured with 1000 cells, see row (c1) in Table~\ref{tab:NumeResult3}, 
is an economic solution to this particular MC problem%
\footnote{The net profit with respect to {\tt TAUOLA}~\cite{Jadach:1993hs,Jezabek:1992qp}
  would  be a three times faster program,
  and more importantly, a significantly simpler code.}.
(Mapping due to resonances is not really necessary in this case.)

\vfill\newpage
\begin{figure}[!ht]
\begin{center}
{\epsfig{file=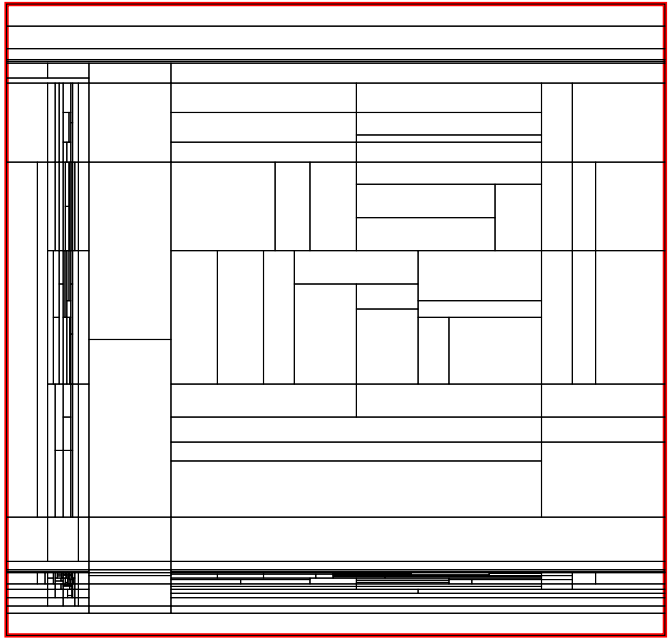,height=105mm}}
\end{center}
\caption{\small\sf
  {\tt Foam} of rectangular cells for the electron--positron beamstrahlung 
  spectrum of {\tt circe2}~\cite{circe2} with 500 cells.}
\label{fig:beastee}
\end{figure}

\subsection{Beamstrahlung spectrum}
In Fig.~\ref{fig:beastee} we show the foam of rectangular cells for
the 2-dimensional beamstrahlung spectrum $D(z_1,z_2)$ of the  electron--positron
collider\cite{Aguilar-Saavedra:2001rg}
at 500~GeV, encoded in the program {\tt circe2} of Refs.~\cite{Ohl:1997fi,circe2}.
It should be stressed that this spectrum is not known analytically but rather
through a numerical fit 
to results of the machine simulation or (in the future) from an experiment.
In order to avoid (integrable) infinite singularities at $z_i=1$ in $D(z_1,z_2)$ 
we use the variables $t_i= (1-z_i)^{0.1},\;\; i=1,2$ in Fig.~\ref{fig:beastee}.
For this exercise we used foam of 500 cells getting the MC efficiency
$\sigma/\langle w \rangle = 0.41$ and $\langle w \rangle/w^\varepsilon_{\max}= 0.64$
(for $\varepsilon = 0.0005$);
enough for practical application (can be easily improved by adding more cells).

Generating $D(z_1,z_2)$ is not really such a very much important and difficult problem.
A more interesting problem is to generate the distribution
$D(z_1,z_2)\sigma(s z_1z_2)$,
where $\sigma(s)$ is the cross section of some physics process, which may have
a strong singularity of its own, such as a resonance or threshold factor.
Such a problem was already treated with the help of the {\tt Foam} program
in KKMC~\cite{Jadach:1999vf} event generator and the study of Ref.\cite{Blair:2001}.

\newpage
\section{Conclusions }

The author hopes that this new adaptive tool for constructing efficient MC programs
will find its way to many applications in high energy physics and beyond.
The main points on the new {\tt Foam} algorithm and the program are the following:
\begin{itemize}
\item
  {\tt Foam} is a versatile {\em adaptive, general-purpose} Monte Carlo simulator.
\item
  The {\tt Foam} algorithm is based on the {\em cellular division} of the integration domain.
\item
  The geometry of the ``foam of cells'' is rather simple, cells of
  {\em simplical and/or hyperrectangular} shape are constructed in the process of a binary split.
\item
  It works in principle for arbitrary distribution --
  {\em no assumption of factorizability} as in {\tt VEGAS} of Ref.\cite{Lepage:1978sw}.
\item
  {\tt Foam} is {\em reducing maximum weight} of the weight distribution,
  it can therefore provide unweighted events.
  The {\em variance reduction}, useful for the integration
  and generating weighted events, is also available.
\item 
  Encoding cells  with {\em memory-efficient methods} allows up to $\sim 10^6$ cells
  to be built in the computer memory of a typical desktop computer.
\item
  The rules for choosing the next cell for division and the division geometry
  are based on rather sophisticated {\em analysis of the projection on the cell edges}.
  This costs CPU time, which becomes the main barrier towards higher MC efficiency.
\item
  {\tt Foam} can deal efficiently with rather strongly peaked distributions,
  up to relatively high dimensions, $\sim 12$ dimensions,
  with today's desktop computers.
\end{itemize}

{\bf\large  Acknowledgements }
\vspace{1mm}

I would like to thank  T.~Ohl, A.~Para, W.~P\l{}aczek, E.~Richter-W\c{a}s,
F.~Tkachov and Z.~W\c{a}s for interesting and useful discussions
and R.~Brun for help in getting persistency using ROOT.
The support and warm hospitality of the CERN Theory Division and DESY Zeuthen is kindly
acknowledged.
The useful assistance of the Parasoft Company in debugging the C++ code
(with the {\tt Insure++} tool) is also acknowledged.

\vfill\newpage
\appendix
\section{Variance optimization}
Suppose we have already constructed the cells $\omega_1,\omega_1,\dots,\omega_N$
and within each cell we have defined the function $\rho'(x)$ constant over the cell,
$\rho'(x)= \rho'_I$.
The integral $R'_I=\int_{\omega_i} \rho'(x) dx^n = \rho'_I V_I$ is known, because
the volume of the cell $V_I$ is known.
The function $\rho(x)$ is not, in general, constant over the cell and the weight
$w=\rho(x)/\rho'(x)$ is used to determine its integral in the usual way:
$R=R' \langle w \rangle_{\rho'}= \sum_I  \langle w \rangle_{\rho'_I} R_I$,
where the average
\begin{displaymath}
  \langle a \rangle_{\rho'_I} =  \frac{1}{R'_I} \int_{\omega_I} \rho'_I(x) a(x) dx^n
\end{displaymath}
is defined for the $I$-th cell alone.

The question is now the following: preserving the geometry of the cells,
can we get smaller variance by simply changing
the probabilities of the generation of the cells?
Rescaling $\rho'_I \to \rho''_I =\lambda_I \rho'_I$
does not affect the integral $R$, because the change of the normalization
of $R'$ and $\langle w \rangle_{\rho'}$ is cancelling.
It is convenient to assume that the above rescaling preserves $R'=R''$
and the total average weight $\langle w \rangle_{\rho'}=\langle w \rangle_{\rho''}$,
that is $\lambda_I$ obey the constraint $\sum_I R'_I = \sum_I R'_I \lambda_I =$~const.
With the above constraint in mind, we now ask: for what values of $\lambda_I$ is
the dispersion of the total weight
$\sigma^2 = \langle w^2 \rangle_{\rho''} -\langle w \rangle_{\rho''}^2 $
minimal?%
\footnote{ This problem was, of course,  often considered in the past, see for instance
  Refs.~\cite{Lepage:1978sw,Kleiss:1994qy}.
  We outline here the solution for the sake of completeness and convenience of the reader.}
Since by construction 
$\langle w \rangle_{\rho''}$ is independent of $\lambda_I$,
we may look only for a minimum of $\langle w^2 \rangle_{\rho''}$.
With the standard methods we get a (local) minimum condition:
\begin{equation}
  \frac{\partial}{\partial \lambda_I} 
  \Big\{ \langle w^2 \rangle_{\rho''} + \Lambda R''  \Big\}
  =\frac{\partial}{\partial \lambda_I}
  \Bigg\{  \frac{1}{R'}
  \sum_I \int_{\omega_I}
  \left(\frac{\rho(x)}{\rho'_I\lambda_I}\right)^2 \rho'_I \lambda_I dx^n
  +\Lambda \left( \sum_I R'_I \lambda_I \right)
  \Bigg\}=0,
\end{equation}
where $\Lambda$ is the Lagrange multiplier.
The solution of the minimum condition
\begin{equation}
  \frac{R'_I}{R'} \langle w^2 \rangle_{\rho'_I} \frac{1}{\lambda_I^2} -\Lambda R'_I = 0
\end{equation}
is simply 
$\lambda_I \simeq {\rm const} \times \sqrt{\langle w^2 \rangle_{\rho'_I}}$,
or more precisely
\begin{equation}
  \label{eq:lambda}
  \lambda_I =
  \frac{\sqrt{\langle w^2 \rangle_{\rho'_I}}}
       {\sum_J \frac{R'_J}{R'}  \sqrt{\langle w^2 \rangle_{\rho'_J}}}.
\end{equation}
The value of  $\langle w^2 \rangle_{\min} \equiv \langle w^2 \rangle_{\rho''}$
calculated at the minimum, that is for $\lambda_I$ of Eq.~(\ref{eq:lambda}),
is also rather simple
\begin{equation}
  \langle w^2 \rangle_{\min} 
  = \sum_I \frac{R'_I}{R'}\;  \langle w^2 \rangle_{\rho'_I}\; \frac{1}{\lambda_I^2}
  = \left( \sum_J \frac{R'_J}{R'}  \sqrt{\langle w^2 \rangle_{\rho'_J}} \right)^2
  = \left( \left\langle \sqrt{\langle w^2 \rangle_{\rho'_J}}\; \right\rangle_{\rho'} \right)^2.
\end{equation}
Let us note that the functional $ \sqrt{\langle w^2 \rangle_{\min}}$,
which we intend to minimize in the process of the evolution of the foam (cell split)
is a simple sum of contributions from all cells.
Consequently, when working out details of
the split of a given cell we may calculate the gain
in terms of the total dispersion independently of other cells, see also below.
This very convenient feature is exploited in the algorithm of the foam build-up.

Adopting (temporarily) the following normalization conventions:
$\rho'_I=1$, $w=\rho(x)$ and
$R'_I=V_I$, $\lambda_I \simeq \sqrt{ \langle \rho^2 \rangle}_I$,
we get
\begin{equation}
  \label{rhoprim}
  \rho''_I = \sqrt{ \langle \rho^2 \rangle_I},\qquad
  R''_I= V_I \sqrt{ \langle \rho^2 \rangle_I},
\end{equation}
where the average $\langle \dots  \rangle_I$ 
is understood as defined for points uniformly distributed within the $I$-th cell.

In the {\tt Foam} algorithm we
do not go, of course, through a judicious adjustment of the relative importance
of the cells, which minimizes the variance as described above;
but instead, we simply declare that
the distribution $\rho'(x)$ is defined by the optimum solution of Eq.~(\ref{rhoprim}).
Once we have done that, for the MC weight $w$ defined with respect to such a new $\rho'(x)$,
we find out that $\langle w^2 \rangle_I=1$ for each cell
and, also for all cells, $\langle w^2 \rangle=1$.
What is then minimized in the process of the cell division is the
ratio of the variance to the average weight%
\footnote{ We exploit here the relation $\langle w \rangle = R/R'$,
  and we should keep in mind that $R$ is constant during the variance minimization.}
\begin{equation}
  \frac{\sigma^2}{\langle w \rangle^2} = \left(\frac{R'}{R}\right)^2-1.
\end{equation}
This quantity is not so convenient to optimize in the process of the cell split (foam evolution)
and we rather chose to minimize a closely related `linearized'' quantity
\begin{equation}
  R_{loss}= R\left( \sqrt{\frac{\sigma^2}{\langle w \rangle^2}+1} -1  \right)
  = \sum_I V_I( \sqrt{ \langle w^2 \rangle_I} - \langle w \rangle_I)
  = \int \rho_{loss}(x) dx^n
  = R'-R,
\end{equation}
which is a sum of contributions from all cells and is a monotonous 
ascending function of $\sigma/\langle w \rangle$.
In the process of the cell division $\omega_I\to \omega_{Ia} \oplus \omega_{Ib}$,
we decrease $\sigma/\langle w \rangle$ step by step, 
by playing with the geometry of the cell split,
so that the gain in the total $R_{loss}=R'-R$  due to a given cell split is as big as possible.
It is convenient that
the contributions from the other cells to $R_{loss}$ are unchanged.
In this way every cell split will lead to a smaller and smaller $\sigma/\langle w \rangle$
in the final MC run.

\vfill
\newpage
\section{Output of the demonstration program in C++}

{\footnotesize
\begin{verbatim}
FFFFFFFFFFFFFFFFFFFFFFFFFFFFFFFFFFFFFFFFFFFFFFFFFFFFFFFFFFFFFFFFFFFFFFFFFFFFFFFF
F                                                                              F
F                   ****************************************                   F
F                   ******     TFOAM::Initialize      ******                   F
F                   ****************************************                   F
F                                                      FoamX                   F
F    Version =            2.05   =               Release date:  2002.02.13     F
F       nDim =          0 =  Dimension of the simplical sub-space              F
F       kDim =          2 =  Dimension of the hyper-cubical sub-space          F
F     nCells =       1000 =  Requested number of Cells (half of them active)   F
F     nSampl =        500 =  No of MC events in exploration of a cell cell     F
F       nBin =          8 =  No of bins in histograms, MC exploration of cell  F
F   EvPerBin =         25 =  Maximum No effective_events/bin, MC exploration   F
F   OptDrive =          2 =  Type of Driver   =1,2 for Sigma,WtMax             F
F    OptEdge =          0 =  Decides whether vertices are included in the MC   F
F    OptPeek =          0 =  Type of the cell Peek =0,1 for maximum, random    F
F     OptOrd =          0 =  Root cell hyp-cub. or simplex, =0,1               F
F   OptMCell =          1 =  MegaCell option, slim memory for hyp-cubes        F
F   OptDebug =          1 =  Additional debug histogram, SetDirectory(1)       F
F   OptCu1st =          1 =  Numbering of dimensions  starts with h-cubic      F
F     OptRej =          0 =  MC rejection on/off for OptRej=0,1                F
F   MaxWtRej =               2   =       Maximum wt in rejection for wt=1 evts F
F                                                                              F
FFFFFFFFFFFFFFFFFFFFFFFFFFFFFFFFFFFFFFFFFFFFFFFFFFFFFFFFFFFFFFFFFFFFFFFFFFFFFFFF
\end{verbatim}

\begin{verbatim}
FFFFFFFFFFFFFFFFFFFFFFFFFFFFFFFFFFFFFFFFFFFFFFFFFFFFFFFFFFFFFFFFFFFFFFFFFFFFFFFF
F                                                                              F
F                   ***  TFOAM::Initialize FINISHED!!!   ***                   F
F     nCalls =     246799 =            Total number of function calls          F
F     XPrime =       1.2748942   =     Primary total integral                  F
F     XDiver =      0.27441356   =     Driver  total integral                  F
F   MCresult =       1.0004806   =     Estimate of the true MC Integral        F
F                                                                              F
FFFFFFFFFFFFFFFFFFFFFFFFFFFFFFFFFFFFFFFFFFFFFFFFFFFFFFFFFFFFFFFFFFFFFFFFFFFFFFFF
\end{verbatim}
\vfill

\newpage
\begin{verbatim}
FFFFFFFFFFFFFFFFFFFFFFFFFFFFFFFFFFFFFFFFFFFFFFFFFFFFFFFFFFFFFFFFFFFFFFFFFFFFFFFF
F                                                                              F
F                   ****************************************                   F
F                   ******      TFOAM::Finalize       ******                   F
F                   ****************************************                   F
F     NevGen =       2000 = Number of generated events in the MC generation    F
F     LastVe =         -1 = Number of vertices (only for simplical option)     F
F     nCalls =     248799 = Total number of function calls                     F
F                   ----------------------------------------                   F
F      AveWt =      0.78836949   =     Average MC weight                       F
F      WtMin =   1.3022085e-06   =     Minimum MC weight (absolute)            F
F      WtMax =       1.0227719   =     Maximum MC weight (absolute)            F
F                   ----------------------------------------                   F
F     XPrime =       1.2748942   =     Primary total integral, R_prime         F
F     XDiver =      0.27441356   =     Driver  total integral, R_loss          F
F                   ----------------------------------------                   F
F      IntMC =       1.0050877 +-    0.0052357675  = Result of the MC Integral F
F   MCerelat =    0.0052092644   =     Relative error of the MC intgral        F
F  <w>/WtMax =      0.80037512   =              MC efficiency, acceptance rate F
F  Sigma/<w> =      0.23296539   =              MC efficiency, variance/ave_wt F
F      WtMax =           0.985   =              WtMax(esp= 0.0005)             F
F      Sigma =       0.1836628   =              variance of MC weight          F
F                                                                              F
FFFFFFFFFFFFFFFFFFFFFFFFFFFFFFFFFFFFFFFFFFFFFFFFFFFFFFFFFFFFFFFFFFFFFFFFFFFFFFFF
\end{verbatim}
\vfill
}

\newpage
%

\end{document}